\documentclass[%
aps,
nofootinbib,amsmath,amssymb,amsfonts,
superscriptaddress,twocolumn
]{revtex4-2}
\usepackage{graphicx}
\usepackage{bm}
\usepackage{xcolor}
\usepackage{txfonts}
\usepackage{comment}
\usepackage{caption}
\usepackage{subcaption}
\usepackage{graphicx,url}
\usepackage[utf8]{inputenc}  
\usepackage{verbatim}
\usepackage{listings}
\usepackage{indentfirst}
\usepackage{cancel}
\usepackage{soul}
\usepackage{xcolor}
\usepackage{bm}
\usepackage{comment}
\usepackage{caption}
\usepackage{subcaption}
\usepackage{mathtools}
 \usepackage{mathrsfs}
\usepackage{hyperref}
\usepackage{float}
\usepackage[section]{placeins}
\hypersetup{
    colorlinks=true,       
    linkcolor=red,          
    citecolor=blue,        
}

\begin{document}
\title{Gravitational radiation close to a black hole horizon: Waveform regularization and the out-going echo}
\author{Manu Srivastava}
\email{manu$_$s@iitb.ac.in}
\affiliation{Department of Physics, Indian Institute of Technology Bombay, Mumbai 400076, India} 
\author{Yanbei Chen} \email{yanbei@caltech.edu}
\affiliation{Theoretical Astrophysics 350-17, California Institute of Technology, Pasadena, California 91125}
\date{\today}
\begin{abstract}
Black hole perturbation theory for Kerr black holes is best studied in the Newman Penrose (NP) Formalism, in which gravitational waves are 
 described as perturbations in the Weyl scalars $\psi_0$ and $\psi_4$, with the governing equation being the well-known Teukolsky equation. 
 Near infinity and near the horizon, $\psi_4$ is dominated by the component that corresponds to waves propagating towards the positive radial direction, while $\psi_0$ is dominated by the component that corresponds to waves that propagate towards the negative radial direction.  
Since gravitational-wave detectors measure out-going waves at infinity, 
research has been mainly focused on $\psi_4$, leaving $\psi_0$ less studied. But
the scenario is reversed in the near horizon region where the in-going-wave boundary condition needs to be imposed. 
{For in-going waves, the components of the tidal tensor measured by observers near the future horizon depend mainly on $\psi_0$. Thus, studying} 
the near horizon phenomena, e.g., tidal heating and gravitational-wave echoes from Extremely Compact Objects (ECOs), requires computing $\psi_0$.  
In this work, we explicitly calculate the source term for the $\psi_0$ Teukolsky equation due to a point particle plunging into a Kerr black hole. We highlight the need to regularize the solution of the $\psi_0$ Teukolsky equation obtained using the usual Green's function techniques. We suggest a regularization scheme for this purpose and go on to compute the $\psi_0$ waveform close to a Schwarzschild horizon for two types of trajectories of the in-falling particle. We compare the $\psi_0$ waveform calculated directly from the Teukolsky equation with the $\psi_0$ waveform obtained by using the Starobinsky-Teukolsky identity on $\psi_4$. %
We also compute the first out-going gravitational-wave echo waveform near infinity, using the near-horizon $\psi_0$ computed directly from the Teukolsky equation, and the Boltzmann boundary condition on the ECO surface. We show that this out-going echo is quantitatively very different (stronger) than the echo obtained using previous prescriptions that did not compute the near-horizon $\psi_0$ directly using the Teukolsky equation. 
\end{abstract}
\maketitle
\section{Introduction}

\vspace*{-10pt}
Detection of gravitational waves \cite{GW150914,GW170817} produced due to mergers of compact astrophysical objects has opened up a new approach towards testing the nature of black holes, in particular the existence of the event horizon, as well as possible deviations from Kerr geometry in the space-time region near the horizon. With these deviations, the gravitational wave sources are not exactly Kerr black holes but some other exotic compact objects (ECO) whose space-times are identical to Kerr black holes except in the region very close to the horizon. The existence of ECOs has been motivated based on effects of Quantum Gravity, exotic matter equations of state, phase transitions etc. in \cite{palenzuela2017gravitational,mazur2004gravitational,pani2009gravitational,giddings2018event,bianchi2020distinguishing,bena2020multipole,mukherjee2020multipole}. For a compact binary coalescence that results in an ECO as its remnant object, the absence of horizon gives rise to `echoes' in the outgoing gravitational waves that appear after the main GR wave~\cite{Xin:2021zir,wang2020echoes,Maggio,micchi2021loud,Cardoso:2016oxy,Mark:2017dnq,Conklin:2017lwb,Maselli:2017tfq,Westerweck:2017hus,Wang:2018gin,Urbano:2018nrs,Lo:2018sep,Nielsen:2018lkf,Tsang:2019zra,Chen:2019hfg,Conklin:2019fcs,LongoMicchi:2020cwm}. A complementary way way to model gravitational waves from ECOs, by parametrizing the compactness of objects, is proposed in \cite{Toubiana:2020lzd}. During the inspiral stage, the absence of horizon in ECOs also modifies the tidal interaction within the binary, and leads to additional signatures in gravitational waves~ \cite{fang2005tidal,li2008generalization,Datta:2019epe,Datta:2020rvo,Chakraborty:2021gdf}. 

In this paper, we shall use Exotic Compact Object, or ECO, to refer to the black-hole candidate that we are studying.   Two related ways of probing the ECO are: (i) to study tidal interactions between an ECO and a companion that spirals around it, and (ii) to search for gravitational wave echoes from the ECO after the plunge of the companion into the ECO.  In both ways, one can apply black-hole perturbation theory to the ECO, assuming that space-time geometry outside the ECO is well described by the Kerr geometry, except for a modified boundary condition on a surface that floats above the horizon.

\vspace*{-2pt}
Perturbations about the Kerr geometry are best studied in the Newman Penrose Formalism \cite{Newman:1961qr}.  Adopting the Kinnersley tetrad, gravitational waves in Kerr geometry are described in terms of perturbations in the Weyl scalars $\psi_0$ and $\psi_4$.  Teukolsky obtained equations~\cite{Teukolsky:1973ha} that describe the radial and angular dependence of the perturbation waveforms.  For vacuum solutions (pure gravitational waves), $\psi_0$ and $\psi_4$ contain the same information about gravitational waves propagating in Kerr~\cite{wald1973perturbations}.  However, both for near null infinity and near the horizon, $\psi_4$ dominate over $\psi_0$ for gravitational waves that propagate along the $+r$ direction, while $\psi_0$ dominate over $\psi_4$ for waves that propagate along the $-r$ direction.  For studies that focus on gravitational waves generated by compact binary coalescence that propagate toward future null infinity, it is natural to compute $\psi_4$.
In further methods  developed to solve the Teukolsky equation more efficiently, e.g. the Sasaki-Nakamura (SN) and the Mano-Suzuki-Takasugi (MST) formalisms \cite{Sasaki:1981sx,Mano:1996vt} were eventually~\cite{Hughes:2000pf} developed for fields with all values of spin, thus incorporating both $\psi_0$ and $\psi_4$.  However, the corresponding equations were mainly applied for $\psi_4$ calculations in the literature since. 
E.g. in Ref~\cite{Sago:2020avw}, the authors have presented the scheme to calculate the in-going $\psi_4$ waveform due to a mass plunging into a Kerr black hole based on the Sasaki-Nakamura formalism. {Such explicit calculations have not been done for in-going $\psi_0$ waves yet.} 
%
%

{As discussed in Ref. \cite{chen2020tidal}, for fiducial observers floating above the horizon (e.g., on the surface of an ECO), the transverse components of the tidal tensors that they measure depend on both  $\psi_0$ and $\psi_4$ [See Eq.~(12) of Ref.~\cite{chen2020tidal}].  For in-going waves, the $\psi_0$ contribution dominates that of  $\psi_4$.  For this reason, to study tidally-induced horizon deformations and  gravitational-wave echoes, one needs to obtain $\psi_0$ near the horizon.} 
For vacuum solutions, it is possible to obtain $\psi_0$ from $\psi_4$ from the Starobinsky-Teukolksy (ST) Identity.  This has indeed been applied by several works ~\cite{wang2020echoes,Maggio,micchi2021loud,chen2020tidal,Xin:2021zir}.    However, in situations where the companion plunges into the ECO, the ST identity may not apply --- and it is the goal of this paper to obtain $\psi_0$ for a plunging particle directly from the Teukolsky equation, and compare with the previous prescriptions that use the ST identity.

\vspace*{-2pt}
In this paper, the source term of the radial Teukolsky equation for $\psi_0$ is explicitly calculated for a point particle of mass $\mu$ plunging into a Kerr black hole. Ref.~\cite{Price_Self_Force} had explored the source term for the simple case of circular trajectories in Schwarzschild background, in the context of metric reconstruction,  but expression for the $\psi_0$ source term for a general geodesic trajectory in an arbitrarily rotating Kerr background has not been available in the literature so far. 
%

Further, it is well known that computing solutions of the Teukolsky equation using the usual Green's Function approach can lead to non-convergent integrals, when the source term does not vanish fast enough near infinity or near the horizon. 
Poisson \cite{Poisson:1996ya} highlighted such an issue for computing $\psi_4$ far away from a Schwarzschild black hole; he subsequently resolved this issue by introducing a regularization prescription.   After obtaining the source term for $\psi_0$ generated by a particle moving along a geodesic orbit in Kerr background, we show that a similar non-convergence issue occurs when we try to compute $\psi_0$ for Kerr geometry.  We introduce a regularization method in which we first insert a $\psi^A_{0}$ that satisfies the in-going boundary condition near the horizon and the out-going boundary condition near infinity,  but corresponds to a source term $S^A$ that only coincides with the source term $S$ generated by our plunging particle for up to two leading orders in $(r-r_+)$.  We then use the usual Green function approach to obtain $\psi_0 -\psi^A_{0}$ from the regular source term $S- S^A$. 

With our regularization method, we go on to compute the $\psi_0$ propagating towards horizon for two kinds of trajectories of the plunging particle in Schwarzschild geometry: a radial in-fall and a quasi-circular plunge from the Effective One-Body (EOB) formalism.  [We emphasize that, even though we have restricted to the simple Schwarzschild case, the source term we obtain and the regularization approach both apply to the general Kerr spacetime.]  We will then compare our directly-obtained horizon-going $\psi_0$  with those obtained by applying the ST identity on $\psi_4$.   We finally apply our horizon-going $\psi_0$ to the computation of gravitational-wave echoes, by using formalism developed by Ref.~ \cite{chen2020tidal}. These echoes will be compared with those obtained from $\psi_4$ and the ST identity. 
%
%
%

This paper is organized as follows. {In Section \ref{sec:SourceTermSetup}, after a brief introduction to the Teukolsky equation, we explicitly calculate the source term of the $\psi_0$ Teukolsky equation for a point particle of mass $\mu$ plunging into a Kerr black hole. We also highlight the need to regularize the solution of the $\psi_0$ equation computed using the Green Function approach. In Section \ref{sec:Ansatz_Regularization}, we introduce our scheme to regularize these solutions. In Section
\ref{sec:Psi0_Waveform_Calc} we go on to compute the $\psi_0$ waveform close to a Schwarzschild horizon for the two kinds of trajectories of the in-falling particle. In Section \ref{sec:Waveform_Comparison}, we show that the $\psi_0$ waveforms obtained directly from the Teukolsky equation are different from those obtained by using the ST identity on $\psi_4$. In Section \ref{sec:Echo_Difference_Starobinky_Transform}, we show that the first echo in the outgoing $\psi_4$ waveform computed with the aid of $\psi_0$ directly computed from the Teukolsky Equation differs from the first echo obtained using $\psi_4$ and the ST identity. Although the echoes calculated in the two ways are qualitatively similar, there are significant quantitative differences.} Throughout the paper, we use geometric units with $G=1=c$.
\section{Teukolsky Equation, its Solution and the $\psi_0$ Source Term}
\label{sec:SourceTermSetup}
\vspace*{-10pt}
In this section, after a brief overview of the Teukolsky equation, we compute the source term of the radial Teukolsky equation for $\psi_0$ due to a point particle plunging into a Kerr black hole. The well-established calculations of the source term for the $\psi_4$ Teukolsky equation can be found in Appendix \ref{sec:App_Calc_Source_Term} and in \cite{Mino:1997bx}. 
\subsection{Teukolsky Equation and Source Term}
\vspace*{-4pt}
For spherically symmetric scenarios like in the case for the Schwarzschild spacetime, 
metric perturbation (Einstein's) equations are separable in radial and angular equations, with the angular sector being described by spherical harmonics, and radial sector divided into axial and polar perturbations \cite{Regge:1957td,Zerilli,Vishveshwara:1970zz}. 
But in the case of Kerr, with spherical symmetry no longer present, separable equations were not found for metric perturbations.   However, in a groundbreaking work \cite{Teukolsky:1973ha}, Teukolsky showed that we can use a single field, either $\psi_0$ or $\psi_4$ in the Newman-Penrose formalism (in the Kinnersley tetrad), to describe gravitational perturbations, and get separate radial and angular perturbation equations for a Kerr background, the so called radial and angular Teukolsky equations.  An underlying reason for such separability was the fact that both the Schwarzschild and the Kerr solutions are  Petrov Type-D spacetimes. 

In the Newman Penrose Formalism, gravitational waves are described in terms of perturbations in the Weyl scalars $\psi_0$ and $\psi_4$. In terms of the Newman Penrose quantities, the decoupled equation \cite{Teukolsky:1973ha} for the $\psi_0$ perturbation (denoted by $\psi_0^B$) is:
\begin{widetext}
\begin{align}
\label{eqn:decoupledNPeqn}
4 \pi T_{0}= 
\Big[ \left( D-3 \epsilon+\epsilon^{*}-4 \rho-\rho^{*}\right) \left(\mathbf{\overline{\Delta}}-4 \gamma+\mu \right) - \left(\delta+\pi^{*}-\alpha^{*}-3 \beta-4 \tau \right) \left(\delta^{*}+\pi-4 \alpha \right)  -3 \psi_{2} \Big] \psi_{0}^{B} 
\end{align}
where
\begin{align}
    T_{0} &=\left(\delta+\pi^{*}-\alpha^{*}-3 \beta-4 \tau\right)\left[\left(D-2 \epsilon-2 \rho^{*}\right) T_{l m}^{B}-\left(\delta+\pi^{*}-2 \alpha^{*}-2 \beta\right) T_{l l}^{B}\right] \\
&+\left(D-3 \epsilon+\epsilon^{*}-4 \rho-\rho^{*}\right)\left[\left(\delta+2 \pi^{*}-2 \beta\right) T_{l m}^{B}-\left(D-2 \epsilon+2 \epsilon^{*}-\rho^{*}\right) T_{m m}^{B}\right]
\end{align}
\end{widetext}
The different Newman Penrose quantities  ($\epsilon$, $\rho$, $\gamma$, $\pi$, $\alpha$, $\beta$, $\tau$, $\Sigma$), and the derivative operators ($D$, $\mathbf{\overline{\Delta}}$, $\delta$), that appear in the above equation are for the background Kerr metric in Boyer Lindquist co-ordinates;  their expressions are given in Appendix \ref{sec:App_Calc_Source_Term}. The expressions for the Stress Tensor projections ($T_{l l}^{B}$, $T_{l m}^{B}$, $T_{m m}^{B}$) are also in Appendix \ref{sec:App_Calc_Source_Term}. 
When Eq. \eqref{eqn:decoupledNPeqn} is separated into radial and angular equations for a Kerr background, the solutions of the angular equation are spin weighted spheroidal harmonics $_{s}S_{lm}(a\omega,\theta)$ \cite{Press:1973zz} with $s=2$ for the $\psi_0$ case. Thus, we can expand the solution and the source in Eq. \eqref{eqn:decoupledNPeqn} in terms of ${}_{2}S_{lm}(a\omega,\theta)$ as:
\begin{equation}
\label{eqn:psi0_r_theta_decomposition}
\psi_0^B=\int d \omega \sum_{l, m} R(r) {}_{2}S_{lm}(a\omega,\theta) e^{i m \varphi} e^{-i \omega t}
\end{equation}
\begin{equation}
\label{eqn:Source_r_theta_decomposition}
    8 \pi \Sigma T_0=\int d \omega \sum_{l, m} T_{lm\omega}^{(0)}(r) {}_{2}S_{lm}(a\omega,\theta) e^{i m \varphi} e^{-i \omega t}
\end{equation} 
In the angular sector, the above decomposition leads to 
{the spin weighted spheroidal harmonics equation \cite{Press:1973zz}:
\begin{equation}
\label{eqn:Spheroidal_Harmonic_EQ}
\begin{split}
&\frac{1}{\sin \theta} \frac{d}{d \theta}\left(\sin \theta \frac{d\, {}_{s}S_{lm}(a\omega,\theta)}{d \theta}\right)+\\
&\Bigg(a^{2} \omega^{2} \cos ^{2} \theta-\frac{m^{2}}{\sin ^{2} \theta}-2 a \omega s \cos \theta-\frac{2 m s \cos \theta}{\sin ^{2} \theta}\\
&-s^{2} \cot ^{2} \theta +\mathcal{E}_{l m}-s^2\Bigg)\, {}_{s}S_{lm}(a\omega,\theta)=0
\end{split}
\end{equation}
{Here $\mathcal{E}_{lm}$  the spheroidal eigenvalue \cite{Press:1973zz} and $s=2$ for the $\psi_0$ case.} Henceforth, for simplicity, we drop the $a\omega$ dependence of $_{s}S_{lm}$.} 
\vspace{3pt}
\newline
In the radial sector the decomposition (\ref{eqn:psi0_r_theta_decomposition}, \ref{eqn:Source_r_theta_decomposition}) leads to the radial Teukolsky Equation:
\begin{align}
\label{eqn:TeukolskyEqn}
   &\frac{d}{d r}\left[\Delta^{s+1} \frac{d R}{d r}\right]
   +\left[\frac{K^{2}-2 i s(r-M) K}{\Delta}+4 i s \omega r-\lambda\right] R \nonumber\\
   =&  \,\Delta^s\,T_{lm\omega}
\end{align}
with, $s=2$, 
\begin{align}
 \Delta&=r^2+a^2-2Mr = (r-r_+)(r-r_-) \\ 
 K&=(r^2+a^2)\omega -am \\
 \lambda&=\mathcal{E}_{lm}+a^2\omega^2-2am\omega-s(s+1)
\end{align}
with $M$ and $a$ being the mass and the specific angular momentum and $r_\pm = M\pm \sqrt{M^2 -a^2}$ the position of the outer and inner horizon. Note that $\Delta=0$ defines the position of the horizon, and that near the horizon, $\Delta \sim (r-r_+)(r_++r_-)$.

 Next, we calculate the source term $T^{(0)}_{lm\omega}$ (0 in the superscript shows that it corresponds to $\psi_0$ perturbation) for a point particle (of mass $\mu$) plunging into the Kerr black hole following an arbitrary geodesic trajectory. Detailed calculation of the $\psi_0$ source term is presented in Appendix \ref{sec:App_Calc_Source_Term}. We will just outline the result here. Using orthogonality relations \eqref{eqn:Spheroidal_Harmonics_Orthogonality}, \eqref{eqn:DiracDeltaDef} in \eqref{eqn:Source_r_theta_decomposition}, we get:
\begin{equation}
\label{eqn:Source_l_m_omega}
    T_{\ell m \omega}^{(0)} (r) =4 \int d \Omega d t \rho^{-1} \rho^{*\,-1} T_0(t,r,\theta,\varphi) e^{-i m \varphi+i \omega t} \frac{_{2}S_{\ell m}(\theta)}{{2 \pi}}
\end{equation}
As shown in Appendix \ref{sec:App_Calc_Source_Term}, \eqref{eqn:Source_l_m_omega} can then be written as:
%
%
\begin{equation}
\label{eqn:Source_delta_derivative}
    \begin{split}
T^{(0)}_{\ell m \omega}(r')= \mu \int_{-\infty}^{\infty} & d t\, e^{i \omega t-i m \varphi(t)}\\
& \Big[\,\, (A_{l l 0}+A_{l m 0}+A_{m m 0}) \delta(r'-r(t))\\
&+\left\{\left(A_{l m 1}+A_{m m 1}\right) \delta(r'-r(t))\right\}_{, r'}\\
&+\left\{A_{m m 2} \delta(r'-r(t))\right\}_{, r' r'}\Big]
\end{split}
\end{equation}
where $\mu$ is the mass of the test particle the $A$ terms are defined in the Appendix \ref{sec:App_Calc_Source_Term}. Here the trajectory in the Boyer-Lindquist coordinate system is parametrized by 
\begin{equation}
    z^\mu(t) = (t,r(t),\theta(t),\varphi(t))
\end{equation}
Similar source term for $\psi_4$ perturbation is available in the literature \cite{Mino:1997bx}. Eq.\eqref{eqn:Source_delta_derivative} can further be written as (after integrating with respect to $dr$ and then renaming $r'$ as $r$):
\begin{widetext}
\begin{equation}
\label{eqn:SourceUsedInNumerics}
\begin{split}
    T^{(0)}_{lm\omega}(r)=\mu e^{i\omega t-im\varphi}&\left[ (A_{ll0}+A_{lm0}+A_{mm0})t'
    +(A_{lm1}+A_{mm1})(t''+i\omega (t')^2-imt'\varphi ')+t'\frac{d}{dr}\left(A_{lm1}+A_{mm1}\right)\right.\\
    &\left. +A_{mm2}(t'''+3i\omega t''t'-2imt''\varphi '-\omega^2(t')^3+2\omega m(t')^2\varphi '-m^2t'(\varphi')^2-imt'\varphi'')\right.\\&
    \left.+2(t''+i\omega (t')^2-im t'\varphi ')\frac{dA_{mm2}}{dr}+t'\frac{d^2(A_{mm2})}{dr^2}\right ]
\end{split}
\end{equation}
\end{widetext}
where $'$ denotes derivative with respect to $r$.  Note here that we have switched to using $r$ as the independent variable along the trajectory, with 
\begin{equation}
    z^\mu(r) = (t(r),r,\theta(r),\varphi(r))
\end{equation}
We use this form of $T^{(0)}_{lm\omega}$ for all our numerical analysis in Sections \ref{sec:Psi0_Waveform_Calc} and \ref{sec:Waveform_Comparison}.

\subsection{Solutions of Radial Teukolsky Equation- Need for Regularization}
\label{sec:Reg_Need}
In this section, we will look at general solutions to the radial Teukolsky equation \eqref{eqn:TeukolskyEqn}. 
We will also highlight that the naive solution that we expect from theory of differential equations, for example in \cite{garfken67:math}, leads to some convergence issues, and that there is a need to regularize the naive solution in order to get meaningful results.
\vspace{\baselineskip}
\newline
The general solution of the Radial Teukolsky equation \eqref{eqn:TeukolskyEqn}, with the appropriate boundary conditions (purely in-going at horizon and purely out-going at infinity), can be written as
\begin{align}
\label{eqn:GeneralSolnTE}
    R_{l m \omega}(r)&=\frac{R^{\infty}_{l m \omega}(r)}{W_{l m \omega}} \int_{r_+}^{r}  R^H_{l m \omega}(r') \Delta^{s} T_{lm\omega}(r') dr'\nonumber\\
    &+\frac{R^H_{l m \omega}(r)}{W_{l m \omega}} \int_{r}^{\infty}  R_{l m \omega}^{\infty}(r') \Delta^{s} T_{lm\omega}(r') dr'
\end{align}
where $R^H_{l m \omega}(r)$ is the solution of the homogeneous Teukolsky equation which satisfies in-going wave boundary condition at the horizon and $R^{\infty}_{l m \omega}(r)$ is the solution of homogeneous Teukolsky equation which satisfies the out-going wave boundary condition at infinity. The Wronskian $W_{lm\omega}$, defined as 
\begin{equation}
\label{eqn:WronskianGeneral}
    W_{lm\omega}=\Delta^{s+1}\left[ R_{lm\omega}^H R_{lm\omega}^{\infty\, \prime}-R_{lm\omega}^{H\,\prime} R_{lm\omega}^{\infty} \right]
\end{equation} 
with $'$ denoting $r$ derivative, is conserved. (Here we have $s=2$ for the $\psi_0$ equation.)  Henceforth we shall often drop the $lm\omega$ dependence of $R$ and $W$ for simplicity.   
%
%
\vspace{\baselineskip}
\newline
The form of the homogeneous solutions near horizon and at infinity is given by (see \cite{Teukolsky:1973ha})
\begin{equation}
    R_{(0)}^{H}(r) \sim\left\{\begin{array}{ll}
\displaystyle \frac{B^{(0)\rm hole}_{lm\omega}}{\Delta^{2}} e^{-i k r^{*}}, & r\rightarrow r_{+} \\
\\
\displaystyle \frac{B^{(0)\rm out}_{lm\omega}}{r^5}e^{i \omega r^{*}}
+\frac{B^{(0)\rm in}_{lm\omega}}{r} e^{-i \omega r^{*}}, & r \rightarrow \infty
\end{array}\right.
\end{equation}
\begin{equation}
R_{(0)}^{\infty}(r) \sim\left\{\begin{array}{ll}
\displaystyle \frac{D^{(0)in}_{lm\omega}}{\Delta^{2}} e^{-i k r^{*}}
+D^{(0)out}_{lm\omega}e^{+i k r^{*}}, & r \rightarrow r_{+} \\
\\
\displaystyle \frac{D^{(0)\infty}_{lm\omega}}{r^5} e^{i \omega r^{*}}, & r \rightarrow \infty
\end{array}\right.
\end{equation}
\begin{align}
\label{eq_RH_psi4}
R^{{H}}_{(4)}(r)
=
\left\{
\begin{array}{ll}
\displaystyle
B^{(4){\rm hole}}_{\ell m\omega}\Delta^2 e^{-ikr_*}, \quad
 & r\rightarrow r_+  \\ 
 \\
 \displaystyle
B^{(4){\rm out}}_{\ell m \omega}r^3 e^{i\omega
	r_*}+\frac{B^{(4){\rm in}}_{\ell m \omega}}{r} e^{-i\omega r_*}, &  r\rightarrow
\infty,
\end{array}
\right.
\end{align}
\begin{align}
\label{eq_Rinf_psi4}
R^{\infty}_{(4)}(r)= \left\{
\begin{array}{ll}
\displaystyle
D^{(4)\rm{out}}_{\ell m \omega} e^{ik r_{*}}+\Delta^2
D^{(4){\rm in}}_{\ell m \omega} e^{-ik r_{*}},  & r\rightarrow r_+ \\
\\
\displaystyle
 D^{(4)\infty}_{\ell m \omega} r^3 e^{i\omega
	r_{*}}, &  r\rightarrow \infty.
\end{array}
\right.
\end{align}
where the subscript $(0)$ or $(4)$ in the LHS indicates that they are the homogeneous solutions of the $\psi_0$ or the $\psi_4$ equation. The other variables above are defined as:
\begin{equation}
\label{eqn:rstarkdef}
\begin{split}
      &r_{*}= r+\frac{2Mr_+}{r_+-r_-}\log\frac{r-r_+}{2M}-\frac{2Mr_-}{r_+-r_-}\log\frac{r-r_-}{2M}\,;\\
      & k= \omega - \frac{am}{2Mr_{+}}.
\end{split}
\end{equation}
Calculating the conserved Wronskian (at $\infty$) for the $\psi_0$ sector using \eqref{eqn:WronskianGeneral}, we get:
\begin{equation}
\label{eqn:Wronskian_psi0}
    W^{(0)}=2 i \omega D_{\ell m \omega}^{ (0)\infty} B_{\ell m \omega}^{(0)\mathrm{in}}\,.
\end{equation}
Putting in all the terms in \eqref{eqn:SourceUsedInNumerics}, it can be shown that $T^{(0)}_{lm\omega}$ is $O(\Delta^{-2})$ near the horizon. We cannot directly use \eqref{eqn:GeneralSolnTE} to calculate the waveform at the horizon because the integrands in both the integrals diverge ($O(\Delta^{-2})$) at the horizon which makes the integrals non-convergent. We suggest a regularization scheme to overcome this issue in Sec.~ \ref{sec:Ansatz_Regularization}.
\vspace{\baselineskip}
\newline
From here on, throughout the paper, all the functions like $R^H$, $R^\infty$, $W$, $T_{l m \omega}$ etc. with no $(0)$ or $(4)$ subscripts/superscripts correspond to the $\psi_0$ sector. We have suppressed the subscripts and superscripts of $(0)$ for ease of notation. Any reference to the $\psi_4$ sector functions will explicitly have a subscript or superscript of $(4)$. All the functions without the $(4)$ subscript or superscript are to be understood to belong to the $\psi_0$ sector from here on.

\section{Method of Regularization}
\label{sec:Ansatz_Regularization}
In this section, we develop a scheme to go around the problem of the non-convergence of integrals in the general solution of the radial Teukolsky equation, when the source plunges into the horizon.
In \cite{Poisson:1996ya}, Poisson developed a regularization scheme to calculate the $\psi_4$ ($s=-2$) waveform at infinity for a Schwarzschild black hole. Regularization was done by keeping the limits of integration arbitrary \eqref{eqn:GeneralSolnTEPoisson} and pushing the divergences in to the homogeneous pieces of the general solution given by:
\begin{align}
\label{eqn:GeneralSolnTEPoisson}
    R_{l m \omega}(r)&=\frac{R^{b}_{l m \omega}(r)}{W_{l m \omega}}\left[H_1+ \int_{a}^{r}  R^a_{l m \omega}(r') \Delta^{s} T_{lm\omega}(r') dr'\right]\nonumber\\
    &+\frac{R^a_{l m \omega}(r)}{W_{l m \omega}} \left[H_2+ \int_{r}^{b}  R_{l m \omega}^{b}(r') \Delta^{s} T_{lm\omega}(r') dr'\right]
\end{align}
where $R^a_{l m \omega}(r)$ is the solution of the homogeneous Teukolsky equation which satisfies a specific  boundary condition at $r=a$, $R^b_{l m \omega}(r)$ is the solution of homogeneous Teukolsky equation which satisfies a specific  boundary condition at $r=b$, $W_{lm\omega}$ is the Wronskian calculated using $R^a$ and $R^{b}$, while $H_1$ and $H_2$ are constants.
In \cite{Poisson:1996ya}, the limits ($a$, $b$) and the homogeneous pieces were fixed at the end using boundary conditions. We found that extending the method of regularization in \cite{Poisson:1996ya} to $s=2$ using the Generalised Sasaki Nakamura equation \cite{Hughes:2000pf}, is not very straight-forward for the analysis near horizon. 
\vspace{\baselineskip}
\newline
{We try and overcome the non-convergence of integrals using a different scheme.} We do so by breaking our general solution into two parts. The first part (the 'Ansatz'- $R_A$) analytically accounts for the leading 2 orders (of $\Delta$) in the source term $T^{(0)}_{l m \omega}$ (recall here that $(0)$ stands for $\psi_0$). The second piece then satisfies the radial Teukolsky equation with a modified source term $\Tilde{T}_{l m \omega}$. The modified source term at the leading order has two powers of $\Delta$ less(-divergent) than the original source term. With this modified source term, none of the integrands in the general solution diverge close to the horizon. The scheme will become clearer through this section. 
\vspace{\baselineskip}
\newline
Let $\alpha$ and $\beta$ denote the right hand side and the left hand side of Eq.~\eqref{eqn:TeukolskyEqn} respectively for $s=2$. The integrands in the general solution of the radial Teukolsky equation are of the form $R^H\alpha$ and $R^{\infty}\alpha$, where $R^H$ and $R^{\infty}$ are the solutions of the homogeneous Teukolsky equation satisfying the in-going boundary condition at the horizon and the outgoing boundary condition at infinity respectively. 
%
For the $\psi_0$ case, the integrands diverge as $O(\Delta^{-2})$ near the horizon. We expand the right hand side of Eq.~\eqref{eqn:TeukolskyEqn} in powers of $\Delta$ near the horizon to get:
\begin{equation}
\label{eqn:TE_RHSdeltaExpansion}
    \alpha=e^{i\omega t - im\varphi}\left[\alpha_0+\alpha_{1}\Delta+ \alpha_{2}\Delta^2+...........\right]
\end{equation}
 To regularize the general solution, we would like to remove the leading 2 orders (of $\Delta$) in $\alpha$. To achieve this, we try and substitute an ansatz, with which the left hand side of Eq. \eqref{eqn:TeukolskyEqn} matches the 2 leading orders of right hand side. We attempt the following ansatz,
\begin{equation}
    \label{eqn:Ansatz_form_Numerics}
     R_{A}(r)=\left( \frac{\mathcal{A}_{0}}{r\Delta^2}+\frac{\mathcal{A}_{1}}{r^3\Delta}\right)e^{i\omega t(r) - im\varphi(r)}\,,
\end{equation}
with $\mathcal{A}_{0}$ and $\mathcal{A}_{1}$ being constants, and $t(r)$, $\varphi(r)$ follow the trajectory of the infalling particle. This is a suitable ansatz in order to compute the $\psi_0$ waveform close to the horizon because it has a behaviour demanded by the boundary condition of a purely in-going wave near the horizon. The different powers of $r$ explicitly put in the denominator of terms in \eqref{eqn:Ansatz_form_Numerics} are to make the ansatz contribution small far away from the  horizon. Substituting \eqref{eqn:Ansatz_form_Numerics} in the left hand side of \eqref{eqn:TeukolskyEqn}, we get:
\begin{align}
\label{eqn:TE_LHSdeltaExpansion}
    \beta=e^{i\omega t - im\varphi} 
    \Big[& \beta_{0 0}\mathcal{A}_{0}+\beta_{0 1}\mathcal{A}_{1}+(\beta_{1 0}\mathcal{A}_{0}+\beta_{1 1}\mathcal{A}_{1})\Delta \nonumber\\
     +&(\beta_{2 0}\mathcal{A}_{0}+\beta_{2 1}\mathcal{A}_{1})\Delta^2+ \dots \Big]
\end{align}
 Equating the two leading order coefficients of $\Delta$ from \eqref{eqn:TE_RHSdeltaExpansion} and \eqref{eqn:TE_LHSdeltaExpansion}, we can fix $\mathcal{A}_{0}$ and $\mathcal{A}_{1}$ to be:

\begin{equation}
\label{eqn:Ansatz_A}
    \mathcal{A}_{0}=\frac{(\beta_{1 1}\alpha_0-\beta_{01}\alpha_1)}{(\beta_{00}\beta_{11}-\beta_{10}\beta_{01})}
\end{equation}
\begin{equation}
\label{eqn:Ansatz_B}
    \mathcal{A}_{1}=\frac{(\beta_{10}\alpha_0-\beta_{00}\alpha_1)}{(\beta_{10}\beta_{01}-\beta_{00}\beta_{11})}
\end{equation}
With the choice of $\mathcal{A}_{0}$ and $\mathcal{A}_{1}$ from equations \eqref{eqn:Ansatz_A} and \eqref{eqn:Ansatz_B}, we now write the general solution of the radial Teukolsky equation \eqref{eqn:TeukolskyEqn} as:
\begin{equation}
\label{eqn:AnsatzPlusSomething}
    R(r)=R_{A}(r)+\widetilde{R}(r)
\end{equation}
{Note that the total solution $R(r)$ in the above equation must satisfy the boundary conditions at the horizon and at infinity. This can be ensured if $\widetilde{R}(r)$ and $R_A(r)$ both satisfy the boundary conditions at horizon and at infinity individually. But it is not clear from Eq. \eqref{eqn:Ansatz_form_Numerics} that $R_A(r)$ satisfies the outgoing boundary condition at infinity. Hence, for the total solution $R(r)$ to satisfy the boundary condition at infinity, apart from $\widetilde{R}(r)$ satisfying the boundary condition, $R_A(r)$ must be cutoff (become negligible) at some finite $r_{\rm max}$. This is automatically ensured if the trajectory of the plunging particle begins at a finite $r_{\rm max}$ (and does not extend to infinity), as then $R_A$ will simply vanish for $r>r_{\rm max}$, and has no  contribution near infinity.  In this work, all the trajectories considered are of this particular nature.} Substituting \eqref{eqn:AnsatzPlusSomething} in \eqref{eqn:TeukolskyEqn}, we get:
\begin{equation}
    \hat{F} [R_A+\widetilde{R}] = \Delta^2\,T_{lm\omega}
\end{equation}
where $\hat{F}$ is the operator that gives the left hand side of Eq.\eqref{eqn:TeukolskyEqn} when applied to $R$. We can define a modified source term as:
\begin{equation}
\label{eq:Ttilde}
\widetilde{T}_{lm\omega} =\frac{\Delta^2 T_{lm\omega} -\hat F(R_A)}{\Delta^2}\,,
\end{equation}
which satisfies the equation:
\begin{equation}
\label{eqn:TeukolskyEqnNew1}
    \hat{F}(\widetilde{R}(r))=\Delta^2\widetilde{T}_{lmw} 
\end{equation}
%
%
%
We observe that \eqref{eqn:TeukolskyEqnNew1} is just the Teukolsky equation \eqref{eqn:TeukolskyEqn} (with $s=2$) for $\widetilde{R}$ with source $\widetilde{T}_{lm\omega}$. Note that for this source, the integrands $R^H\Delta^2\widetilde{T}_{lm\omega}$ and $R^{\infty}\Delta^2\widetilde{T}_{lm\omega}$  don't diverge near horizon. This is because $\mathcal{A}_{0}$ and $\mathcal{A}_{1}$ are chosen as in \eqref{eqn:Ansatz_A}, \eqref{eqn:Ansatz_B} to make sure that $\widetilde{T}_{lm\omega}$ is suppressed by two orders of $\Delta$ when compared to $T_{lm\omega}$. Also, as discussed above, $\widetilde{R}(r)$ must satisfy the same boundary conditions as the total solution. Therefore the general solution for $\widetilde{R}$ is:
\begin{align}
    \widetilde{R}_{l m \omega}(r)&=\frac{1}{W}R^{\infty}(r)\left[\int_{r_{+}}^{r}  R^H(r') \Delta^{2} \widetilde{T}_{lm\omega}(r') dr'\right]\nonumber\\
    &+\frac{1}{W}R^H(r) \left[\int_{r}^{\infty}  R^{\infty}(r') \Delta^{2} \widetilde{T}_{lm\omega}(r') dr'\right]
\end{align}
where W is given by \eqref{eqn:Wronskian_psi0}. Also as $r\rightarrow r_+$, the first integral becomes negligible and \eqref{eqn:AnsatzPlusSomething} can be expressed as:
%
%
%
\begin{equation}
\label{eqn:Regularize_Solution_at_Horizon}
\begin{split}
    & R(r\rightarrow r_+)\sim \left( \frac{\mathcal{A}_{0}}{r\Delta^2}+\frac{\mathcal{A}_{1}}{r^3\Delta}\right)e^{i\omega t(r) - im\varphi(r)}\\&
    +\frac{B^{(0)hole}_{lm\omega}}{W\,\Delta^2} e^{-i k r^{*}} \left[\int_{r_+}^{\infty}  R^{\infty}(r') \Delta^{2} \widetilde{T}_{lm\omega}(r') dr'\right]
\end{split}
\end{equation}
\newline
Thus we have been able to regularize the radial Teukolsky equation (for $s=2$) that involves sources near the horizon. Further, \eqref{eqn:psi0_r_theta_decomposition} and \eqref{eqn:Regularize_Solution_at_Horizon} can be used to calculate  the $\psi_0$ perturbation close to the horizon.
\newline
\section{$\psi_0$ waveforms close to the horizon of a Schwarzschild black hole}
\label{sec:Psi0_Waveform_Calc}
In this section, we display results of the near horizon solution of the radial Teukolsky equation for two kinds of trajectories of a test particle plunging into a Schwarzschild black hole. First trajectory in a radial plunge in which the test particle falls along a radial geodesic into the black hole. The second is an Effective One Body (EOB) trajectory that transitions from a quasi-circular orbit into a plunge~\cite{Buonanno:2000ef}. The regularization ansatz that we have used is given in \eqref{eqn:Ansatz_form_Numerics}.
\subsection{Radial in-fall}

\begin{figure}[htb!]
    \centering
    \includegraphics[width=0.45\textwidth]{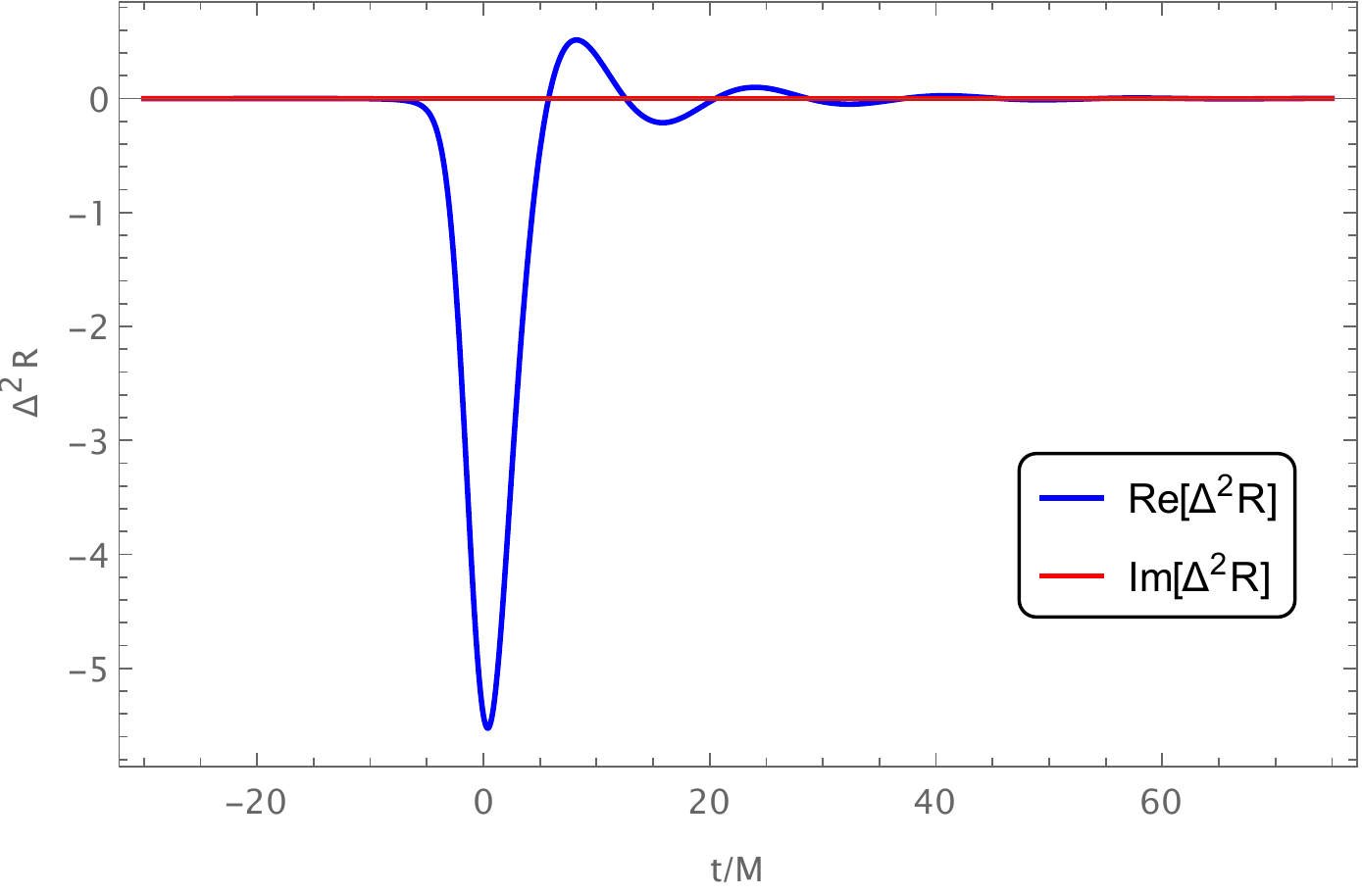}
    \caption{$\psi_0$ waveform for radial in-fall trajectory. The time coordinate here is advanced time, with $t=0$ corresponding to the plunge of the particle into the horizon.}
    \label{fig:radialin-fallinTime}
\end{figure}

Let us consider a particle that falls radially from infinity, initially at rest.  The total solution of the radial Teukolsky equation (i.e., the sum of the Ansatz part and the integral part) at the horizon (for $l=m=2$) in this case has been plotted in Fig.~\ref{fig:radialin-fallinTime} in advanced time.  
The in-falling particle follows a radial geodesic of the Schwarzschild background, described by $\theta=\pi/2$, $\varphi=0$, and
\begin{equation}
\label{eqn:radial_geodesic}
    \frac{dt}{dr}=-\frac{1}{(1-\frac{2M}{r})}\left(\frac{r}{2M}\right)^{1/2}
\end{equation}
{To obtain the waveform in Fig.~\ref{fig:radialin-fallinTime}, we made the following modification to the geodesic trajectory  \eqref{eqn:radial_geodesic} to cutoff the contribution in the integral in \eqref{eqn:Regularize_Solution_at_Horizon} from $r^{\prime}>r_{\rm max} = 300M$:
\begin{equation}
    r_{\rm mod}=\frac{r-80 \log \left(\cosh \left(\frac{300-r}{80}\right)\right)}{1+\tanh \left(\frac{149}{40}\right)}+2-\frac{2-80 \log \left(\cosh \left(\frac{149}{40}\right)\right)}{1+\tanh \left(\frac{149}{40}\right)};
\end{equation}
with $r$ being the radial coordinate in the geodesic trajectory.
\begin{figure}[htb!]
    \centering
    \includegraphics[width=0.45\textwidth]{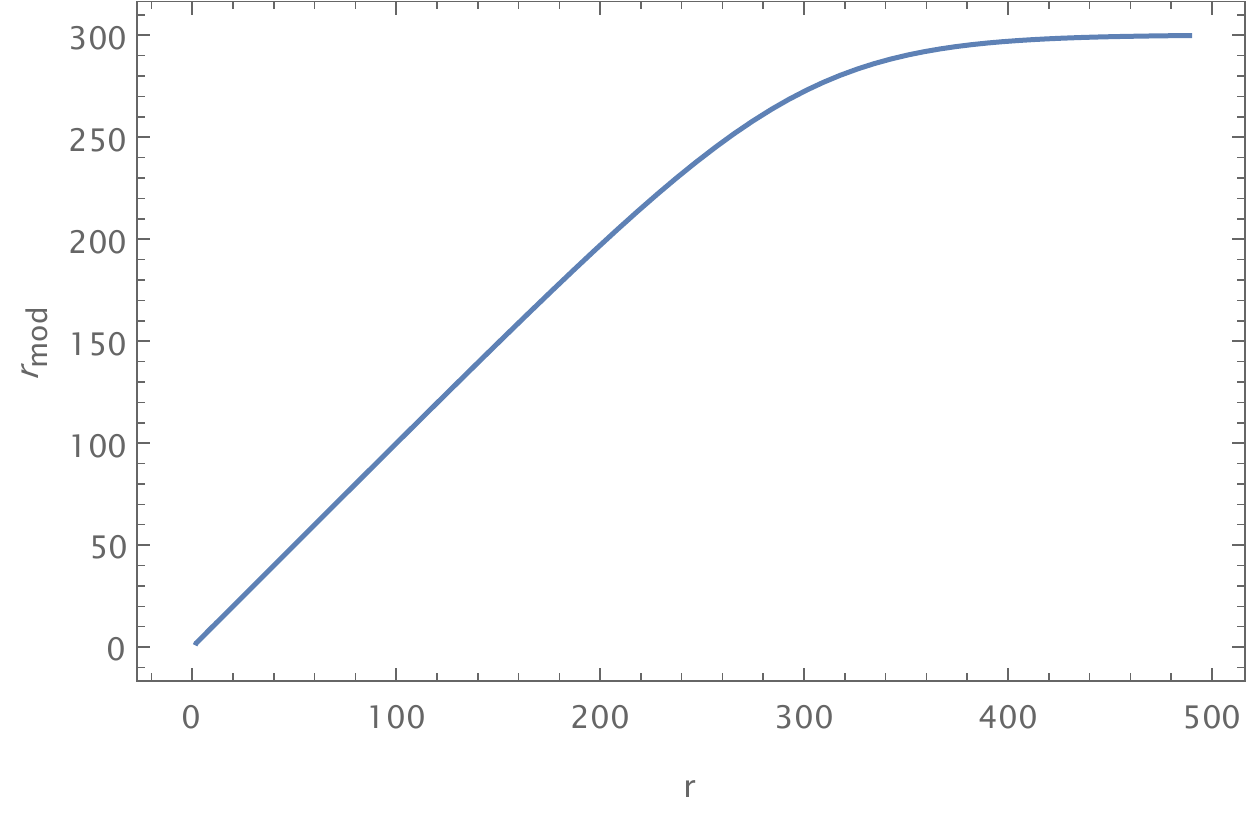}
    \caption{Variation of $r_{\rm mod}$ with the geodesic radial co-ordinate.}
    \label{fig:rmod_vs_r}
\end{figure}
Fig. \ref{fig:rmod_vs_r} shows how $r_{\rm mod}$ depends on $r$.  For small values of $r$, $r_{\rm mod}$ is the same as $r$; $r_{\rm mod}$ agrees very well with $r$ when $r\sim 2$.  But as $r$ increases to larger and larger values, $r_{\rm mod}$ asymptotically becomes $r_{\rm max} = 300M$. Using $r_{\rm mod}$ as radial co-ordinate for the integral in \eqref{eqn:Regularize_Solution_at_Horizon} has enabled us to cutoff the contribution from $r^{\prime}>r_{\rm max} = 300M$ in the integral in \eqref{eqn:Regularize_Solution_at_Horizon}.}

Note that for a geodesic trajectory, $t+r_*$ becomes a constant, say $v_0$, as the particle reaches very close to the horizon. This constant, $v_0$, is property of the trajectory and represents the location of the plunge, in advanced time, of the in-falling particle into the black hole horizon. In Fig. \ref{fig:radialin-fallinTime}, and in subsequent Figs.\ \ref{fig:Psi_0_Plots}--\ref{fig:EOBSTvsDirect}, we have shifted the location of the plunge to zero on the horizontal axis. [In the frequency domain, this corresponded to pulling out a phase factor $e^{i\omega v_0}$ from the total frequency-domain waveform.]
%
%

\subsection{Effective One Body (EOB) Quasi-Circular Plunge}
\label{sec:EOB_Psi0}
\begin{figure*}[htb!]
	\centering
	\includegraphics[width=\columnwidth]{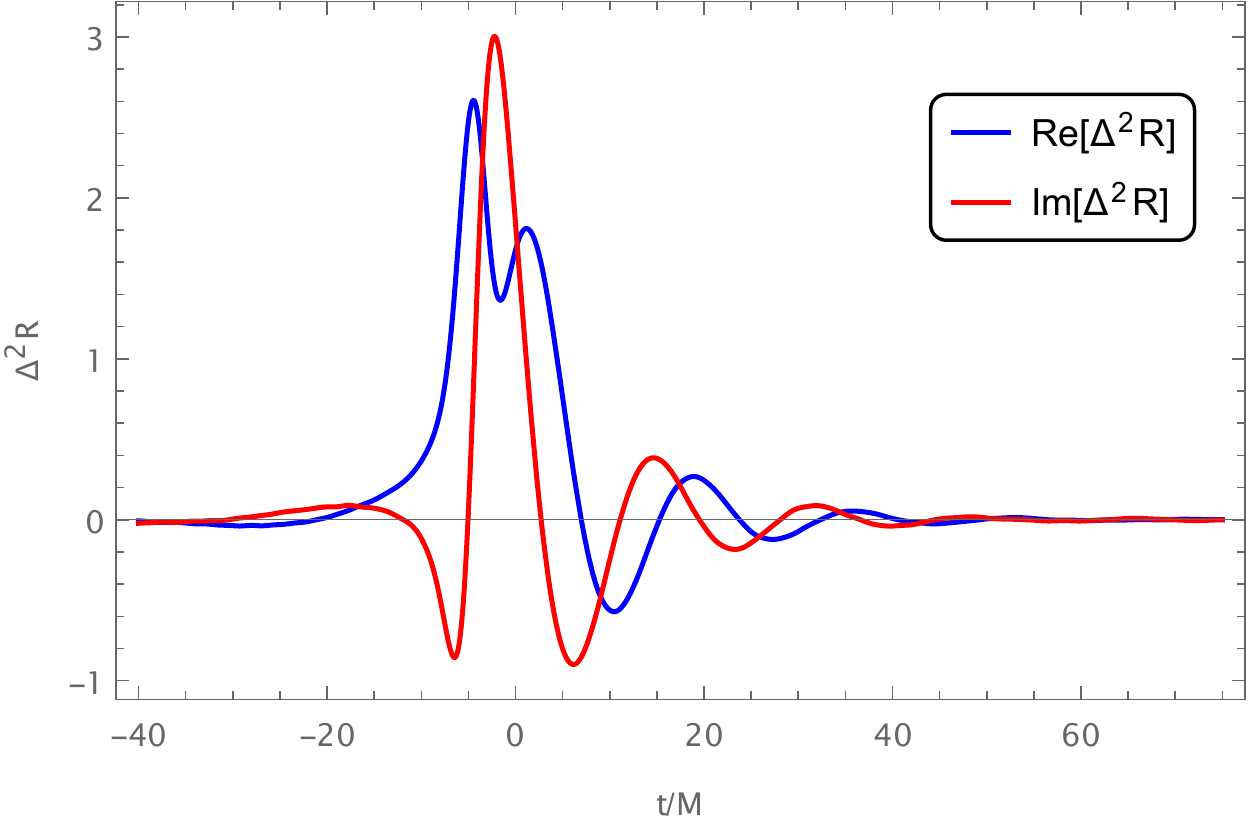}
	\includegraphics[width=\columnwidth]{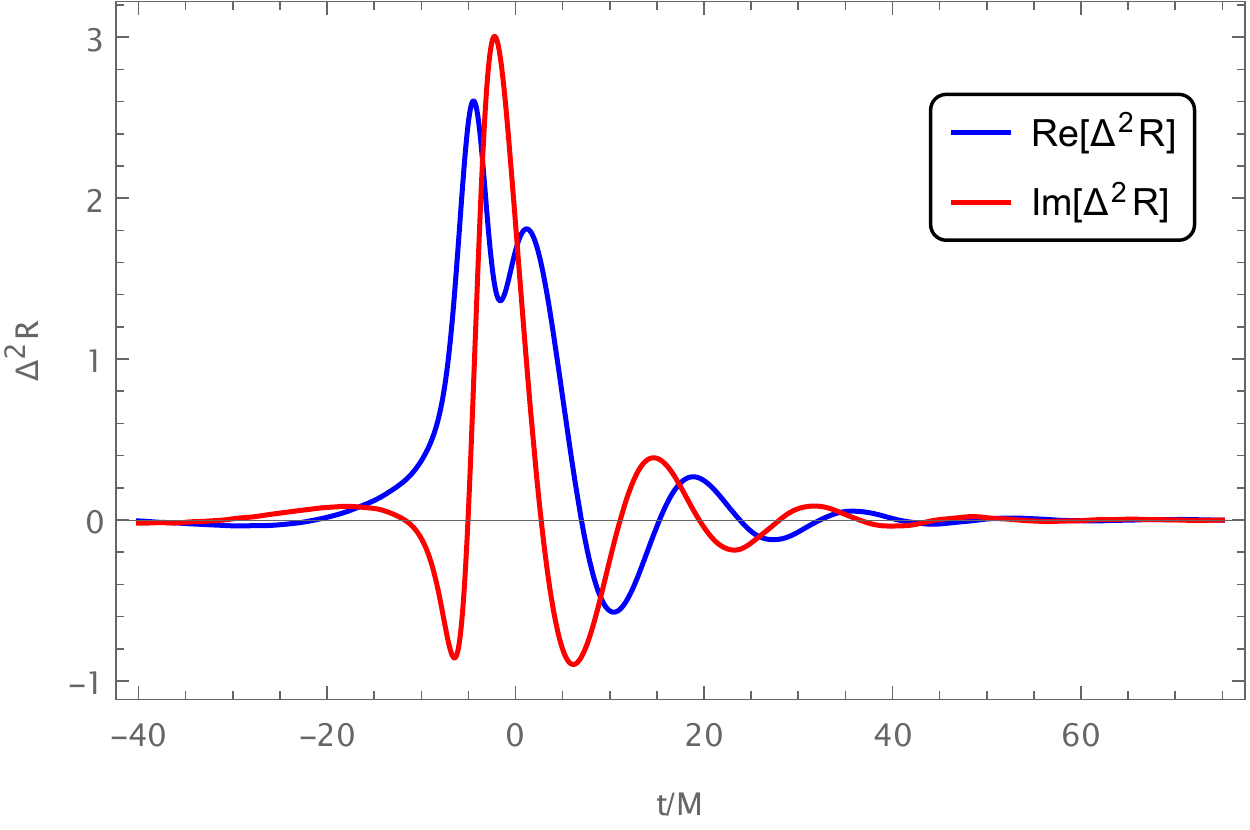}
	\includegraphics[width=\columnwidth]{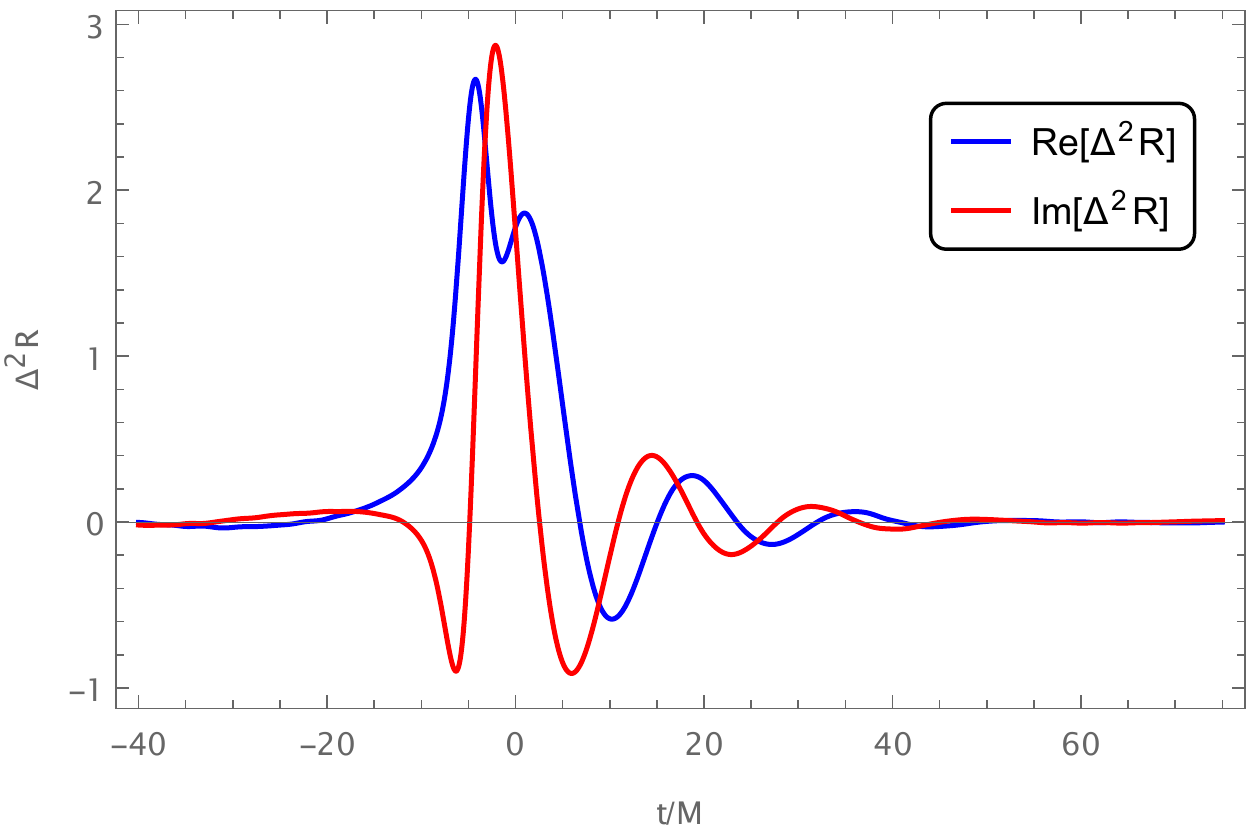}
	\includegraphics[width=\columnwidth]{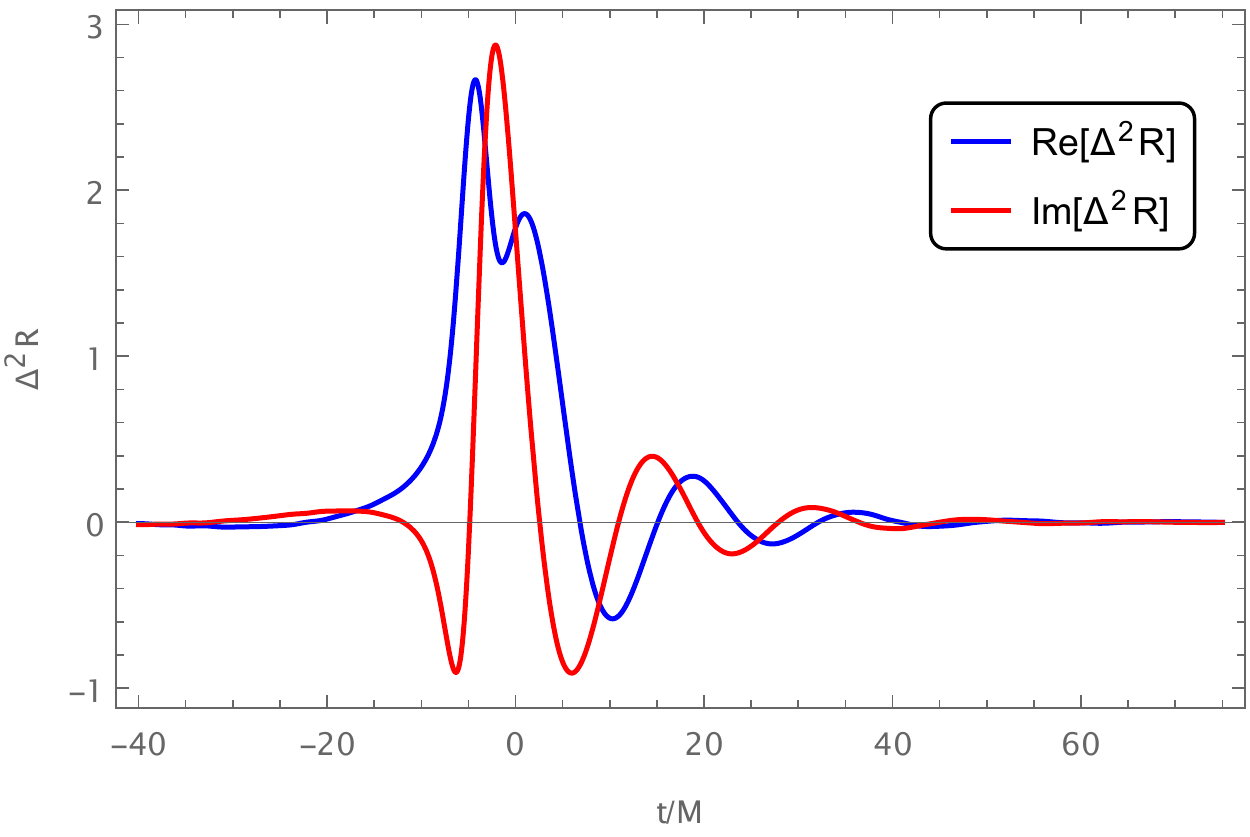}
	\includegraphics[width=\columnwidth]{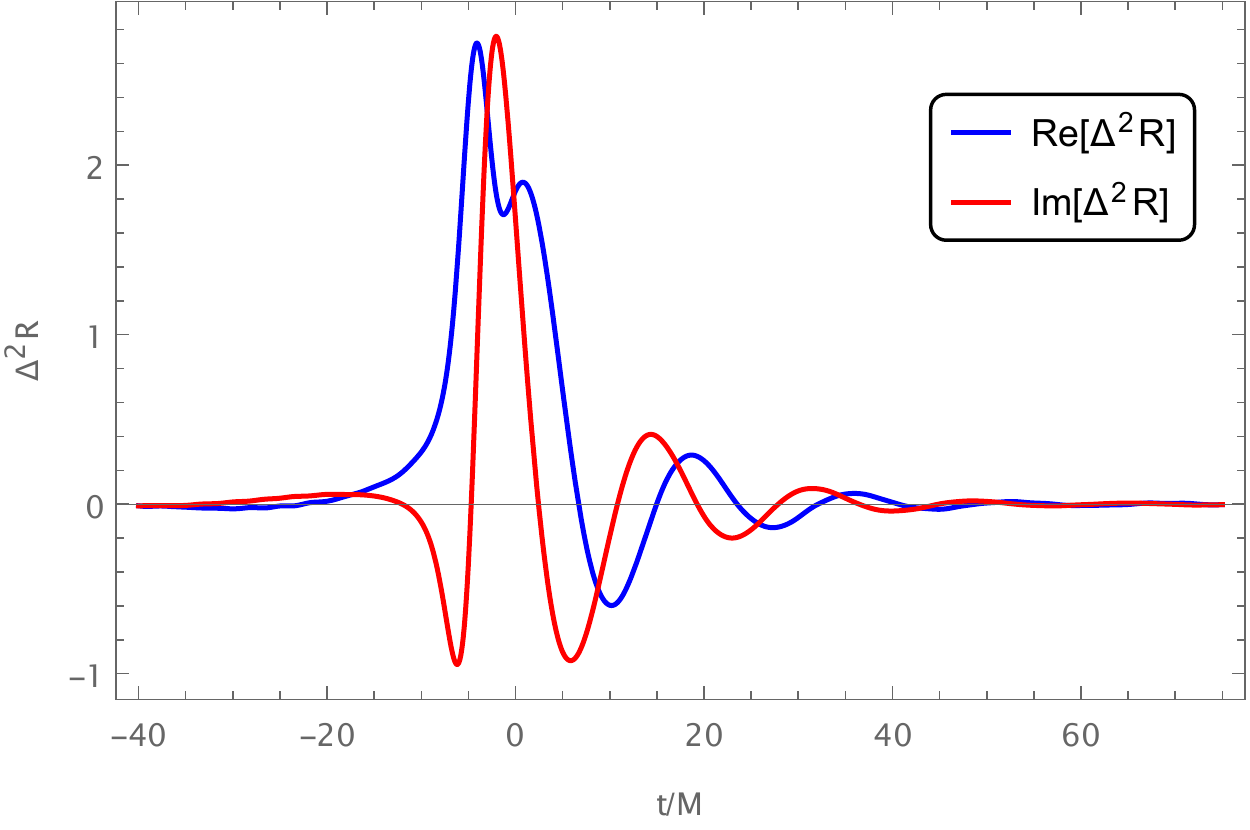}	
	\includegraphics[width=\columnwidth]{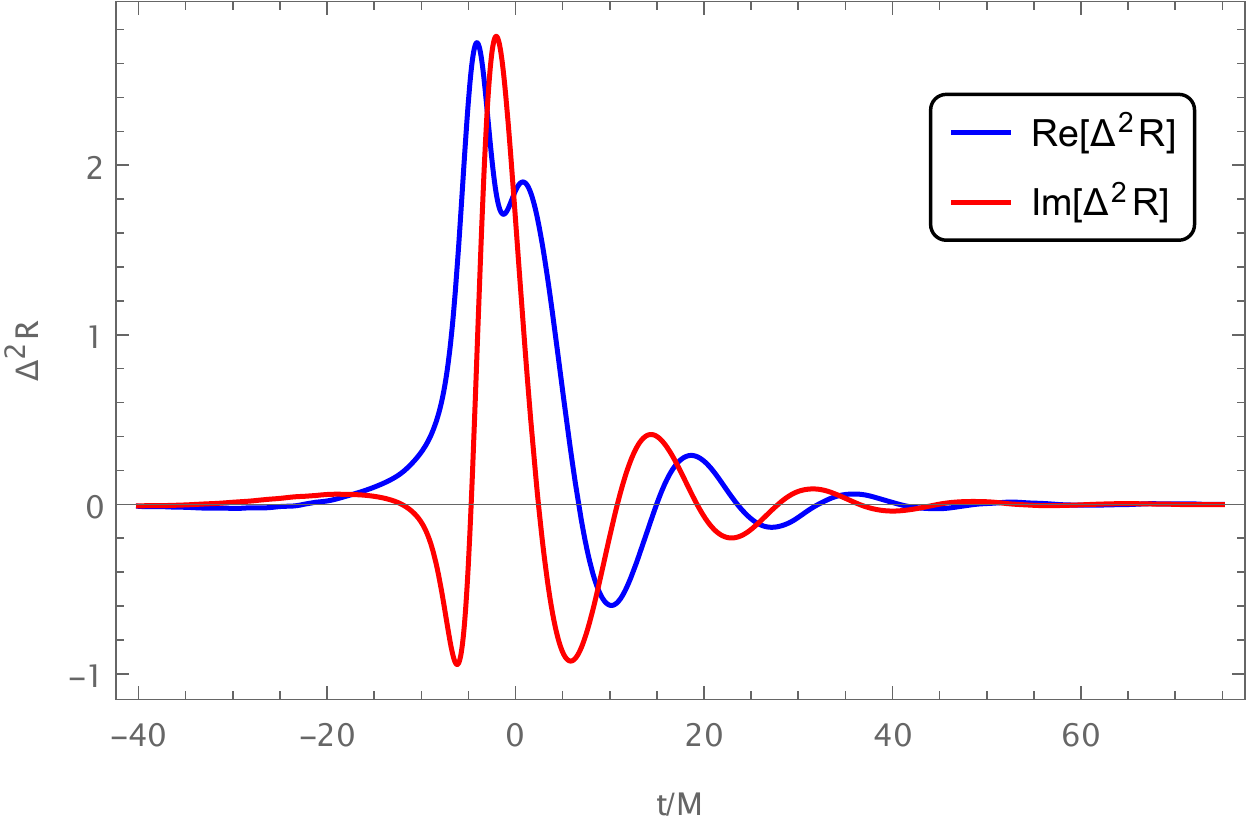}	
	\caption{$\psi_0$ Waveforms for EOB quasi-circular trajectories. The time coordinate here is advanced time, with $t=0$ corresponding to the plunge of the particle into the horizon. The top panel is for $\eta=0.1$, the middle panel is for $\eta=0.16$ and the bottom panel is for $\eta=0.22$ trajectories. [Right column- Trajectory cutoff at $r_{\rm max}=7M$; Left column- Trajectory cutoff at $r_{\rm max}=10M$.]}
	\label{fig:Psi_0_Plots}
\end{figure*}
In this section, we look at results for a more realistic and general trajectory of the in-falling particle. The trajectory we use is the Effective One Body (EOB) quasi-circular plunge (decaying circular orbits) following the analysis in \cite{Buonanno:2000ef}.  The trajectory is not exactly a geodesic because of the radiation reaction (which causes the decay of the circular orbits). The parameters used to generate the EOB trajectory are: start point at $r_0=15M$. The other initial conditions have been fixed using adiabatic approximation conditions \cite{Buonanno:2000ef}. We have discussed results for trajectories with symmetric mass ratios of $\eta=0.1,\,0.16,\,0.22$. The expression for $\eta$ is terms of the masses ($m_1$ and $m_2$) of the two merging objects is:
\begin{equation}
    \eta=\frac{m_1m_2}{(m_1+m_2)^2}
\end{equation}
The dimensionless EOB Hamiltonian (rescaled using the reduced mass ($M_{red}$) of black hole and the test particle) used for the evolution of the trajectory is:
\begin{equation}
  \widehat{H}^{\text {improved }}=\frac{H_{\mathrm{improved}}(\mathbf{r}, \mathbf{p})}{M_{red}}=\frac{1}{\eta} \sqrt{1+2 \eta\left(\frac{H_{\text {eff }}-M_{red}}{M_{red}}\right)}
\end{equation}
where
\begin{equation} \frac{H_{\mathrm{eff}}(\mathbf{r}, \mathbf{p})}{M_{red}}=\sqrt{F(r)\left[1+\left(\frac{\mathbf{p}}{M_{red}}\right)^{2}+\left(F(r)-1\right)\left({\frac{\mathbf{\widehat{r}} \cdot \mathbf{p}}{M_{red}}}\right)^{2}\right]}
\end{equation}
and
\begin{equation}
    F(r)=1-\frac{2(M+\mu)}{r};
\end{equation}
We then use the following Hamilton's equations to evolve the trajectory:
\begin{equation}
\frac{d r}{d t}=\frac{\partial \widehat{H}^{\text {improved }}}{\partial p_{r}}\left(r, p_{r}, p_{\varphi}\right)
\end{equation}
\begin{equation}\frac{d \varphi}{d \widehat{t}} \equiv \widehat{\omega}=\frac{\partial \widehat{H}^{\text {improved }}}{\partial p_{\varphi}}\left(r, p_{r}, p_{\varphi}\right)\end{equation}
\begin{equation}\frac{d p_{r}}{d t}=-\frac{\partial \widehat{H}^{\text {improved }}}{\partial r}\left(r, p_{r}, p_{\varphi}\right)\end{equation}
\begin{equation}\frac{d p_{\varphi}}{d \widehat{t}}=\widehat{F}^{\varphi}\left[\widehat{\omega}\left(r, p_{r}, p_{\varphi}\right)\right]\end{equation}
Here $\widehat{F}^{\varphi}$ is the radiation reaction that causes the slow decay of the orbit. The procedure to calculate the radiation reaction is highlighted in \cite{Yanbei:EOBref}. Note that $\widehat{t}=t/(M+\mu)$, $\widehat{\omega}=\omega (M+
\mu)$.
\vspace{\baselineskip{}}
\newline
In this case, we have implemented the same cut-off strategy, namely cutting off the source term $T$ at $r>r_{\rm max}$ ($r_{\rm max}$ being different than the radial infall value), in a slightly different way, without having to modify the EOB trajectory.  As we impose this cut-off on the full source term $T$, the Ansatz part $R_A$, the integral part $\tilde T_{lm\omega}$ will both be cut off at $r>r_{\rm max}$.  In our practical calculation though,  the Ansatz part will not be evaluated at locations away from the horizon, therefore an explicit cutoff on $R_A$ is not necessary.  We will only need to use a window function
when evaluating the integral(-part) of \eqref{eqn:Regularize_Solution_at_Horizon} to remove any contribution from $r > r_{\rm max}$ in $\tilde T_{lm\omega}$ [Eq.~\eqref{eq:Ttilde}] . We specifically used a window of  $ (1-\tanh (2 (r^{\prime}-r_{\rm max}))/M)/2$. Such windowing also helped eliminate numerical errors that accumulate while integrating highly oscillatory functions.
\vspace{\baselineskip}
\newline
%
%
Fig.~\ref{fig:Psi_0_Plots} contains the $\psi_0$ waveforms (for $l=m=2$) corresponding to the test particle following an EOB quasi-circular trajectory while plunging into the Schwarzschild black hole. The different rows are plots for trajectories with different mass ratios.  The two columns in Fig. \ref{fig:Psi_0_Plots} correspond to different distances ($r_{\rm max}=10M$ and $r_{\rm max}=7M$) at which the integral part of the solution was smoothly cutoff. Note that to evaluate the $\psi_0$ waveform at horizon, the ansatz piece doesn't need to be evaluated at infinity, hence there is no windowing required for the ansatz contribution.
\newline
%
%
%
%
We found almost no difference in the waveform in the relevant region (relevant advanced time interval) for the two cases with different cut-off radii. This can be seen by comparing the figures of the two columns in Fig.~\ref{fig:Psi_0_Plots}. This reinforces our opinion that maximum contribution to the integral part in the waveform should come from the region close to horizon (justifying the use of windowing functions). 
\section{Comparing our $\psi_0$ results with those obtained using Starobinsky-Teukolsky Identities}
\label{sec:Waveform_Comparison}
In this section we first highlight how $\psi_4$ close to the horizon is related to the in-going $\psi_0$ at the horizon via the ST identity for a Kerr background. We then compare the $\psi_0$ described in Section \ref{sec:Psi0_Waveform_Calc} to the $\psi_0$ computed from $\psi_4$ via the ST identity for the case of a Schwarzschild background.

\subsection{$\psi_0$ Computation using the Starobinsky-Teukolsky Identity}
We can find the in-going $\psi_0$ from the $\psi_4$ waveform at the horizon via the Starobinsky-Teukolsky (ST) identity 
\begin{equation}
\label{eq:ST}
    Y^{\rm in}_{lm\omega}= {\sigma}_{lm\omega} Z^{\rm in}_{lm\omega},
\end{equation}
where
\begin{equation}
    \sigma_{lm\omega}= \frac{64(2Mr_+)^4 ik (k^2+4\epsilon^2)(-ik+4\epsilon)}{C_{lm\omega}}.
\end{equation}

\begin{equation}
\label{eqn:Yin_Def}
    Y^{\rm in}_{lm\omega}=R^{(0) \rm{in}}\Delta^2e^{ikr^*};\quad\quad Z^{\rm in}_{lm\omega}=R^{(4) \rm in}\Delta^{-2}e^{ikr^*}
\end{equation}
{Note that $R^{(0)\rm in}$ and $R^{(4)\rm in}$ are the in-going components of $R^{(0)}$ and $R^{(4)}$ respectively close to the horizon, where $R^{(0)}$ and $R^{(4)}$ are related to the full $\psi_0$ and $\psi_4$ waveforms and are defined in \eqref{eqn:Psi0_r_theta_decomposition} and \eqref{eqn:Psi4_r_theta_decomposition} respectively.} $C_{lm\omega}$ is the Starobinsky constant, given by 
\begin{align}
    |C_{lm\omega}|^2 &= (Q^2+4 a\omega m - 4 a^2\omega^2)[(Q-2)^2 +36 a\omega m\nonumber\\& - 36 a^2\omega^2]+(2Q-1)(96a^2\omega^2-48 a\omega m)\nonumber\\
    &+144\omega^2(M^2-a^2),
    \nonumber\\
    \mathrm{Im} \; C_{lm\omega} &=12 M \omega, \nonumber\\
    \mathrm{Re} \; C_{lm\omega}& = +\sqrt{|C_{lm\omega}|^2-(\mathrm{Im}\; C_{lm\omega})^2},
\end{align}
with $Q= \mathcal{E}_{lm} +a^2\omega^2-2a\omega m$, where $\mathcal{E}_{lm}$ is the spheroidal eigenvalue \cite{Press:1973zz}.  Note that to compute $\psi_0$ at horizon via the ST identity, we need the $\psi_4$ waveform at the horizon. For the radial in-fall case in a Schwarzschild background, the $\psi_4$ waveform expression can be found in \cite{Poisson:1996ya}. For the EOB trajectory case, we use the general solution of the form \eqref{eqn:GeneralSolnTE} which results in an expression of the following form (See \cite{Mino:1997bx}):
\begin{align}
\label{eqn:Zin_Psi4}
     Z^{\rm in}_{lm\omega}=\frac{B^{(4)\rm  hole}_{lm\omega}\mu}{2i\omega B^{(4)\rm in}_{lm\omega}D^{(4)\rm \infty}_{lm\omega}}\int_{-\infty}^{\infty}& dt\, e^{i\omega t- im \varphi(t)}\nonumber\\
     &
     \Bigg[\left(A_{n n 0}+A_{n m^{*} 0}+A_{m^{*} m^{*} 0}\right)R^{\infty}_{(4)}
    \nonumber\\
    &-(A_{n m^{*} 0}+A_{m^{*} m^{*} 1})\frac{dR^{\infty}_{(4)}}{dr}\nonumber\\
    &   +(A_{m^{*} m^{*} 2})\frac{d^2R^{\infty}_{(4)}}{dr^2}\Bigg]
\end{align}
\subsection{Waveform Comparison}
\begin{figure}[!htb]
    \centering
    \includegraphics[width=0.45\textwidth]{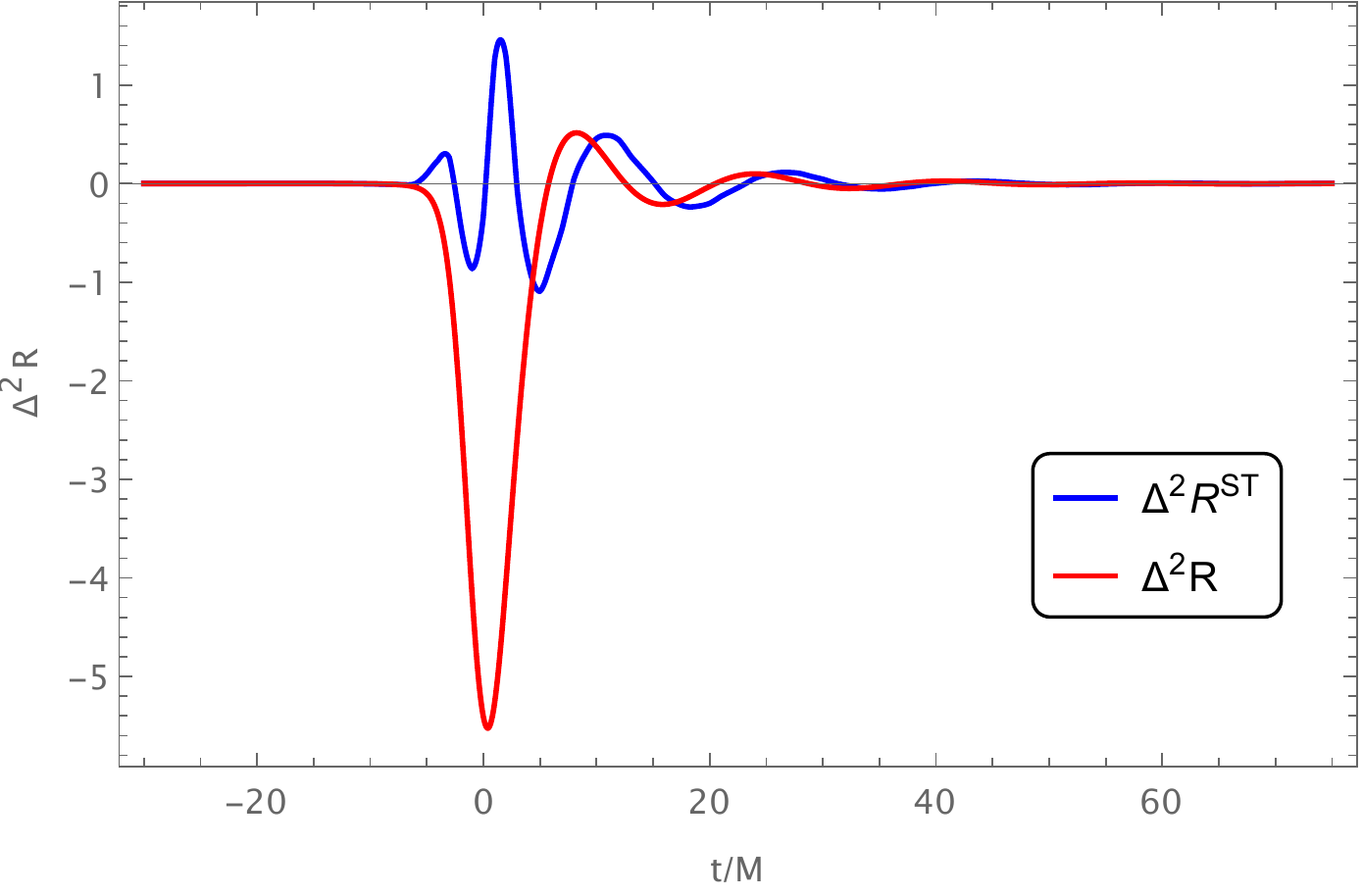}
    \caption{$\psi_0$ waveform comparison for radial in-fall trajectory. The time coordinate here is advanced time, with $t=0$ corresponding to the plunge of the particle into the horizon.}
    \label{fig:radialin-fallSTvsDirect}
\end{figure}
\begin{figure*}[htb!]
	\centering
	\includegraphics[width=\columnwidth]{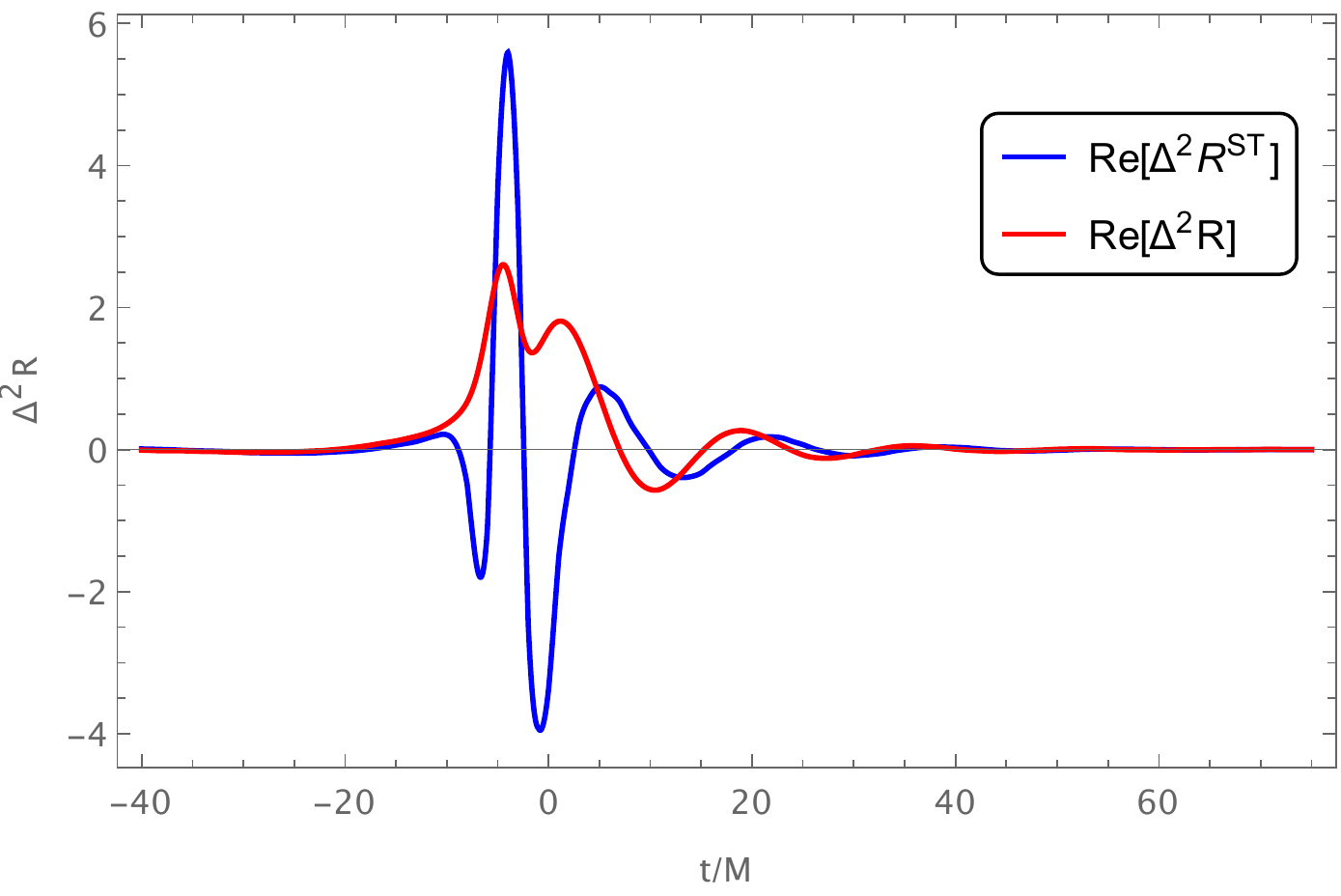}
	\includegraphics[width=\columnwidth]{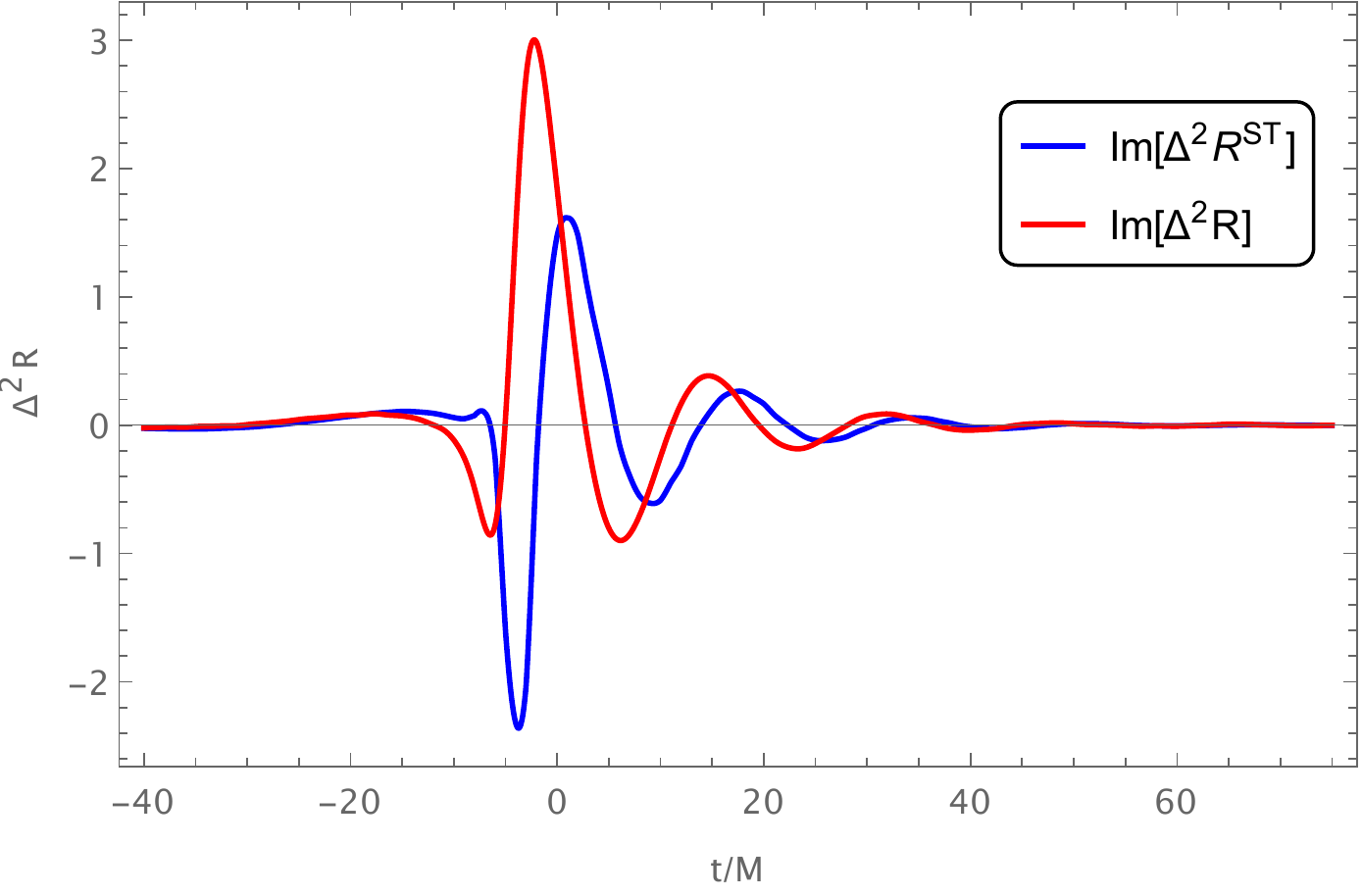}
	\includegraphics[width=\columnwidth]{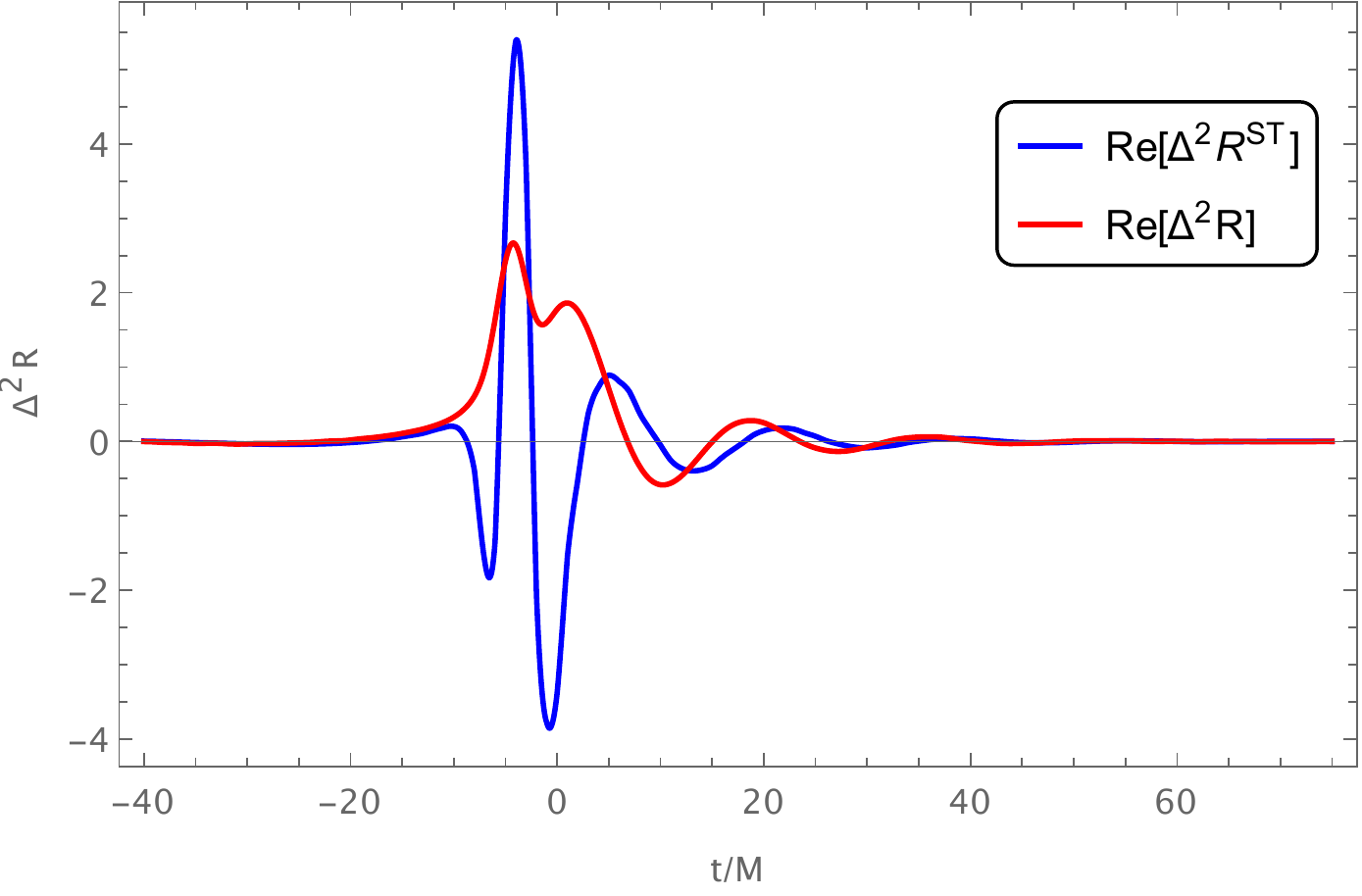}
	\includegraphics[width=\columnwidth]{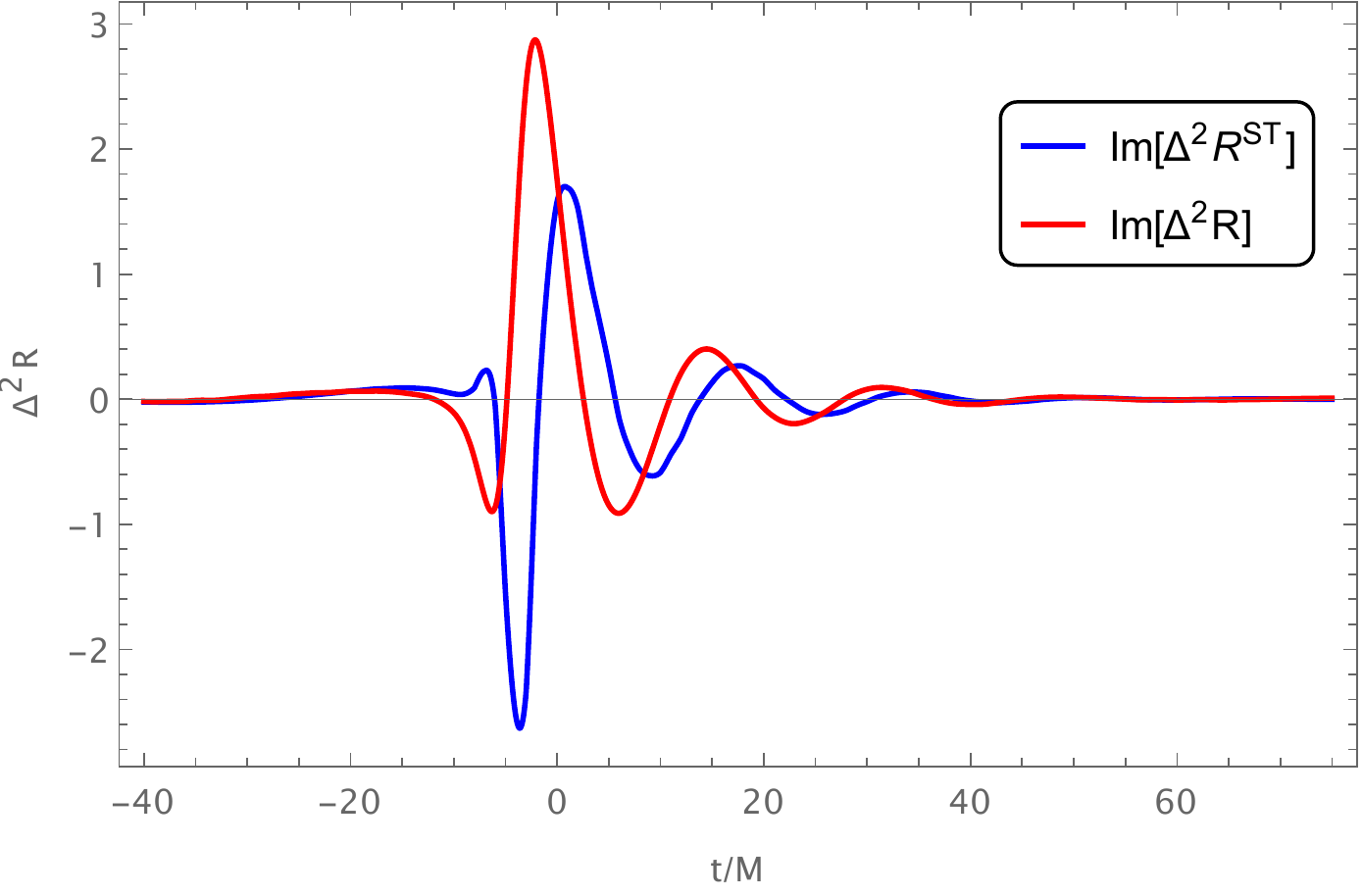}
	\includegraphics[width=\columnwidth]{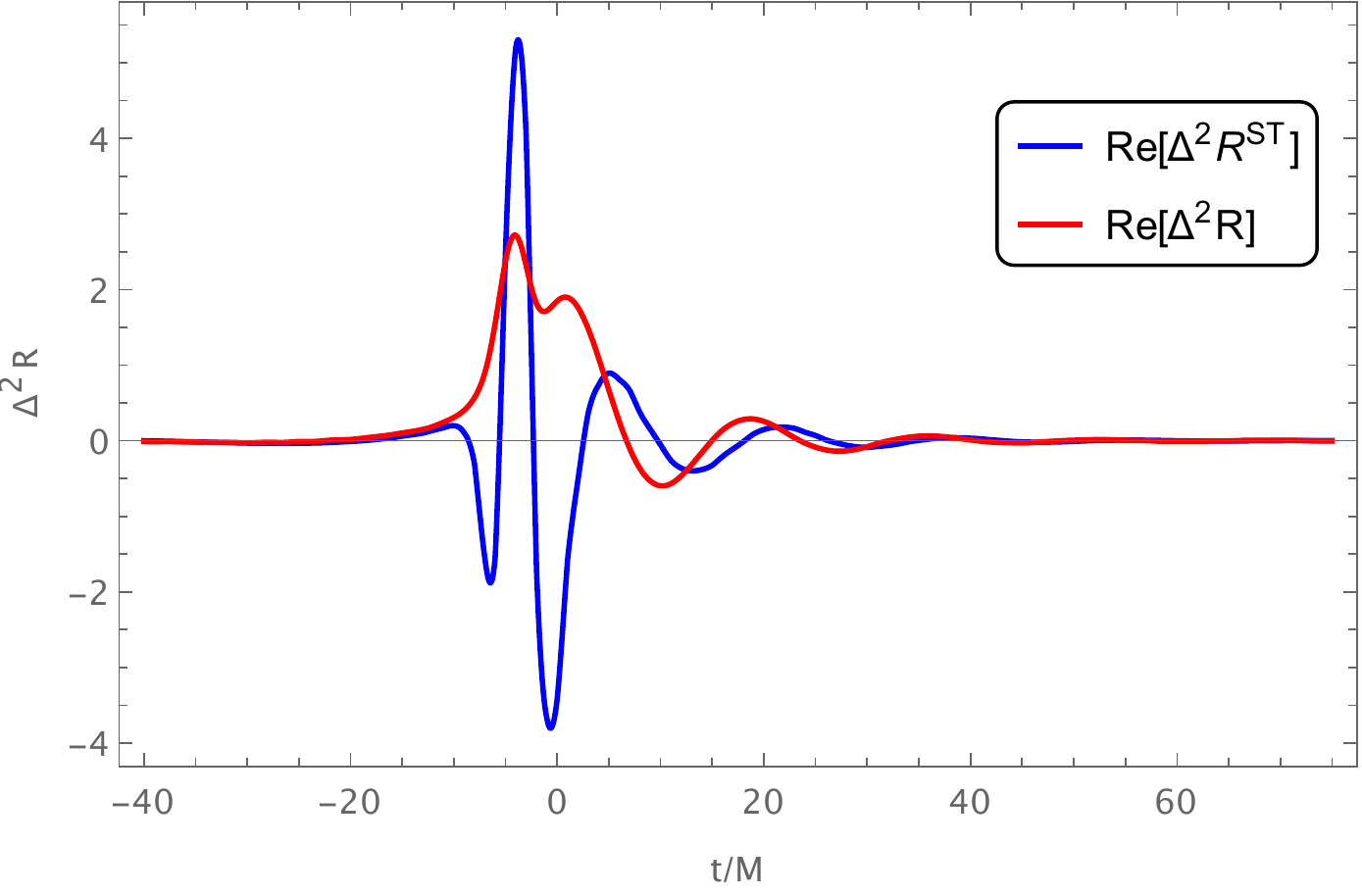}	
	\includegraphics[width=\columnwidth]{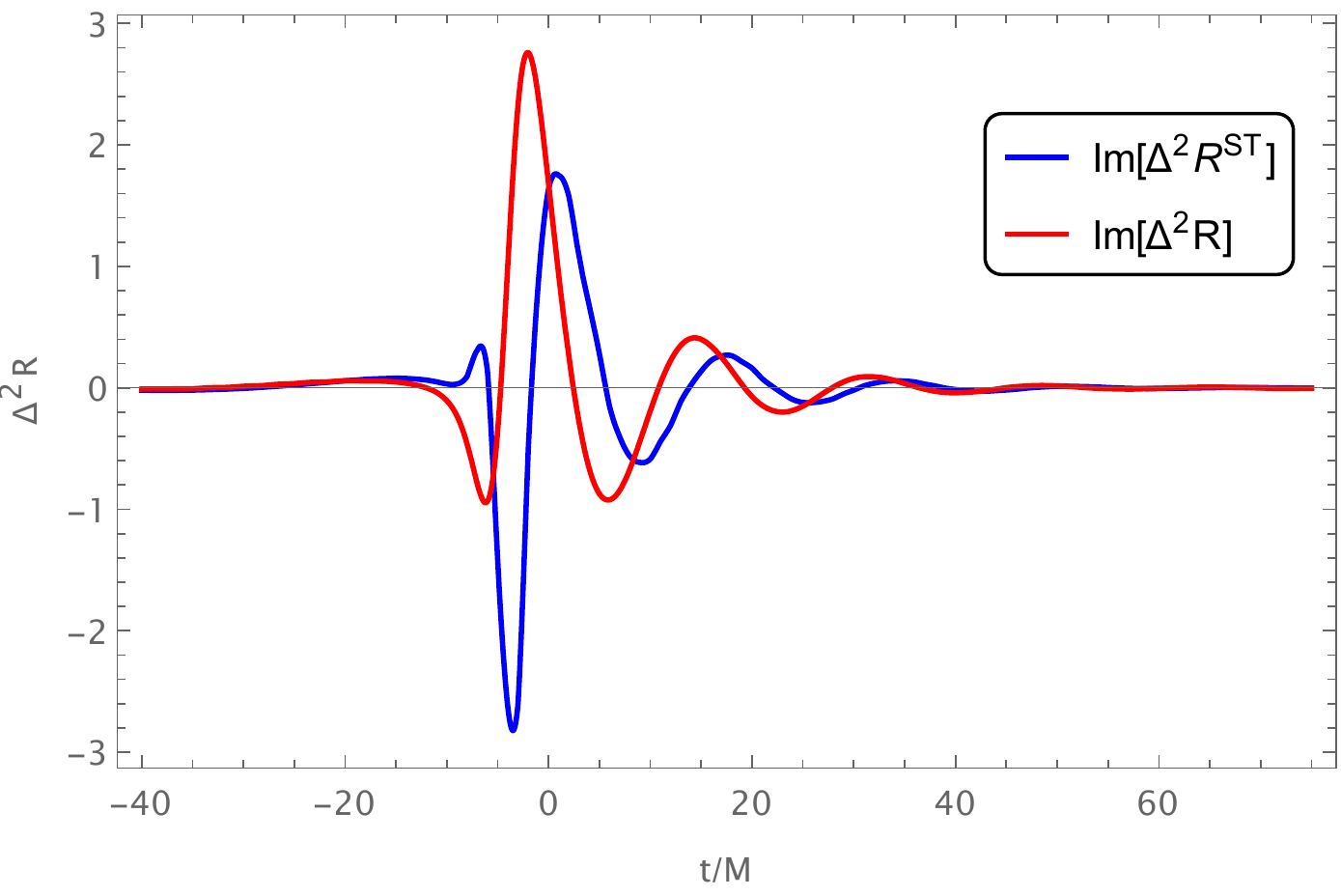}	
	\caption{$\psi_0$ waveform comparison for EOB quasi-circular trajectories. The time coordinate here is advanced time, with $t=0$ corresponding to the plunge of the particle into the horizon. The left column contains the real part comparisons while the right column contains the imaginary part comparisons. The top panel is for $\eta=0.1$, the middle panel is for $\eta=0.16$ and the bottom panel is for $\eta=0.22$ trajectories.}
	\label{fig:EOBSTvsDirect}
\end{figure*}
%
In this subsection we present our results, comparing the $\psi_0$ waveform \eqref{eqn:Regularize_Solution_at_Horizon} close to the horizon computed directly from the Teukolsky equation with the $\psi_0$ waveform computed from the $\psi_4$ waveform via the ST identity. As it turns out, waveforms obtained using the two approaches are quite different from each other. Fig.~\ref{fig:radialin-fallSTvsDirect} shows the $\psi_0$ waveforms (for $l=m=2$) computed using these two different approaches for the radial in-fall trajectory in Schwarzschild background. The superscript ``ST'' in the legends implies that the corresponding waveform is computed using the ST identity. The other legend without the superscript corresponds to the waveform directly computed from the Teukolsky equation. It is quite clear that the two approaches predict very different $\psi_0$ waveforms towards the horizon. Similar result is seen for the EOB trajectory case. Fig. \ref{fig:EOBSTvsDirect} highlights the differences in the real and imaginary parts of the $\psi_0$ waveform (for $l=m=2$) computed using the two approaches for EOB trajectories of the in-falling particle. Once again, the predictions of the two approaches are very different.
\vspace{\baselineskip}
\newline
Our results in this section indicate that using the Starobinsky-Teukolsky identity on $\psi_4$ is not equivalent to computing $\psi_0$ directly from the Teukolsky equation. This is mainly due to the fact that the Staorbinsky-Teukolsky identity holds only for vacuum spacetimes in black hole backgrounds. With a particle plunging into the black hole, the spacetime in the near-horizon region is no longer vacuum and the predictions of the Starobinky-Teukolsky identity are no longer valid --- although during the inspiral, the particle is far away from the horizon and one would still expect the Starobinsky-Teukolsky identity to hold.  Thus, the right way to compute the $\psi_0$ waveform, during the plunge, is to begin with the Teukolsky equation and compute its solution employing a relevant regularization scheme as done in Section \ref{sec:Psi0_Waveform_Calc}. 
\newline

%
%
\section{Comparison of Echo Waveforms}
\label{sec:Echo_Difference_Starobinky_Transform}
\begin{figure*}[htb!]
	\centering
	\includegraphics[width=5.5cm]{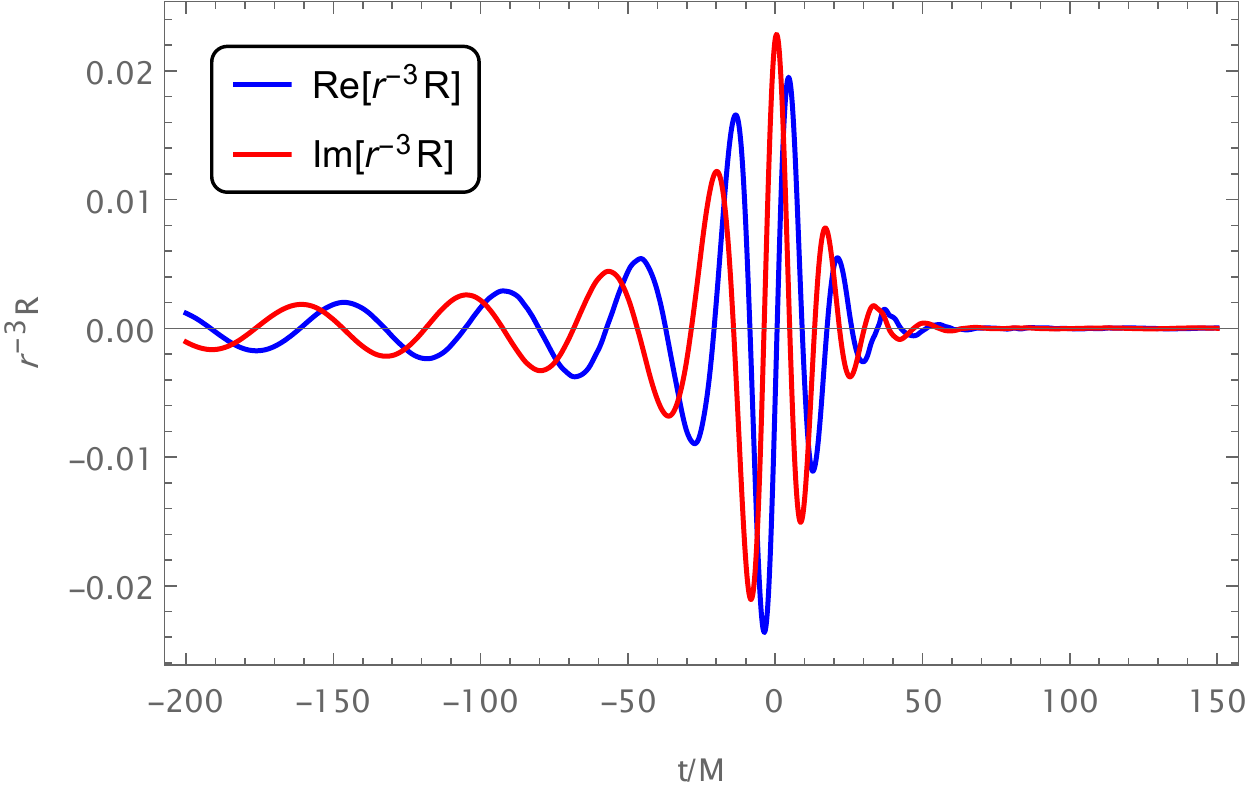}
	\includegraphics[width=5.5cm]{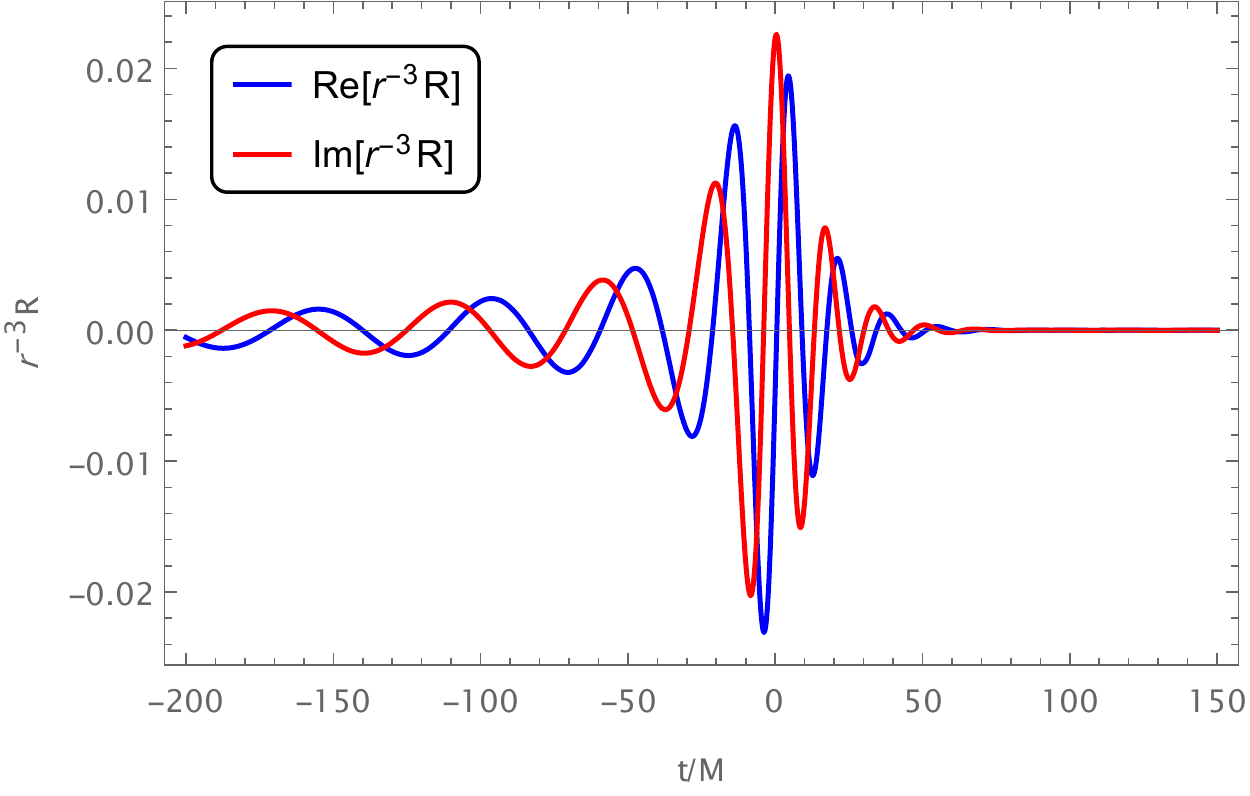}
	\includegraphics[width=5.5cm]{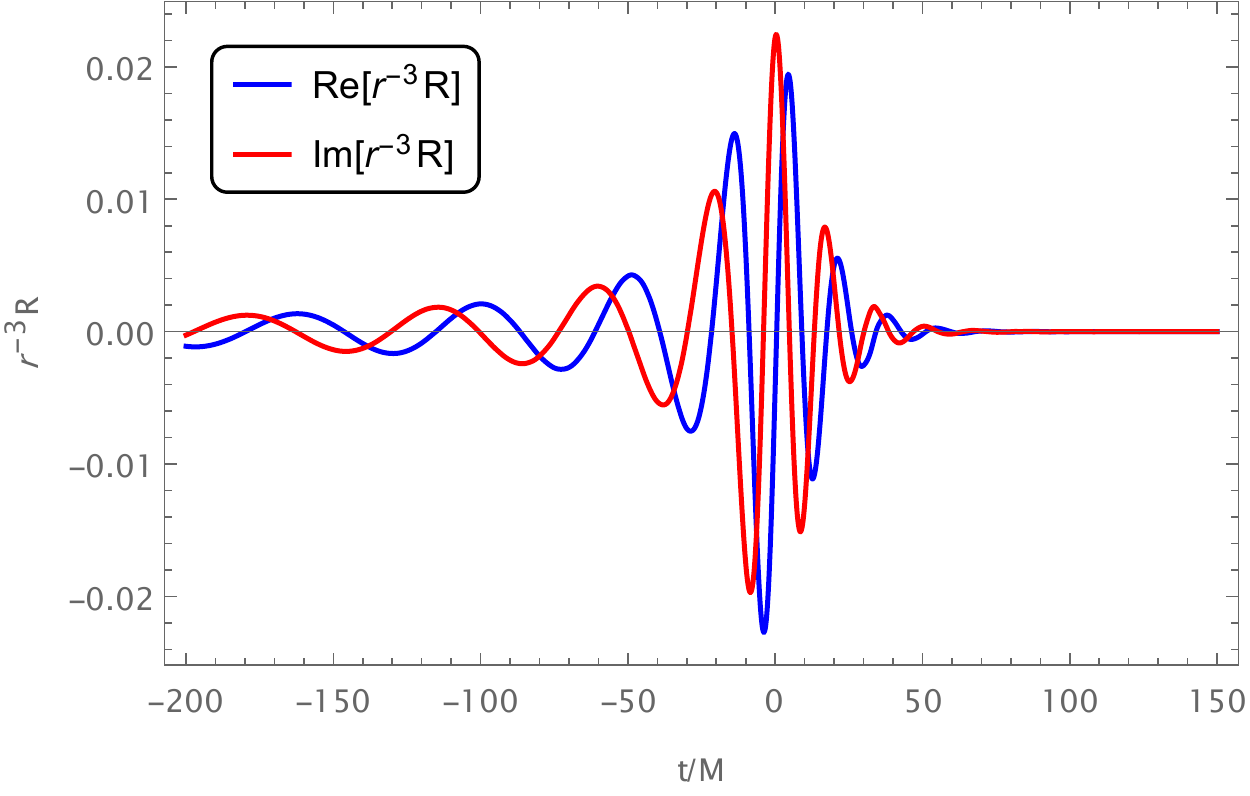}	
	\caption{$\psi_4$ Waveform at infinity for EOB trajectory. The left figure is for $\eta=0.1$, the middle figure is for $\eta=0.16$ and the right figure is for $\eta=0.22$}
	\label{fig:EOB_Psi4_inf}
\end{figure*}
%
%
%
\begin{figure*}[htb!]
	\centering
	\includegraphics[width=\columnwidth]{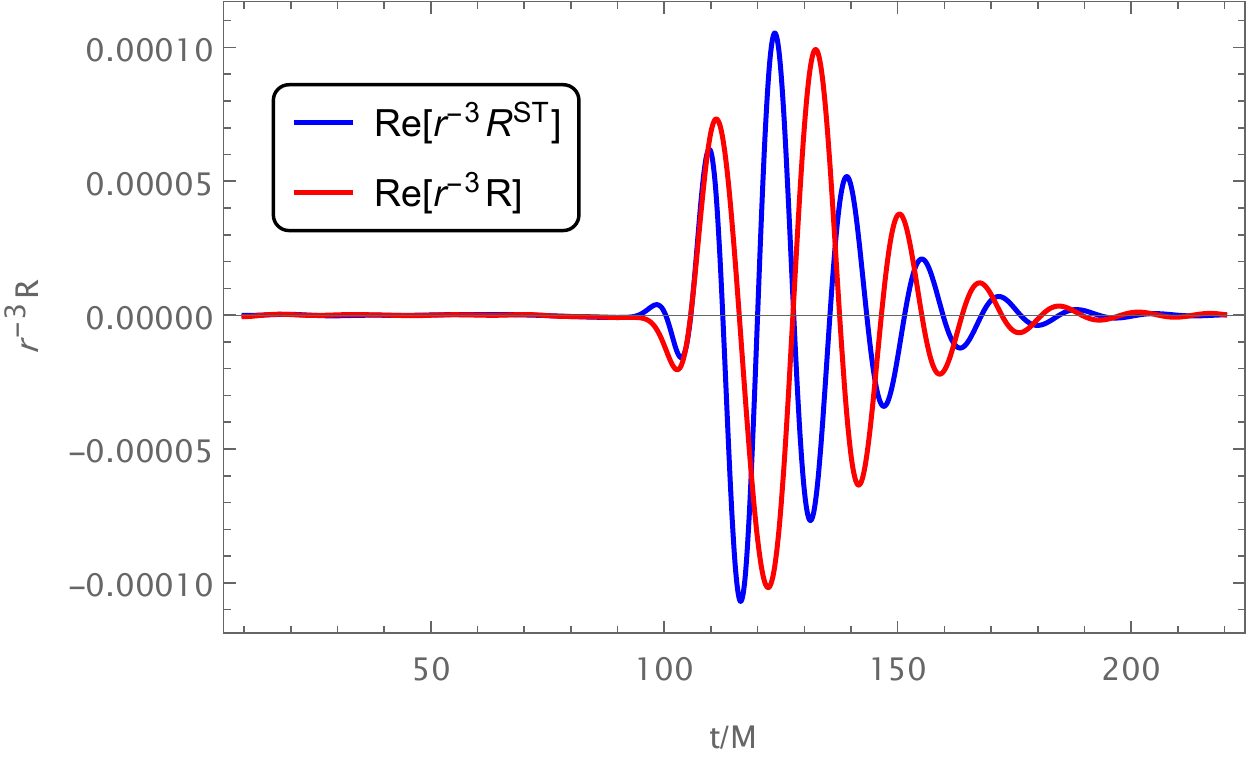}
	\includegraphics[width=\columnwidth]{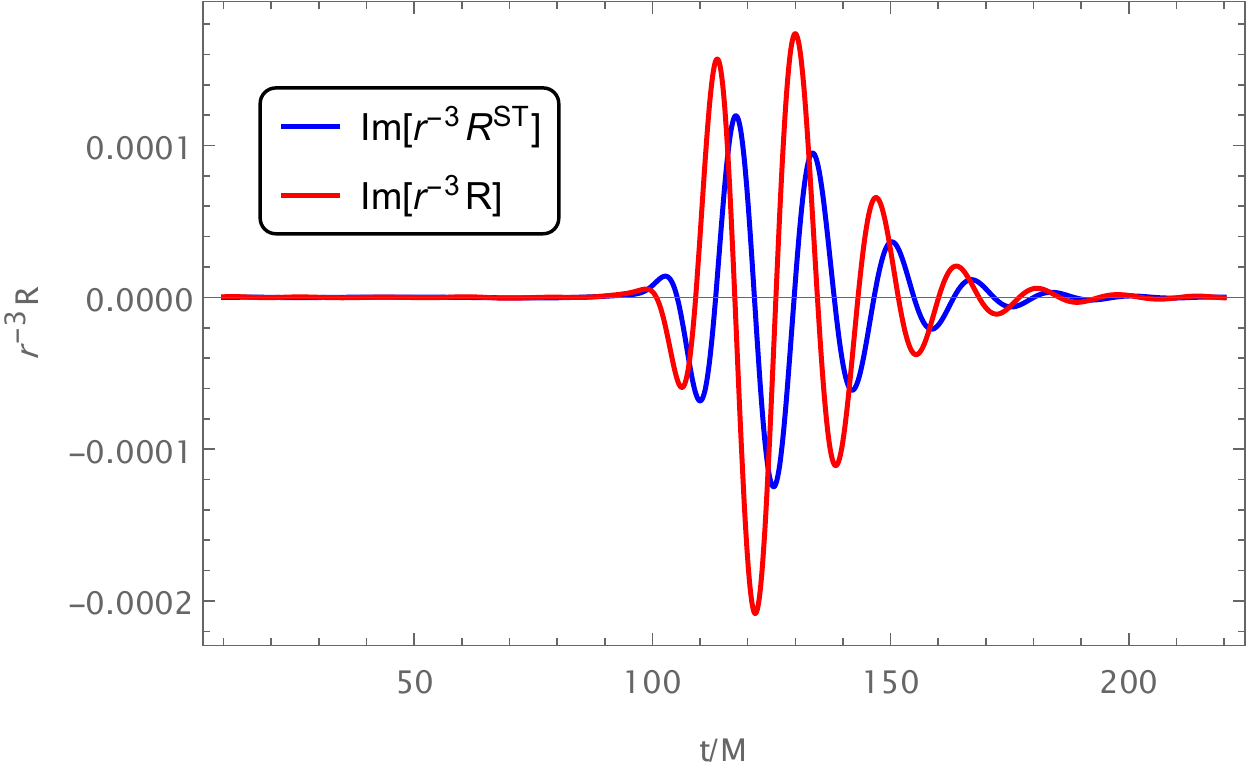}
	\includegraphics[width=\columnwidth]{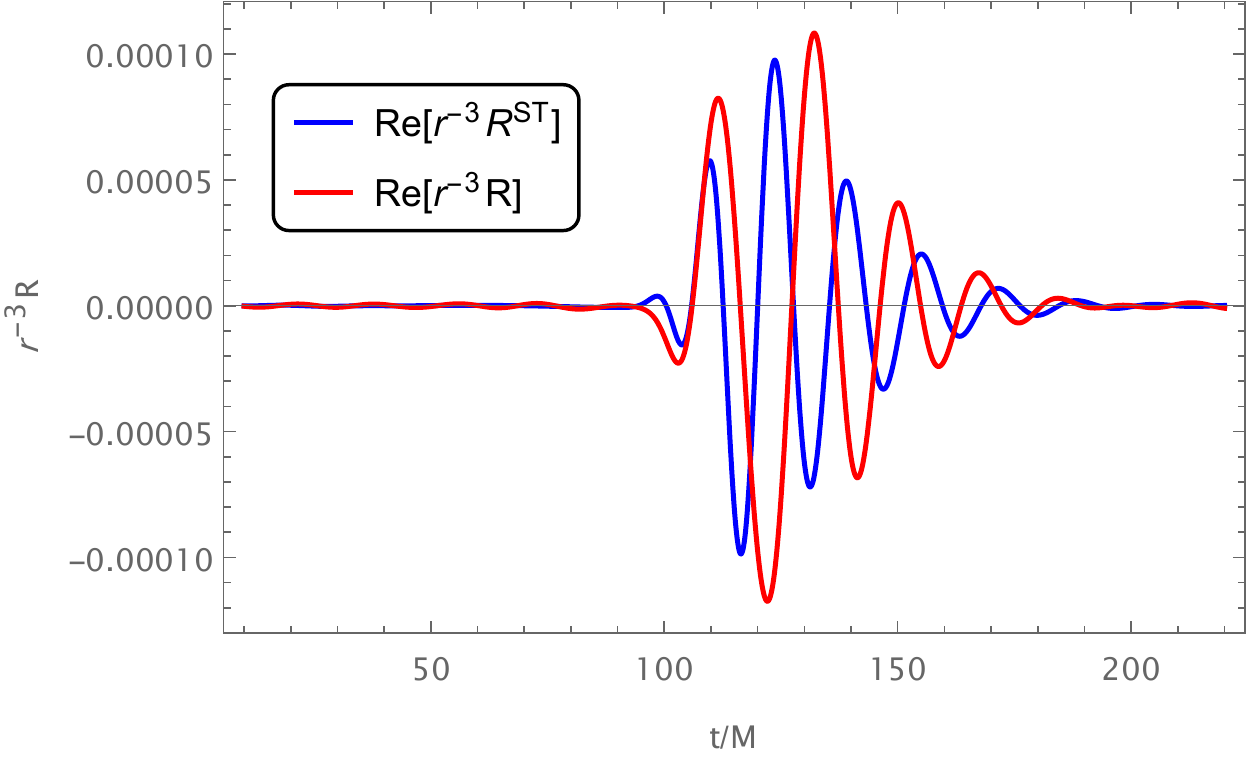}
	\includegraphics[width=\columnwidth]{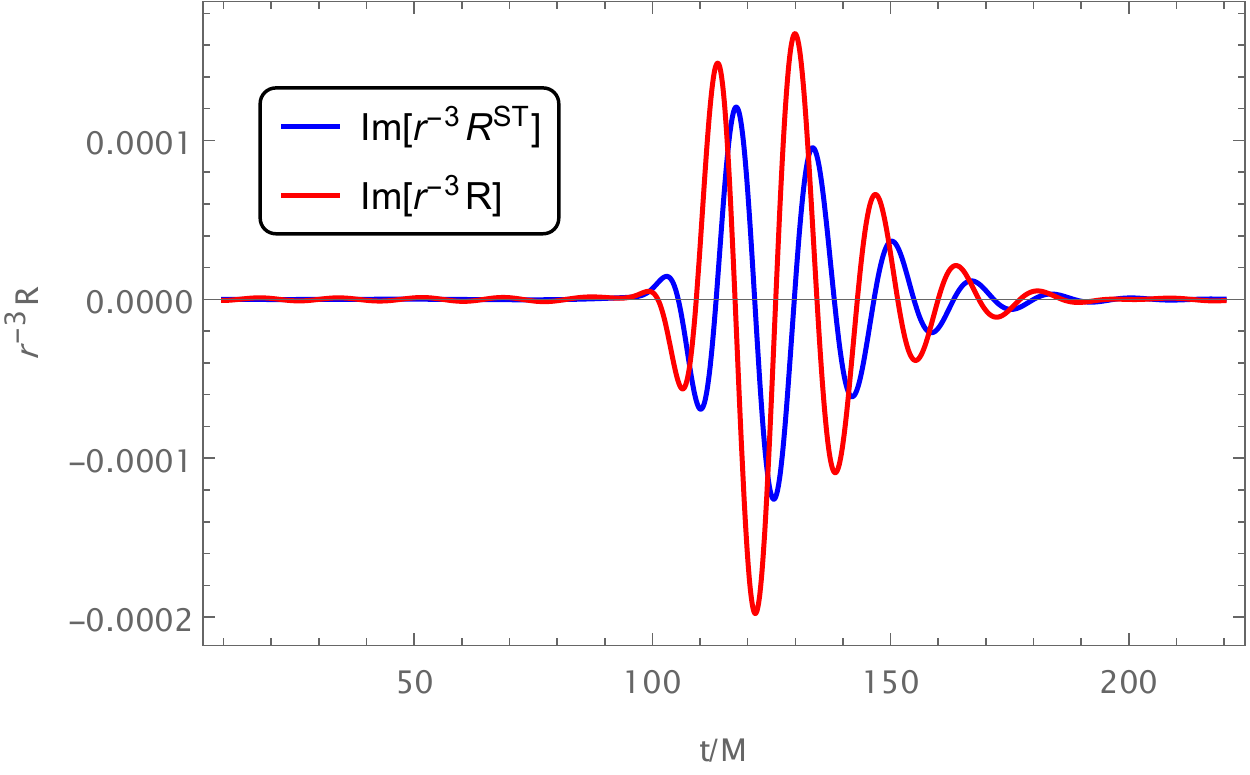}
	\includegraphics[width=\columnwidth]{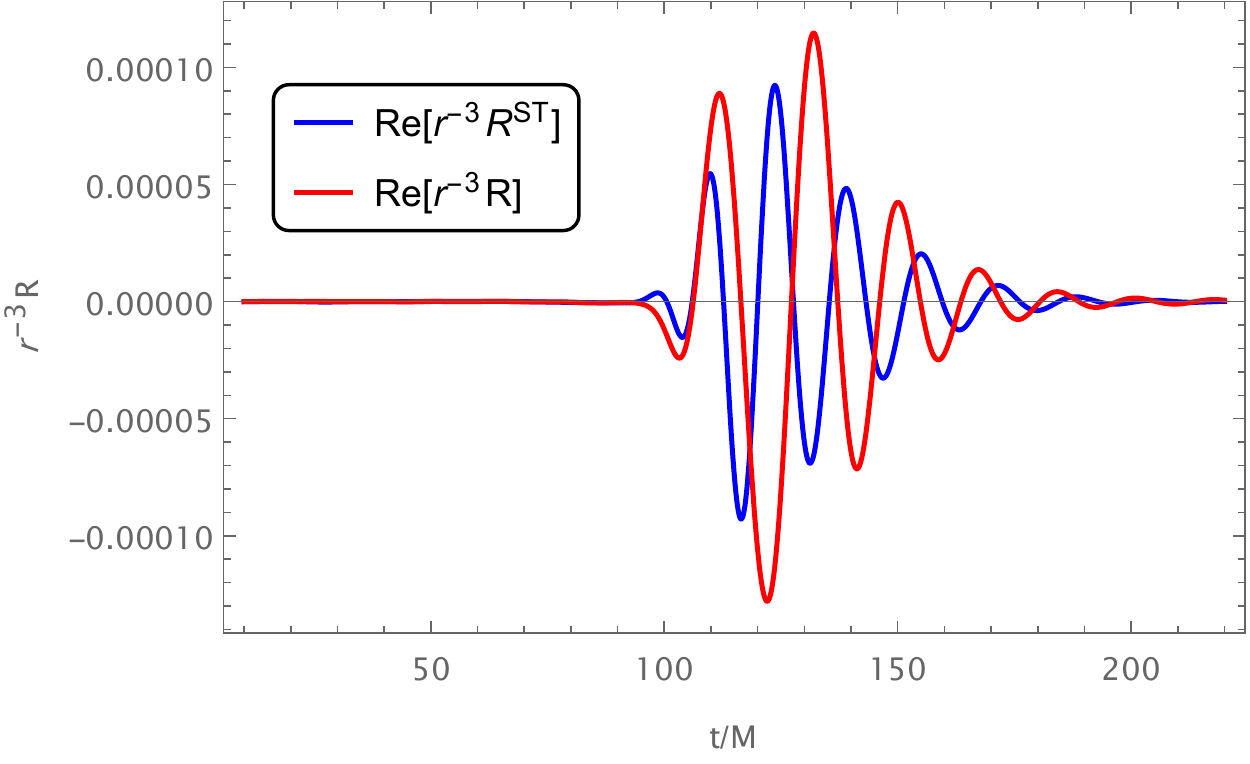}	
	\includegraphics[width=\columnwidth]{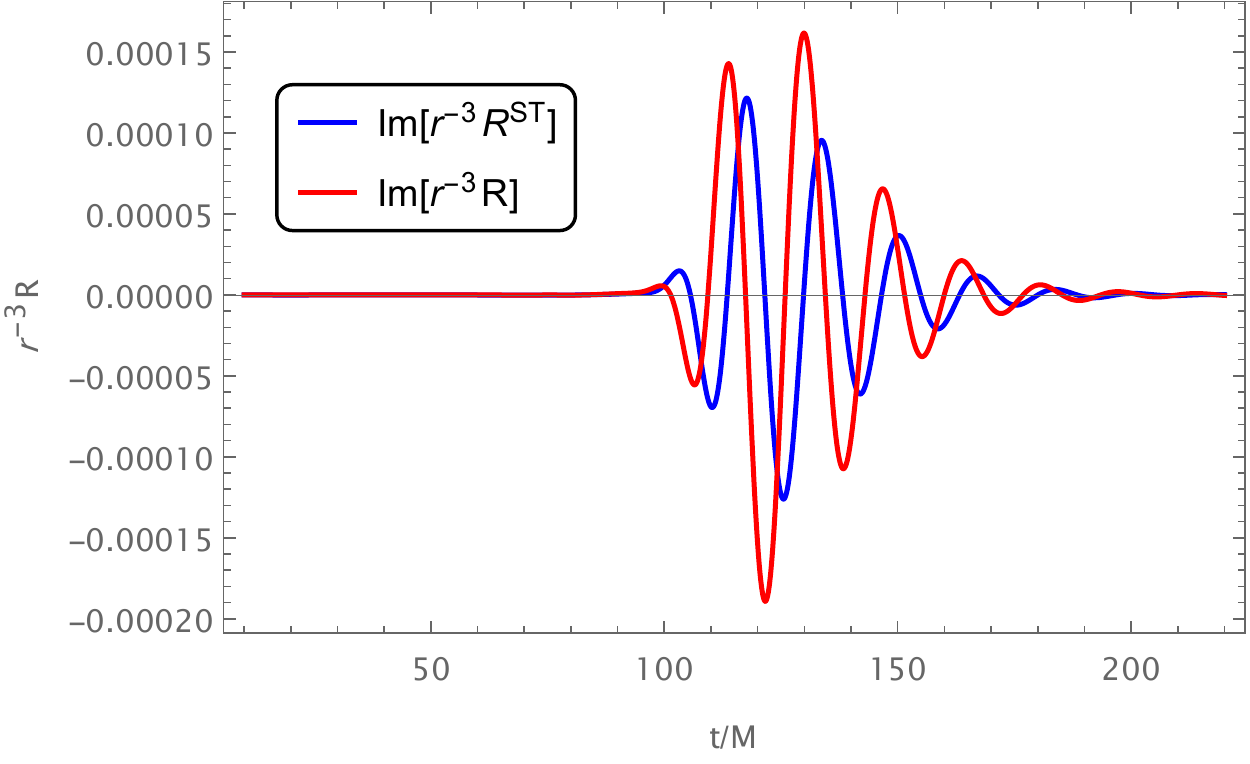}	
	\caption{Out-going echo comparison ($\tilde{\gamma}/M=10^{-6}$) for EOB quasi-circular trajectories. The left column contains the real part comparisons while the right column contains the imaginary part comparisons. The top panel is for $\eta=0.1$, the middle panel is for $\eta=0.16$ and the bottom panel is for $\eta=0.22$}
	\label{fig:EchoCompare}
\end{figure*}

In several works \cite{Xin:2021zir,wang2020echoes,Maggio,micchi2021loud}, echoes in out-going $\psi_4$ waveforms are computed with the aid of $\psi_0$ waveform at the horizon computed via the Starobinksy-Teukolsky identity. This is mainly done due to two obstacles the unavailability of the explicit source term for the $\psi_0$ Teukolsky equation and the convergence issues of solutions of the Teukolsky equation. In this paper, we have derived the explicit $\psi_0$ source term and have suggested a regularization scheme to get convergent solutions of the Teukolsky equation, thus overcoming both the difficulties. Hence we can now compute the out-going $\psi_4$ echoes at infinity using the correct horizon $\psi_0$ waveform. In the subsequent sections, we will do some comparative analysis of such echoes in Schwarzschild background.


In Ref.~\cite{chen2020tidal}, the boundary condition for gravitational waves near the ECO surface was formulated as a relation between in-going $\psi_0$ and out-going $\psi_4$ in that region: both are directly connected to the tidal fields experienced by zero-angular momentum, near-horizon fiducial observers staying at constant redshift (with respect to infinity) in Kerr spacetime.  More specifically, 
the first (also the most dominant) echo waveform in the outgoing $\psi_4$ can be expressed as \cite{Xin:2021zir}:
\begin{equation}
    \label{first_echo}
       Z^{\infty \, {\rm Echo}}_{\ell m \omega} =
      \mathcal{R}_{\ell m \omega}^{\rm ECO}  \mathcal{J}_{\ell m \omega} Y^{\rm in}_{\ell m \omega}
\end{equation}
where
\begin{equation}
    \mathcal{J}_{\ell m \omega}=\frac{(-1)^{m+1}}{4}
    \frac{D_{\ell m \omega}^{(4)\infty}}{D_{\ell m \omega}^{(4)\rm out}} .
\end{equation}
$D_{\ell m \omega}^{(4)\infty}$ and $D_{\ell m \omega}^{(4)\rm out}$ are defined in Eq. \eqref{eq_Rinf_psi4}. Here $Z^{\infty \, {\rm Echo}}_{\ell m \omega}$ are components of out-going $\psi_4$ at infinity, while $Y^{\rm in}_{\ell m \omega}$, defined in Eq. \eqref{eqn:Yin_Def}, are components of $\psi_0$ on the horizon.  The quantity $\mathcal{R}_{\ell m \omega}^{\rm ECO}$ is the reflectivity directly associated with the tidal response of the ECO~\cite{chen2020tidal}, and is modelled in several ways \cite{wang2020echoes,Oshita_2020}. In this work, we use the Boltzmann reflectivity model \cite{wang2020echoes,Oshita_2020}  which is given by: %
\begin{equation}
\label{eq:RB}
    \mathcal{R}_{\ell m \omega}^{\rm ECO}=\mathcal{R}_{\ell m \omega}^{B} =\exp\left(-\frac{|k|}{2 T_H}\right)
    \exp\left[
    -i\frac{k}{\pi T_H}\log(\tilde{\gamma}|k|)\right],
\end{equation}
where $T_H$ is the Hawking temperature of the Kerr black hole and $k$ is defined in \eqref{eqn:rstarkdef}. The quantity $\tilde{\gamma}$ is a free parameter of the model that determines the separation between the main wave and the echo. Note that $T_H=1/(8\pi M)$ and $k=\omega$ for a Schwarzschild black hole. 
\vspace{\baselineskip}
\newline
In this section, we have computed the first echo in $\psi_4$ waveform at infinity in two ways. In the first way, we use the in-going $\psi_0$ waveform at horizon, $Y^{\rm in}_{\ell m \omega}$, computed directly from the Teukolsky equation as in \eqref{eqn:Regularize_Solution_at_Horizon}, insert it into Eq.~\eqref{first_echo} to obtain the first echo at infinity.  In the second way, we first obtain the $\psi_0$  waveform at horizon, $Y^{\rm in}_{\ell m \omega}$, by applying the Starobinsky-Teukolsky~\eqref{eq:ST} to the $\psi_4$ waveform at horizon, $Z^{\rm in}_{\ell m \omega}$, and then insert $Y^{\rm in}_{\ell m \omega}$ into Eq.~\eqref{first_echo}).  The latter approach is more commonly followed in literature \cite{Xin:2021zir,wang2020echoes,Maggio,micchi2021loud}. 

We use the same EOB trajectories for the particle plunging in to the black holes as in section \eqref{sec:EOB_Psi0}.
Fig.~\ref{fig:EOB_Psi4_inf} contains the waveforms of the main outgoing $\psi_4$ wave (for $l=m=2$) at infinity (given in Eq. \eqref{eqn:Zin_Psi4}) for the three different trajectories considered in this paper. The echo in Eq. \eqref{first_echo} adds to this main wave \eqref{eqn:Yin_Def} to give the total $\psi_4$ waveform. The comparative plots for the first echoes (for $l=m=2$) computed using $\psi_0$ calculated directly from the Teukolsky equation and using $\psi_4$ with the ST identity are illustrated in Fig.~\ref{fig:EchoCompare}. The ``ST'' in the superscript corresponds to the use of Starobinsky-Teukolsky identity. It is evident that although the echoes are qualitatively similar, there are significant quantitative differences between the two. We believe the first way of using $\psi_0$ computed directly from the Teukolsky equation is the  correct way of computing echoes. As discussed in Section \ref{sec:Waveform_Comparison}, this is because the Starobinsky-Teukolsky identity holds for vacuum spacetimes and a particle plunging in the black hole background makes the spacetime non vacuum.


Our results also suggest that the echoes calculated using the  $\psi_0$ waveforms computed directly from the Teukolsky equation are slightly stronger than the ones calculated using $\psi_4$ waveform and the ST identity. For a quantitative analysis, let:
\begin{equation}
   \chi= \frac{\sqrt{\int |Z^{\infty \, {\rm Echo}}_{\ell m \omega}|^{2}dt }}{\sqrt{\int |Z^{\infty\,ST \, {\rm Echo}}_{\ell m \omega}|^{2}dt }}
\end{equation}
where the numerator in the above definition has the echo waveform calculated using $\psi_0$ waveform directly from the Teukolsky equation while the denominator has the echo waveform calculated employing Starobinsky-Teukolsky identity. For the three different trajectories, we get the  following values of $\chi$:
\begin{equation}
    \label{eqn:power comparison}
    \begin{split}
    &\chi(\eta=0.10)=1.40222\\
    &\chi(\eta=0.16)=1.43018\\
    &\chi(\eta=0.22)=1.44997\\
    \end{split}
\end{equation}
This indicates that the echoes obtained using the $\psi_0$ waveform directly computed from the Teukolsky equation are stronger than those whose computation involved the ST identity. Eqs.~\eqref{eqn:power comparison} also indicates a trend that echoes computed using $\psi_0$ directly computed from the Teukolsky equation become stronger and stronger in comparison to the echoes computed using $\psi_4$ and the ST identity with increasing values of $\eta$.
%
%
%
%
\newline
%
\section{Discussions and Conclusions}
\label{sec:Conclusion}
In this paper, we have computed the source term of the $\psi_0$ Teukolsky equation explicitly in terms of the trajectory parameters for a point particle plunging into a Kerr black hole. An analytic expression for the $\psi_0$ source term enables us to compute the $\psi_0$ waveform directly from Teukolsky equation. We have also proposed a regularization scheme to handle the divergences appearing in the integrals of Teukolsky equation solutions for the $\psi_0$ sector. %
\vspace{\baselineskip}
\newline
We have computed the $\psi_0$ waveform close to the horizon due to a particle plunging into a Schwarzschild black hole via different trajectories including a purely radial in-fall and other quasi-circular trajectories. We have also computed the first echoes in the outgoing $\psi_4$ waveform employing the directly computed $\psi_0$ waveforms. We have explicitly shown that, for a particle plunging into a Schwarzschild black hole, there are significant quantitative differences in the $\psi_0$ waveforms (and corresponding $\psi_4$ echoes) when computed directly from the Teukolsky equation compared to when they are computed employing the Starobinsky-Teukolsky identity on $\psi_4$. We believe using the Teukolsky equation directly to compute these waveforms is the more accurate way in non-vacuum spacetimes, for example, in the scenario of a particle plunging into the horizon. Our calculation is not required when the particle stays far away from the horizon, for example, when studying tidal interactions between binary black holes during the inspiral stage. 
\vspace{\baselineskip}
\newline
Note that the derivation of the source term as well as the proposition of the regularization scheme have been done for a general arbitrarily rotating Kerr black hole. But for numerical simplicity, the numerical computation of waveforms and echoes has been done for a Schwarzschild background. We aim to extend the computational analysis to the case of a Kerr background in a future publication. 
\section*{Acknowledgements} \label{sec:acknowledgements}
The authors thank Baoyi Chen, Sizheng Ma, and Qingwen Wang for valuable inputs and discussions. We are especially grateful to Shuo Xin for his help. This work is part of the Dual Degree Thesis project of MS. We thank S. Shankaranarayanan for his constant support during the project. Y.C.\ is supported by the Simons Foundation (Award Number 568762) and the National Science Foundation (Grants PHY--2011968, PHY--2011961 and PHY--1836809). 

\appendix

\begin{widetext}

\section{Calculating the Source Term}
\label{sec:App_Calc_Source_Term}


We would like to calculate the source terms for the radial inhomogeneous Teukolsky Equation for $\psi^B_0$. Similar source term for $\psi^B_4$ is available in the literature \cite{Mino:1997bx}. The decoupled equations for the Weyl scalar perturbations are \cite{Teukolsky:1973ha}:
\begin{equation}
\label{eqn:psi0DecoupledEqn}
    \begin{aligned}
\left[\left(D-3 \epsilon+\epsilon^{*}-4 \rho-\rho^{*}\right)(\mathbf{\overline{\Delta}}-4 \gamma+\mu)\right.\\
&\left.-\left(\delta+\pi^{*}-\alpha^{*}-3 \beta-4 \tau\right)\left(\delta^{*}+\pi-4 \alpha\right)-3 \psi_{2}\right] \psi_{0}^{B} &=4 \pi T_{0}
\end{aligned}
\end{equation}
where
\begin{equation}
\label{eqn:T0sourceinNP}
   \begin{split}
T_{0} =&\left(\delta+\pi^{*}-\alpha^{*}-3 \beta-4 \tau\right)\left[\left(D-2 \epsilon-2 \rho^{*}\right) T_{l m}^{B}-\left(\delta+\pi^{*}-2 \alpha^{*}-2 \beta\right) T_{l l}^{B}\right] \\& 
+\left(D-3 \epsilon+\epsilon^{*}-4 \rho-\rho^{*}\right)\left[\left(\delta+2 \pi^{*}-2 \beta\right) T_{l m}^{B}-\left(D-2 \epsilon+2 \epsilon^{*}-\rho^{*}\right) T_{m m}^{B}\right]
\end{split}
\end{equation}
and 
\begin{equation}
\label{eqn:psi4DecoupledEqn}
\begin{aligned}
\left[\left(\mathbf{\overline{\Delta}}+3 \gamma-\gamma^{*}+4 \mu+\mu^{*}\right)(D+4 \epsilon-\rho)\right.\\&
\left.-\left(\delta^{*}-\tau^{*}+\beta^{*}+3 \alpha+4 \pi\right)(\delta-\tau+4 \beta)-3 \psi_{2}\right] \psi_{4}^{B} &=4 \pi T_{4}
\end{aligned}
\end{equation}
where
\begin{equation}
\label{eqn:T4souceinNP}
    \begin{split}
T_{4}=\left(\mathbf{\overline{\Delta}}+3 \gamma-\gamma^{*}+4 \mu+\mu^{*}\right)\left[\left(\delta^{*}-2 \tau^{*}+2 \alpha\right) T^B_{n m *}-\left(\mathbf{\overline{\Delta}}+2 \gamma-2 \gamma^{*}+\mu^{*}\right) T^B_{m * m *}\right] \\
\quad+\left(\delta^{*}-\tau^{*}+\beta^{*}+3 \alpha+4 \pi\right)\left[\left(\mathbf{\overline{\Delta}}+2 \gamma+2 \mu^{*}\right) T^B_{n m *}-\left(\delta^{*}-\tau^{*}+2 \beta^{*}+2 \alpha\right) T^B_{n n}\right]
\end{split}
\end{equation}
When we separate the decoupled equations into a radial and angular equations, the solutions of the angular equation are spin weighted spheroidal harmonics $_{s}S_{lm}(\theta)$. To get the source term for the radial equation we write:
\begin{equation}
\label{eqn:App_Source_r_theta_decomposition}
    4 \pi \Sigma T=\int d \omega \sum_{l, m} T_{lm\omega}(r)_{s} S_{l}^{m}(\theta) e^{i m \varphi} e^{-i \omega t}
\end{equation}
\begin{equation}
\label{eqn:Psi0_r_theta_decomposition}
\psi_0^B=\int d \omega \sum_{l, m} R^{(0)}(r)_{2} S_{lm}(\theta) e^{i m \varphi} e^{-i \omega t}
\end{equation}
\begin{equation}
\label{eqn:Psi4_r_theta_decomposition}
\psi_4^B=\rho^4 \int d \omega \sum_{l, m} R^{(4)}(r)_{(-2)} S_{lm}(\theta) e^{i m \varphi} e^{-i \omega t}
\end{equation}
We know $T$ (related to $T_0$ and $T_4$ in \eqref{eqn:TwithT0T4}),   from which we need to find the form of $T_{lm\omega}(r)$ using \eqref{eqn:App_Source_r_theta_decomposition}. 
\subsection{Orthogonality Relations and Normalization}
Following \cite{Teukolsky:1974yv}, the spin weighted spheroidal harmonics are normalized as:
 \begin{equation}
 \label{eqn:Slm_normalisation}
     \int_{0}^{\pi} {}_{s}S_{l m}^{2}(\theta) \sin \theta d \theta=1
 \end{equation}
 So the orthogonality relation considered is:
 \begin{equation}
 \label{eqn:Spheroidal_Harmonics_Orthogonality}
     \int_{S^{2}} {}_{s}S_{l m}e^{im\varphi} ({}_{s}{S}_{l^{\prime} m^{\prime}})e^{-im^{\prime}\varphi} d \Omega=2\pi\delta_{l l^{\prime}} \delta_{m m^{\prime}}
 \end{equation}
 The 2$\pi$ in the RHS is required to satisfy the normalization \eqref{eqn:Slm_normalisation}. The other relation to be used is the Dirac-delta definition:
 \begin{equation}
 \label{eqn:DiracDeltaDef}
     \delta(x-\alpha)=\frac{1}{2 \pi} \int_{-\infty}^{\infty} e^{i t(x-\alpha)} d t
 \end{equation}
 \subsection{General expression for $T_{lm\omega}(r)$}
 Using the orthogonality relations with \eqref{eqn:Source_r_theta_decomposition}, we get:
 \begin{equation}
     T_{\ell m \omega}=4 \int d \Omega d t\, \rho^{-1} \rho^{*\,-1}\left(\frac{T}{2}\right) e^{-i m \varphi+i \omega t} \frac{_{s}S_{\ell m}^{a \omega}}{{2 \pi}}
 \end{equation}
 where 
 \begin{equation}
 \label{eqn:TwithT0T4}
     \begin{split}
          &  T=2T_0 \qquad\qquad \text{for} \quad s=2\\
          & T=2\rho^{-4}T_4\qquad \text{for} \quad s=-2
     \end{split}
 \end{equation}
This gives:
\begin{equation}
\label{eqn:T4lmomega}
     T_{\ell m \omega}^{(4)}=4 \int d \Omega d t\, \rho^{-5} \rho^{*\,-1}\left(T_4\right) e^{-i m \varphi+i \omega t} \frac{_{-2}S_{\ell m}^{a \omega}}{{2 \pi}}
\end{equation}
\begin{equation}
\label{eqn:T0lmomega}
    T_{\ell m \omega}^{(0)}=4 \int d \Omega d t\, \rho^{-1} \rho^{*\,-1}\left(T_0\right) e^{-i m \varphi+i \omega t} \frac{_{2}S_{\ell m}^{a \omega}}{{2 \pi}}
\end{equation}
\subsection{Some Definitions and Algebraic Tricks}
\label{sec:AlgebraicTricks}
The Boyer Lindquist form of the Kerr metric is:
\begin{equation}
\label{eqn:KerrMetric}
    \begin{split}
d s^{2}=&(1-2 M r / \Sigma) d t^{2}+\left(4 M a r \sin ^{2}(\theta) / \Sigma\right) d t d \varphi-(\Sigma / \Delta) d r^{2}-\Sigma d \theta^{2} \\
&-\sin ^{2}(\theta)\left(r^{2}+a^{2}+2 M a^{2} r \sin ^{2}(\theta) / \Sigma\right) d \varphi^{2}
\end{split}
\end{equation}
The NP tetrad basis chosen for this metric is:
\begin{equation}
\label{eqn:tetradBasis}
    \begin{split}
&l^{\mu} =\left[\left(r^{2}+a^{2}\right) / \Delta, 1,0, a / \Delta\right], \quad n^{\mu}=\left[r^{2}+a^{2},-\Delta, 0, a\right] /(2 \Sigma) \\
& m^{\mu} =[i a \sin \theta, 0,1, i / \sin \theta] /\left[2^{1 / 2}(r+i a \cos \theta)\right]
\end{split}
\end{equation}
\begin{equation}
    \begin{split}
&l_{\mu} =\left[1, -\left(\frac{\Sigma}{\Delta} \right), 0, -a\sin^2\theta \right], \quad n_{\mu}=\left[\Delta, \Sigma, 0, -a\Delta\sin^2\theta\right] /(2 \Sigma) \\
& m_{\mu} =[i a \sin \theta, 0,-\Sigma, -i (r^2+a^2) \sin \theta] /\left[2^{1 / 2}(r+i a \cos \theta)\right]
\end{split}
\end{equation}
with non-zero spin coefficients being:
\begin{equation}
    \begin{split}
&\rho=-1 /(r-i a \cos \theta), \quad \beta=-\rho^{*} \cot \theta /(2 \sqrt{2}), \quad \pi=i a \rho^{2} \sin \theta / \sqrt{2}\\
&\tau=-i a \rho \rho^{*} \sin \theta / \sqrt{2}, \quad \mu=\rho^{2} \rho^{*} \Delta / 2,\quad \gamma=\mu+\rho \rho^{*}(r-M) / 2\\ &\alpha=\pi-\beta^{*}
\end{split}
\end{equation}
The directional derivatives are given by:
\begin{equation}
    \label{eqn:Directional_Derivatives}
    D=l^{\mu}\partial_{\mu}\,;\qquad \mathbf{\overline{\Delta}}=n^{\mu}\partial_{\mu}\,;\qquad \delta=m^{\mu}\partial_{\mu}\,;
\end{equation}
The only non-zero Weyl Scalar for the Kerr metric is:
\begin{equation}
    \psi_2=M\rho^3
\end{equation}
While the other four Weyl scalars zero. Here we have used the notation:
\begin{equation}
    \Sigma=r^2+a^2\cos^2\theta=\rho^{-1}\rho^{*\,-1}
\end{equation}
\begin{equation}
    \Delta=r^2+a^2-2Mr
\end{equation}
We also define the following differential operators:
\begin{equation}
    \mathscr{L}_{n}=\partial_{\theta}+\frac{m}{\sin \theta}-a\, \omega \sin \theta+n \cot \theta
\end{equation}
\begin{equation}
    \mathscr{L}_{n}^{\dagger}=\mathscr{L}_{n}(-\omega,-m)
\end{equation}
\begin{equation}
    \mathscr{D}=\partial_{r}-i K / \Delta
\end{equation}
\begin{equation}
    \mathscr{D}^{\dagger}=\mathscr{D}(-\omega,-m)=\partial_{r}+i K / \Delta
\end{equation}
Now some of the algebraic tricks with these differential operators that are useful in calculating the form of $T_{lmw}$ are:
\begin{equation}
    \left [ \mathscr{L}_i + ia\sin\theta\left ( n\rho^{*}-m\rho \right ) \right ]\zeta=(\rho^{*})^n\rho^m \mathscr{L}_i\left [(\rho^{*})^{-n}\rho^{-m}\zeta  \right ]
\end{equation}
\begin{equation}
    \Delta ^{-j}\rho^{-m}(\rho^{*})^{-n}\mathscr{D}\left ( \Delta ^{j}\rho^{m}(\rho^{*})^{n}\zeta \right )=\left [ \mathscr{D}+m\rho+n\rho^{*}+ \frac{2j(r-M)}{\Delta} \right ]\zeta
\end{equation}
\begin{equation}
    \mathscr{L}_n\left ( \frac{\zeta}{\sin\theta} \right )=\frac{\mathscr{L}_{n-1}(\zeta)}{\sin\theta}
\end{equation}
Note that the above three equations are valid even for $\mathscr{L}^{\dagger}$ and $\mathscr{D}^{\dagger}$. 
And finally an integration by parts identity when $\zeta_1$ vanishes fast enough near $\theta=0$ and $\theta=\pi$: 
\begin{equation}
    \int_{0}^{\pi}d\theta \zeta_1\mathscr{L}_n(\zeta_2)=-\int_{0}^{\pi}d\theta \zeta_2\mathscr{L}_{-n}^{\dagger}(\zeta_1)
\end{equation}
\subsection{Form of $T_4$ and $T_0$ in the Metric}
Using the equations, definitions, tricks mentioned in section \eqref{sec:AlgebraicTricks}, we can write $T_4$ given by equation \eqref{eqn:T4souceinNP} as:
\begin{equation}
\label{eqn:T4Step1}
    \begin{split}
T_4=&-\frac{1}{2} \rho^{8} \rho^{*} \mathscr{L}_{-1}\left[\rho^{-4} \mathscr{L}_{0}\left(\rho^{-2} \rho^{*\,-1} T_{n n}^B\right)\right]
+\frac{1}{2 \sqrt{2}} \rho^{8} \rho^{*} \Delta^{2} \mathscr{L}_{-1}\left[\rho^{-4} \rho^{*\,2} \mathscr{D}^{\dagger}\left(\rho^{-2} \rho^{*\,-2} \Delta^{-1} T_{m^{*} n}^B\right)\right]\\
&-\frac{1}{4} \rho^{8} \rho^{*} \Delta^{2} \mathscr{D}^{\dagger}\left[\rho^{-4} \mathscr{D}^{\dagger}\left(\rho^{-2} \rho^{*} T_{m^{*} m^{*}}^B\right)\right]
+\frac{1}{2 \sqrt{2}} \rho^{8} \rho^{*} \Delta^{2} \mathscr{D}^{\dagger}\left[\rho^{-4} \rho^{*\,2} \Delta^{-1} \mathscr{L}_{-1}\left(\rho^{-2} \rho^{*\,-2} T_{m^{*} n}^B\right)\right]
\end{split}
\end{equation}
The sign difference in 2 terms (with prefactors $\frac{1}{2\sqrt{2}}$)  when compared with equation (2.15) of \cite{Mino:1997bx} is due to different sign in the definition of $\rho$. Now using the equations, definitions, tricks mentioned in section \eqref{sec:AlgebraicTricks}, we can write $T_0$ given by equation \eqref{eqn:T0sourceinNP} as:
\begin{equation}
\label{eqn:T0Step2}
\begin{split}
    T_0=&-\frac{1}{2} \rho^{4} \rho^{*} \mathscr{L}_{-1}^{\dagger}\left[\rho^{-4} \mathscr{L}_{0}^{\dagger}\left( \rho^{*} T_{l l}^B\right)\right]
-\frac{1}{\sqrt{2}} \rho^{4} \rho^{*} \mathscr{L}_{-1}^{\dagger}\left[\rho^{-4} \rho^{*\,2} \mathscr{D}\left( \rho^{*\,-2} T_{l m}^B\right)\right]\\
&- \rho^{4} \rho^{*} \mathscr{D}\left[\rho^{-4} \mathscr{D}\left( \rho^{*\,-1} T_{m m}^B\right)\right]
-\frac{1}{\sqrt{2}} \rho^{4} \rho^{*} \mathscr{D}\left[\rho^{-4} \rho^{*\,2} \mathscr{L}_{-1}^{\dagger}\left(\rho^{*\,-2} T_{l m}^B\right)\right]
\end{split}
\end{equation}
\subsection{Stress Energy Tensor components}
The perturbation stress energy tensor is due to a test particle of mass $\mu$ moving with four velocity $u^\alpha$ in the black hole spacetime. Let $x'$ be an event in spacetime and $x(\tau)$ be the test particle's world line, i.e. the geodesic along which the particle is moving. The Stress Energy Tensor is given by ($\tau$ is the proper time along the geodesic) \cite{Poisson:1996ya}:
\begin{equation}
\label{eqn:StressTensorTestParticle}
   T^{\alpha \beta}\left(x^{\prime}\right)=\mu \int d \tau\, u^{\alpha} u^{\beta} \delta\left[x^{\prime}-x(\tau)\right]
\end{equation}
where the delta function is normalised as: 
\begin{equation}
    \int \delta^4(x) \sqrt{-g}\, d^{4} x=1
\end{equation}
where
\begin{equation}
    g=-\Sigma ^2\sin^2\theta
\end{equation}
Therefore the delta fuction can be written as:
\begin{equation}
    \delta^4(x'-x(\tau))=\frac{\delta(r'-r(\tau))\delta(\theta'-\theta(\tau))\delta(\varphi'-\varphi(\tau))\delta(t'-t(\tau))}{\Sigma \sin\theta}
\end{equation}
Now integrating \eqref{eqn:StressTensorTestParticle} with respect to $t'$, we get:
\begin{equation}
\label{eqn:StressTensorSimplified}
    T^{\alpha \beta}=\frac{\mu}{\Sigma \sin \theta d t / d \tau} \frac{d x^{\alpha}}{d \tau} \frac{d x^{\beta}}{d \tau} \delta(r'-r(t)) \delta(\theta'-\theta(t)) \delta(\varphi'-\varphi(t))
\end{equation}
In the Newman Penrose formalism:
\begin{equation}
\label{eqn:TllTnn}
\begin{split}
    &T^B_{n n}=T^{\alpha \beta}n_{\alpha} n_{\beta}, \quad T^B_{n m^{*}}=T^{\alpha \beta}n_{\alpha} m^{*}_{\beta}, \quad T^B_{m^{*} m^{*}}=T^{\alpha \beta}m^{*}_{\alpha} m^{*}_{\beta} \\
    & T^B_{l l}=T^{\alpha \beta}l_{\alpha} l_{\beta}, \,\,\,\quad T^B_{l m}=T^{\alpha \beta}l_{\alpha} m_{\beta}, \quad\,\,\,\, T^B_{m m}=T^{\alpha \beta}m_{\alpha} m_{\beta}
    \end{split}
\end{equation}
We define:
\begin{equation}
\label{eqn:CtermsDef}
    T_{a b}=\left ( \frac{\mu}{\sin\theta} \right )C_{a b}\delta(r'-r(t)) \delta(\theta'-\theta(t)) \delta(\varphi'-\varphi(t))
\end{equation}
where $a$, $b$ can be $l$, $n$, $m$ or $m^{*}$. 

To calculate the C coefficients we make use of the following geodesic equations \cite{Mino:1997bx} (Kerr metric): 
\begin{equation}
\label{eqn:GedesicEq}
    \begin{split}
&\Sigma \frac{d \theta}{d \tau}=\pm\left[C-\cos ^{2} \theta\left\{a^{2}\left(1-E^{2}\right)+\frac{l_{z}^{2}}{\sin ^{2} \theta}\right\}\right]^{1 / 2} \equiv \Theta(\theta)\\
&\Sigma \frac{d \varphi}{d \tau}=-\left(a E-\frac{l_{z}}{\sin ^{2} \theta}\right)+\frac{a}{\Delta}\left(E\left(r^{2}+a^{2}\right)-a l_{z}\right) \equiv \varphi\\
&
\Sigma \frac{d t}{d \tau}=-\left(a E-\frac{l_{z}}{\sin ^{2} \theta}\right) a \sin ^{2} \theta+\frac{r^{2}+a^{2}}{\Delta}\left(E\left(r^{2}+a^{2}\right)-a l_{z}\right) \equiv T \\
&\Sigma \frac{d r}{d \tau}=\pm \sqrt{\mathscr{R}}
\end{split}
\end{equation}
where E, $l_z$, C are energy, z component of angular momentum and Cartar constant of the test particle respectively. And $\mathscr{R}$ is:
\begin{equation}
    \mathscr{R}=\left[E\left(r^{2}+a^{2}\right)-a l_{z}\right]^{2}-\Delta\left[\left(E a-l_{z}\right)^{2}+r^{2}+C\right]
\end{equation}
Using \eqref{eqn:KerrMetric}, \eqref{eqn:tetradBasis}, \eqref{eqn:StressTensorSimplified}, \eqref{eqn:TllTnn}, \eqref{eqn:CtermsDef} we can deduce that:
\begin{equation}
\label{eqn:Cnn}
    C_{n n}=\frac{1}{4 \Sigma^{3} \dot{t}}\left[E\left(r^{2}+a^{2}\right)-a l_{z}+\Sigma \frac{d r}{d \tau}\right]^{2}
\end{equation}
\begin{equation}
\label{eqn:Cmbarn}
    C_{m^{*} n}=\frac{\rho}{2 \sqrt{2} \Sigma^{2} \dot{t}}\left[E\left(r^{2}+a^{2}\right)-a l_{z}+\Sigma \frac{d r}{d \tau}\right]\left[i \sin \theta\left(a E-\frac{l_{z}}{\sin ^{2} \theta}\right)+\Sigma\frac{d\theta}{d\tau}\right]
\end{equation}
\begin{equation}
\label{eqn:Cmbarmbar}
    C_{m^{*} m^{*}}=\frac{\rho^{2}}{2 \Sigma \dot{t}}\left[i \sin \theta\left(a E-\frac{l_{z}}{\sin ^{2} \theta}\right)+\Sigma\frac{d\theta}{d\tau}\right]^{2}
\end{equation}
\begin{equation}
\label{eqn:Cll}
    C_{l l}=\frac{1}{ \Sigma \Delta^2 \dot{t}}\left[E\left(r^{2}+a^{2}\right)-a l_{z}-\Sigma \frac{d r}{d \tau}\right]^{2}
\end{equation}
\begin{equation}
\label{eqn:Clm}
    C_{l m}=-\frac{\rho^{*}}{ \sqrt{2} \Sigma\Delta \dot{t}}\left[E\left(r^{2}+a^{2}\right)-a l_{z}-\Sigma \frac{d r}{d \tau}\right]\left[i \sin \theta\left(a E-\frac{l_{z}}{\sin ^{2} \theta}\right)-\Sigma\frac{d\theta}{d\tau}\right]
\end{equation}
\begin{equation}
\label{eqn:Cmm}
    C_{m m}=\frac{\rho^{*\,2}}{2 \Sigma \dot{t}}\left[i \sin \theta\left(a E-\frac{l_{z}}{\sin ^{2} \theta}\right)-\Sigma\frac{d\theta}{d\tau}\right]^{2}
\end{equation}
Here $\dot{t}=\frac{dt}{d\tau}$. {The equations (\ref{eqn:Cnn} - \ref{eqn:Cmbarmbar}) match with those obtained in \cite{Mino:1997bx} for constant $\theta$ modulo the opposite sign of $\rho$.}
\subsection{Getting back to $T_{lm\omega}$}
We first focus on $T_{l m \omega}^{(4)}$. Substituting \eqref{eqn:T4Step1} back in \eqref{eqn:T4lmomega} and integrating with respect to $\varphi'$, we get:
\begin{equation}
\begin{split}
     T_{l m \omega}^{(4)}=&\frac{4\mu}{2\pi}\int dt e^{i\omega t - im\varphi(t)}\int d\theta'(\sin\theta' \rho^3 (_{-2}S_{lm}(\theta')))\\
    &\left\{-\frac{1}{2}  \mathscr{L}_{-1}\left[\rho^{-4} \mathscr{L}_{0}\left(\rho^{-2} \rho^{*\,-1} \left ( \frac{C_{n n}}{\sin\theta'} \right )\delta(r'-r(t)) \delta(\theta'-\theta(t))\right)\right]\right.\\
&\left.+\frac{1}{2 \sqrt{2}} \Delta^{2} \mathscr{L}_{-1}\left[\rho^{-4} \rho^{*\,2} \mathscr{D}^{\dagger}\left(\rho^{-2} \rho^{*\,-2} \Delta^{-1} \left ( \frac{C_{n m^{*}}}{\sin\theta'} \right )\delta(r'-r(t)) \delta(\theta'-\theta(t))\right)\right]\right.\\
&\left.-\frac{1}{4}  \Delta^{2} \mathscr{D}^{\dagger}\left[\rho^{-4} \mathscr{D}^{\dagger}\left(\rho^{-2} \rho^{*} \left ( \frac{C_{m^{*} m^{*}}}{\sin\theta'} \right )\delta(r'-r(t)) \delta(\theta'-\theta(t))\right)\right]\right.\\
&\left.+\frac{1}{2 \sqrt{2}}  \Delta^{2} \mathscr{D}^{\dagger}\left[\rho^{-4} \rho^{*\,2} \Delta^{-1} \mathscr{L}_{-1}\left(\rho^{-2} \rho^{*\,-2} \left ( \frac{C_{n m^{*}}}{\sin\theta'} \right )\delta(r'-r(t)) \delta(\theta'-\theta(t))\right)\right]\right\}
    \end{split}
\end{equation}
Now integrating by parts and using some of the algebraic tricks, we get:
\begin{equation}
\label{eqn:T4lmwStep2}
\begin{split}
    T_{l m \omega}^{(4)}=&
\frac{4 \mu}{{2 \pi}} \int_{-\infty}^{\infty} d t \int d \theta' e^{i \omega t-i m \varphi(t)} \\
& \times\left[-\frac{1}{2} \mathscr{L}_{1}^{\dagger}\left\{\rho^{-4} \mathscr{L}_{2}^{\dagger}\left(\rho^{3} S\right)\right\} C_{n n} \rho^{-2} \rho^{*\,-1} \delta(r'-r(t)) \delta(\theta'-\theta(t))\right.\\
&\left.-\frac{\Delta^{2} \rho^{*\,2}}{\sqrt{2} \rho}\left(\mathscr{L}_{2}^{\dagger} (S)+i a(\rho -\rho^{*}) \sin \theta S\right) \mathscr{D}^{\dagger}\left\{C_{m^{*} n} \rho^{-2} \rho^{*\,-2} \Delta^{-1} \delta(r'-r(t)) \delta(\theta'-\theta(t))\right\}\right. \\
&\left.-\frac{1}{2 \sqrt{2}} \mathscr{L}_{2}^{\dagger}\left\{\rho^{3} S\left(\rho^{*\,2} \rho^{-4}\right)_{, r'}\right\} C_{m^{*} n} \Delta \rho^{-2} \rho^{*\,-2} \delta(r'-r(t)) \delta(\theta'-\theta(t))\right. \\
&\left.-\frac{1}{4} (\rho^{3} \Delta^{2} S) \mathscr{D}^{\dagger}\left\{\rho^{-4} \mathscr{D}^{\dagger}\left(\rho^{*} \rho^{-2} C_{m^{*} m^{*}} \delta(r'-r(t)) \delta(\theta'-\theta(t))\right)\right\}\right]
\end{split}
\end{equation}
In \eqref{eqn:T4lmwStep2}, S is $_{-2}S_{lm}$. In this form of \eqref{eqn:T4lmwStep2} we can readily integrate the delta function in $\theta$. Again there is a sign difference in 2 terms when compared to \cite{Mino:1997bx} due to different definition of $\rho$.
\vspace{\baselineskip}
\newline
Similarly for $T_{l m \omega}^{(0)}$, substituting \eqref{eqn:T0Step2} back in \eqref{eqn:T0lmomega} and integrating with respect to $\varphi'$, we get:
\begin{equation}
    \begin{split}
    T_{l m \omega}^{(0)}=& \frac{4\mu}{2\pi}\int dt e^{i\omega t - im\varphi(t)}\int d\theta'(\sin\theta' \rho^3 (_{2}S_{lm}(\theta')))\\
    &\left\{-\frac{1}{2}  \mathscr{L}_{-1}^{\dagger}\left[\rho^{-4} \mathscr{L}_{0}^{\dagger}\left( \rho^{*} \left ( \frac{C_{l l}}{\sin\theta'} \right )\delta(r'-r(t)) \delta(\theta'-\theta(t))\right)\right]\right.\\
&\left.-\frac{1}{ \sqrt{2}} \mathscr{L}_{-1}^{\dagger}\left[\rho^{-4} \rho^{*\,2} \mathscr{D}\left(\rho^{*\,-2} \left ( \frac{C_{l m}}{\sin\theta'} \right )\delta(r'-r(t)) \delta(\theta'-\theta(t))\right)\right]\right.\\
&\left.- \mathscr{D}\left[\rho^{-4} \mathscr{D}\left( \rho^{*\,-1} \left ( \frac{C_{m m}}{\sin\theta'} \right )\delta(r'-r(t)) \delta(\theta'-\theta(t))\right)\right]\right.\\
&\left.-\frac{1}{\sqrt{2}} \mathscr{D}\left[\rho^{-4} \rho^{*\,2}  \mathscr{L}_{-1}^{\dagger}\left( \rho^{*\,-2} \left ( \frac{C_{l m}}{\sin\theta'} \right )\delta(r'-r(t)) \delta(\theta'-\theta(t))\right)\right]\right\}
    \end{split}
\end{equation}
In this case, integrating by parts and using the algebraic tricks gives:
\begin{equation}
\label{eqn:T0lmwStep2}
    \begin{split}
    T_{l m \omega}^{(0)}=&
\frac{4 \mu}{{2 \pi}} \int_{-\infty}^{\infty} d t \int d \theta' e^{i \omega t-i m \varphi(t)} \\
& \times\left[-\frac{1}{2} \mathscr{L}_{1}\left\{\rho^{-4} \mathscr{L}_{2}\left(\rho^{3} S\right)\right\} C_{l l} \rho^{*} \delta(r'-r(t)) \delta(\theta'-\theta(t))\right.\\
&\left. +\frac{2 \rho^{*\,2}}{\sqrt{2} \rho}\left(\mathscr{L}_{2}(S)+i a(\rho -\rho^{*}) \sin \theta S\right) \mathscr{D}\left\{C_{l m} \rho^{*\,-2}  \delta(r'-r(t)) \delta(\theta'-\theta(t))\right\}\right. \\
&\left. +\frac{1}{ \sqrt{2}} \mathscr{L}_{2}\left\{\rho^{3} S\left(\rho^{*\,2} \rho^{-4}\right)_{, r'}\right\} C_{l m} \rho^{*\,-2} \delta(r'-r(t)) \delta(\theta'-\theta(t))\right. \\
&\left.- (\rho^{3} S) \mathscr{D}\left\{\rho^{-4} \mathscr{D}\left(\rho^{*\,-1} C_{m m} \delta(r'-r(t)) \delta(\theta'-\theta(t))\right)\right\}\right.\bigg]
\end{split}
\end{equation}
In \eqref{eqn:T0lmwStep2}, $S$ is $_{2}S_{lm}$. In this form of \eqref{eqn:T0lmwStep2} we can readily integrate the delta function in $\theta$.
\subsection{Calculating the A terms}
After integrating \eqref{eqn:T4lmwStep2} with respect to $\theta'$, we can express it as:
\begin{equation}
    \begin{split}
T^{(4)}_{\ell m \omega}(r')=& \mu \int_{-\infty}^{\infty} d t e^{i \omega t-i m \varphi(t)} \Delta^{2}\left.\bigg[\left(A_{n n 0}+A_{m^{*} n 0}+A_{m^{*} m^{*} 0}\right) \delta(r'-r(t))\right.\\
&\left.+\left\{\left(A_{m^{*} n 1}+A_{m^{*} m^{*} 1}\right) \delta(r'-r(t))\right\}_{, r'}+\left\{A_{m^{*} m^{*} 2} \delta(r'-r(t))\right\}_{, r' r'}\right.\bigg]
\end{split}
\end{equation}
where the following A terms are evaluated at $\theta'=\theta(t)$ with $S= _{-2}S_{l m}$. 
\begin{equation}
    A_{n n 0}=\left(\frac{4}{2 \pi}\right)\left(\frac{-1}{2}\right)\frac{C_{n n} \rho^{-2} \rho^{*\,-1}}{\Delta^{2}}   \mathscr{L}_{1}^{\dagger}\left\{\rho^{-4} \mathscr{L}_{2}^{\dagger}\left(\rho^{3} S\right)\right\}
\end{equation}
\begin{equation}
    A_{n m^{*} 0}=\left(\frac{4}{2 \pi}\right)\left(\frac{-1}{\sqrt{2}}\right)\frac{C_{n m^{*}} \rho^{-3} }{\Delta} \left[\left(\mathscr{L}_{2}^{\dagger} S\right)\left(\frac{i K}{\Delta}-\rho-\rho^{*}\right)-a \sin \theta S \frac{K}{\Delta}(\rho-\rho^{*})\right]
\end{equation}
\begin{equation}
    A_{m^{*} m^{*} 0}=\left(\frac{4}{2 \pi}\right)\left(\frac{-1}{4}\right)\rho^{-3} \rho^{*} C_{m^{*} m^{*}} S\left[-i\left(\frac{K}{\Delta}\right)_{, r'}-\frac{K^{2}}{\Delta^{2}}-2 i \rho \frac{K}{\Delta}\right]
\end{equation}
\begin{equation}
    A_{n m^{*} 1}=\left(\frac{4}{2 \pi}\right)\left(\frac{-1}{\sqrt{2}}\right)\frac{C_{n m^{*}} \rho^{-3} }{\Delta} \left[\mathscr{L}_{2}^{\dagger} S+i a \sin \theta(\rho - \rho^{*}) S\right]
\end{equation}
\begin{equation}
    A_{m^{*} m^{*} 1}=\left(\frac{4}{2 \pi}\right)\left(\frac{-1}{2}\right) \rho^{-3} \rho^{*} C_{m^{*} m^{*}} S\left(i \frac{K}{\Delta}-\rho\right)
\end{equation}
\begin{equation}
    A_{m^{*} m^{*} 2}=\left(\frac{4}{2 \pi}\right)\left(\frac{-1}{4}\right) \rho^{-3} \rho^{*} C_{m^{*} m^{*}} S
\end{equation}
{The above equations match with those obtained in \cite{Mino:1997bx} modulo the opposite sign of $\rho$ and an overall normalization factor.}

Similarly integrating \eqref{eqn:T0lmwStep2} w.r.t. $\theta'$, we can express it as:
\begin{equation}
\label{eqn:SourcewithDiracDeltaDerivatives}
\begin{split}
T^{(0)}_{\ell m \omega}(r')=& \mu \int_{-\infty}^{\infty} d t e^{i \omega t-i m \varphi(t)} \bigg[\left(A_{l l 0}+A_{l m 0}+A_{m m 0}\right) \delta(r'-r(t))\\
&+\left\{\left(A_{l m 1}+A_{m m 1}\right) \delta(r'-r(t))\right\}_{, r'}+\left\{A_{m m 2} \delta(r'-r(t))\right\}_{, r' r'}\bigg]
\end{split}
\end{equation}
where the following A terms are evaluated at $\theta'=\theta(t)$ with $S= _{2}S_{l m}$. 
\begin{equation}
    A_{l l 0}=\left(\frac{4}{2 \pi}\right)\left(\frac{-1}{2}\right)(C_{l l} \rho^{*}) \mathscr{L}_{1}\left\{\rho^{-4} \mathscr{L}_{2}\left(\rho^{3} S\right)\right\}
\end{equation}
\begin{equation}
    A_{l m 0}=\left(\frac{4}{2 \pi}\right)\left(\frac{2}{\sqrt{2}}\right)\frac{C_{l m}  }{\rho} \left[\left(\mathscr{L}_{2} S\right)\left(\frac{-i K}{\Delta}-\rho-\rho^{*}\right)+a \sin \theta S \frac{K}{\Delta}(\rho-\rho^{*})\right]
\end{equation}
\begin{equation}
    A_{m m 0}=\left(\frac{-4}{2 \pi}\right)\rho^{-1} \rho^{*\,-1} C_{m m} S\left[+i\left(\frac{K}{\Delta}\right)_{, r'}-\frac{K^{2}}{\Delta^{2}}+2 i \rho \frac{K}{\Delta}\right]
\end{equation}
\begin{equation}
    A_{l m 1}=\left(\frac{4}{2 \pi}\right)\left(\frac{2}{\sqrt{2}}\right)\frac{C_{l m} }{\rho} \left[\mathscr{L}_{2}S + i a \sin \theta(\rho - \rho^{*}) S\right]
\end{equation}
\begin{equation}
    A_{m m 1}=\left(\frac{8}{2 \pi}\right)\rho^{-1} \rho^{*\,-1} C_{m m} S\left(i \frac{K}{\Delta}+\rho\right)
\end{equation}
\begin{equation}
    A_{m m 2}=\left(\frac{-4}{2 \pi}\right)\rho^{-1} \rho^{*\,-1} C_{m m} S
\end{equation}
\eqref{eqn:SourcewithDiracDeltaDerivatives} can further be written as (after integrating with respect to dr and then renaming $r'$ as $r$):
\begin{equation}
\begin{split}
    T^{(0)}_{lm\omega}(r)=\mu e^{i\omega t-im\varphi}&\bigg[ (A_{ll0}+A_{lm0}+A_{mm0})t'\\& +(A_{lm1}+A_{mm1})(t''+i\omega (t')^2-imt'\varphi ')+t'\frac{d}{dr}\left(A_{lm1}+A_{mm1}\right)\\
    & +A_{mm2}(t'''+3i\omega t''t'-2imt''\varphi '-\omega^2(t')^3+2\omega m(t')^2\varphi '-m^2t'(\varphi')^2-imt'\varphi'')\\ &+2(t''+i\omega (t')^2-im t'\varphi ')\frac{dA_{mm2}}{dr}+t'\frac{d^2(A_{mm2})}{dr^2} \bigg]
\end{split}
\end{equation}
where $'$ denotes derivative with respect to $r$.
%
\end{widetext}



\bibliography{apssamp}

\begin{thebibliography}{52}%
\makeatletter
\providecommand \@ifxundefined [1]{%
 \@ifx{#1\undefined}
}%
\providecommand \@ifnum [1]{%
 \ifnum #1\expandafter \@firstoftwo
 \else \expandafter \@secondoftwo
 \fi
}%
\providecommand \@ifx [1]{%
 \ifx #1\expandafter \@firstoftwo
 \else \expandafter \@secondoftwo
 \fi
}%
\providecommand \natexlab [1]{#1}%
\providecommand \enquote  [1]{``#1''}%
\providecommand \bibnamefont  [1]{#1}%
\providecommand \bibfnamefont [1]{#1}%
\providecommand \citenamefont [1]{#1}%
\providecommand \href@noop [0]{\@secondoftwo}%
\providecommand \href [0]{\begingroup \@sanitize@url \@href}%
\providecommand \@href[1]{\@@startlink{#1}\@@href}%
\providecommand \@@href[1]{\endgroup#1\@@endlink}%
\providecommand \@sanitize@url [0]{\catcode `\\12\catcode `\$12\catcode
  `\&12\catcode `\#12\catcode `\^12\catcode `\_12\catcode `\%12\relax}%
\providecommand \@@startlink[1]{}%
\providecommand \@@endlink[0]{}%
\providecommand \url  [0]{\begingroup\@sanitize@url \@url }%
\providecommand \@url [1]{\endgroup\@href {#1}{\urlprefix }}%
\providecommand \urlprefix  [0]{URL }%
\providecommand \Eprint [0]{\href }%
\providecommand \doibase [0]{https://doi.org/}%
\providecommand \selectlanguage [0]{\@gobble}%
\providecommand \bibinfo  [0]{\@secondoftwo}%
\providecommand \bibfield  [0]{\@secondoftwo}%
\providecommand \translation [1]{[#1]}%
\providecommand \BibitemOpen [0]{}%
\providecommand \bibitemStop [0]{}%
\providecommand \bibitemNoStop [0]{.\EOS\space}%
\providecommand \EOS [0]{\spacefactor3000\relax}%
\providecommand \BibitemShut  [1]{\csname bibitem#1\endcsname}%
\let\auto@bib@innerbib\@empty
\bibitem [{\citenamefont {Abbott}\ \emph {et~al.}(2016)\citenamefont {Abbott},
  \citenamefont {Abbott}, \citenamefont {Abbott}, \citenamefont {Abernathy},
  \citenamefont {Acernese}, \citenamefont {Ackley}, \citenamefont {Adams},
  \citenamefont {Adams}, \citenamefont {Addesso}, \citenamefont {Adhikari}
  \emph {et~al.}}]{GW150914}%
  \BibitemOpen
  \bibfield  {author} {\bibinfo {author} {\bibfnamefont {B.~P.}\ \bibnamefont
  {Abbott}}, \bibinfo {author} {\bibfnamefont {R.}~\bibnamefont {Abbott}},
  \bibinfo {author} {\bibfnamefont {T.}~\bibnamefont {Abbott}}, \bibinfo
  {author} {\bibfnamefont {M.}~\bibnamefont {Abernathy}}, \bibinfo {author}
  {\bibfnamefont {F.}~\bibnamefont {Acernese}}, \bibinfo {author}
  {\bibfnamefont {K.}~\bibnamefont {Ackley}}, \bibinfo {author} {\bibfnamefont
  {C.}~\bibnamefont {Adams}}, \bibinfo {author} {\bibfnamefont
  {T.}~\bibnamefont {Adams}}, \bibinfo {author} {\bibfnamefont
  {P.}~\bibnamefont {Addesso}}, \bibinfo {author} {\bibfnamefont
  {R.}~\bibnamefont {Adhikari}}, \emph {et~al.},\ }\bibfield  {title} {\bibinfo
  {title} {Observation of gravitational waves from a binary black hole
  merger},\ }\href@noop {} {\bibfield  {journal} {\bibinfo  {journal} {Physical
  review letters}\ }\textbf {\bibinfo {volume} {116}},\ \bibinfo {pages}
  {061102} (\bibinfo {year} {2016})}\BibitemShut {NoStop}%
\bibitem [{\citenamefont {Abbott}\ \emph {et~al.}(2017)\citenamefont {Abbott},
  \citenamefont {Abbott}, \citenamefont {Abbott}, \citenamefont {Acernese},
  \citenamefont {Ackley}, \citenamefont {Adams}, \citenamefont {Adams},
  \citenamefont {Addesso}, \citenamefont {Adhikari}, \citenamefont {Adya} \emph
  {et~al.}}]{GW170817}%
  \BibitemOpen
  \bibfield  {author} {\bibinfo {author} {\bibfnamefont {B.~P.}\ \bibnamefont
  {Abbott}}, \bibinfo {author} {\bibfnamefont {R.}~\bibnamefont {Abbott}},
  \bibinfo {author} {\bibfnamefont {T.}~\bibnamefont {Abbott}}, \bibinfo
  {author} {\bibfnamefont {F.}~\bibnamefont {Acernese}}, \bibinfo {author}
  {\bibfnamefont {K.}~\bibnamefont {Ackley}}, \bibinfo {author} {\bibfnamefont
  {C.}~\bibnamefont {Adams}}, \bibinfo {author} {\bibfnamefont
  {T.}~\bibnamefont {Adams}}, \bibinfo {author} {\bibfnamefont
  {P.}~\bibnamefont {Addesso}}, \bibinfo {author} {\bibfnamefont
  {R.}~\bibnamefont {Adhikari}}, \bibinfo {author} {\bibfnamefont
  {V.}~\bibnamefont {Adya}}, \emph {et~al.},\ }\bibfield  {title} {\bibinfo
  {title} {{GW170817}: observation of gravitational waves from a binary neutron
  star inspiral},\ }\href@noop {} {\bibfield  {journal} {\bibinfo  {journal}
  {Physical Review Letters}\ }\textbf {\bibinfo {volume} {119}},\ \bibinfo
  {pages} {161101} (\bibinfo {year} {2017})}\BibitemShut {NoStop}%
\bibitem [{\citenamefont {Palenzuela}\ \emph {et~al.}(2017)\citenamefont
  {Palenzuela}, \citenamefont {Pani}, \citenamefont {Bezares}, \citenamefont
  {Cardoso}, \citenamefont {Lehner},\ and\ \citenamefont
  {Liebling}}]{palenzuela2017gravitational}%
  \BibitemOpen
  \bibfield  {author} {\bibinfo {author} {\bibfnamefont {C.}~\bibnamefont
  {Palenzuela}}, \bibinfo {author} {\bibfnamefont {P.}~\bibnamefont {Pani}},
  \bibinfo {author} {\bibfnamefont {M.}~\bibnamefont {Bezares}}, \bibinfo
  {author} {\bibfnamefont {V.}~\bibnamefont {Cardoso}}, \bibinfo {author}
  {\bibfnamefont {L.}~\bibnamefont {Lehner}},\ and\ \bibinfo {author}
  {\bibfnamefont {S.}~\bibnamefont {Liebling}},\ }\bibfield  {title} {\bibinfo
  {title} {{Gravitational Wave Signatures of Highly Compact Boson Star
  Binaries}},\ }\href {https://doi.org/10.1103/PhysRevD.96.104058} {\bibfield
  {journal} {\bibinfo  {journal} {Phys. Rev. D}\ }\textbf {\bibinfo {volume}
  {96}},\ \bibinfo {pages} {104058} (\bibinfo {year} {2017})},\ \Eprint
  {https://arxiv.org/abs/1710.09432} {arXiv:1710.09432 [gr-qc]} \BibitemShut
  {NoStop}%
\bibitem [{\citenamefont {Mazur}\ and\ \citenamefont
  {Mottola}(2004)}]{mazur2004gravitational}%
  \BibitemOpen
  \bibfield  {author} {\bibinfo {author} {\bibfnamefont {P.~O.}\ \bibnamefont
  {Mazur}}\ and\ \bibinfo {author} {\bibfnamefont {E.}~\bibnamefont
  {Mottola}},\ }\bibfield  {title} {\bibinfo {title} {{Gravitational vacuum
  condensate stars}},\ }\href {https://doi.org/10.1073/pnas.0402717101}
  {\bibfield  {journal} {\bibinfo  {journal} {Proc. Nat. Acad. Sci.}\ }\textbf
  {\bibinfo {volume} {101}},\ \bibinfo {pages} {9545} (\bibinfo {year}
  {2004})},\ \Eprint {https://arxiv.org/abs/gr-qc/0407075}
  {arXiv:gr-qc/0407075} \BibitemShut {NoStop}%
\bibitem [{\citenamefont {Pani}\ \emph {et~al.}(2009)\citenamefont {Pani},
  \citenamefont {Berti}, \citenamefont {Cardoso}, \citenamefont {Chen},\ and\
  \citenamefont {Norte}}]{pani2009gravitational}%
  \BibitemOpen
  \bibfield  {author} {\bibinfo {author} {\bibfnamefont {P.}~\bibnamefont
  {Pani}}, \bibinfo {author} {\bibfnamefont {E.}~\bibnamefont {Berti}},
  \bibinfo {author} {\bibfnamefont {V.}~\bibnamefont {Cardoso}}, \bibinfo
  {author} {\bibfnamefont {Y.}~\bibnamefont {Chen}},\ and\ \bibinfo {author}
  {\bibfnamefont {R.}~\bibnamefont {Norte}},\ }\bibfield  {title} {\bibinfo
  {title} {{Gravitational wave signatures of the absence of an event horizon.
  I. Nonradial oscillations of a thin-shell gravastar}},\ }\href
  {https://doi.org/10.1103/PhysRevD.80.124047} {\bibfield  {journal} {\bibinfo
  {journal} {Phys. Rev. D}\ }\textbf {\bibinfo {volume} {80}},\ \bibinfo
  {pages} {124047} (\bibinfo {year} {2009})},\ \Eprint
  {https://arxiv.org/abs/0909.0287} {arXiv:0909.0287 [gr-qc]} \BibitemShut
  {NoStop}%
\bibitem [{\citenamefont {Giddings}\ and\ \citenamefont
  {Psaltis}(2018)}]{giddings2018event}%
  \BibitemOpen
  \bibfield  {author} {\bibinfo {author} {\bibfnamefont {S.~B.}\ \bibnamefont
  {Giddings}}\ and\ \bibinfo {author} {\bibfnamefont {D.}~\bibnamefont
  {Psaltis}},\ }\bibfield  {title} {\bibinfo {title} {{Event Horizon Telescope
  Observations as Probes for Quantum Structure of Astrophysical Black Holes}},\
  }\href {https://doi.org/10.1103/PhysRevD.97.084035} {\bibfield  {journal}
  {\bibinfo  {journal} {Phys. Rev. D}\ }\textbf {\bibinfo {volume} {97}},\
  \bibinfo {pages} {084035} (\bibinfo {year} {2018})},\ \Eprint
  {https://arxiv.org/abs/1606.07814} {arXiv:1606.07814 [astro-ph.HE]}
  \BibitemShut {NoStop}%
\bibitem [{\citenamefont {Bianchi}\ \emph {et~al.}(2020)\citenamefont
  {Bianchi}, \citenamefont {Consoli}, \citenamefont {Grillo}, \citenamefont
  {Morales}, \citenamefont {Pani},\ and\ \citenamefont
  {Raposo}}]{bianchi2020distinguishing}%
  \BibitemOpen
  \bibfield  {author} {\bibinfo {author} {\bibfnamefont {M.}~\bibnamefont
  {Bianchi}}, \bibinfo {author} {\bibfnamefont {D.}~\bibnamefont {Consoli}},
  \bibinfo {author} {\bibfnamefont {A.}~\bibnamefont {Grillo}}, \bibinfo
  {author} {\bibfnamefont {J.~F.}\ \bibnamefont {Morales}}, \bibinfo {author}
  {\bibfnamefont {P.}~\bibnamefont {Pani}},\ and\ \bibinfo {author}
  {\bibfnamefont {G.}~\bibnamefont {Raposo}},\ }\bibfield  {title} {\bibinfo
  {title} {{Distinguishing fuzzballs from black holes through their multipolar
  structure}},\ }\href {https://doi.org/10.1103/PhysRevLett.125.221601}
  {\bibfield  {journal} {\bibinfo  {journal} {Phys. Rev. Lett.}\ }\textbf
  {\bibinfo {volume} {125}},\ \bibinfo {pages} {221601} (\bibinfo {year}
  {2020})},\ \Eprint {https://arxiv.org/abs/2007.01743} {arXiv:2007.01743
  [hep-th]} \BibitemShut {NoStop}%
\bibitem [{\citenamefont {Bena}\ and\ \citenamefont
  {Mayerson}(2020)}]{bena2020multipole}%
  \BibitemOpen
  \bibfield  {author} {\bibinfo {author} {\bibfnamefont {I.}~\bibnamefont
  {Bena}}\ and\ \bibinfo {author} {\bibfnamefont {D.~R.}\ \bibnamefont
  {Mayerson}},\ }\bibfield  {title} {\bibinfo {title} {{Multipole Ratios: A New
  Window into Black Holes}},\ }\href
  {https://doi.org/10.1103/PhysRevLett.125.221602} {\bibfield  {journal}
  {\bibinfo  {journal} {Phys. Rev. Lett.}\ }\textbf {\bibinfo {volume} {125}},\
  \bibinfo {pages} {22} (\bibinfo {year} {2020})},\ \Eprint
  {https://arxiv.org/abs/2006.10750} {arXiv:2006.10750 [hep-th]} \BibitemShut
  {NoStop}%
\bibitem [{\citenamefont {Mukherjee}\ and\ \citenamefont
  {Chakraborty}(2020)}]{mukherjee2020multipole}%
  \BibitemOpen
  \bibfield  {author} {\bibinfo {author} {\bibfnamefont {S.}~\bibnamefont
  {Mukherjee}}\ and\ \bibinfo {author} {\bibfnamefont {S.}~\bibnamefont
  {Chakraborty}},\ }\bibfield  {title} {\bibinfo {title} {{Multipole moments of
  compact objects with NUT charge: Theoretical and observational
  implications}},\ }\href {https://doi.org/10.1103/PhysRevD.102.124058}
  {\bibfield  {journal} {\bibinfo  {journal} {Phys. Rev. D}\ }\textbf {\bibinfo
  {volume} {102}},\ \bibinfo {pages} {124058} (\bibinfo {year} {2020})},\
  \Eprint {https://arxiv.org/abs/2008.06891} {arXiv:2008.06891 [gr-qc]}
  \BibitemShut {NoStop}%
\bibitem [{\citenamefont {Xin}\ \emph {et~al.}(2021)\citenamefont {Xin},
  \citenamefont {Chen}, \citenamefont {Lo}, \citenamefont {Sun}, \citenamefont
  {Han}, \citenamefont {Zhong}, \citenamefont {Srivastava}, \citenamefont {Ma},
  \citenamefont {Wang},\ and\ \citenamefont {Chen}}]{Xin:2021zir}%
  \BibitemOpen
  \bibfield  {author} {\bibinfo {author} {\bibfnamefont {S.}~\bibnamefont
  {Xin}}, \bibinfo {author} {\bibfnamefont {B.}~\bibnamefont {Chen}}, \bibinfo
  {author} {\bibfnamefont {R.~K.~L.}\ \bibnamefont {Lo}}, \bibinfo {author}
  {\bibfnamefont {L.}~\bibnamefont {Sun}}, \bibinfo {author} {\bibfnamefont
  {W.-B.}\ \bibnamefont {Han}}, \bibinfo {author} {\bibfnamefont
  {X.}~\bibnamefont {Zhong}}, \bibinfo {author} {\bibfnamefont
  {M.}~\bibnamefont {Srivastava}}, \bibinfo {author} {\bibfnamefont
  {S.}~\bibnamefont {Ma}}, \bibinfo {author} {\bibfnamefont {Q.}~\bibnamefont
  {Wang}},\ and\ \bibinfo {author} {\bibfnamefont {Y.}~\bibnamefont {Chen}},\
  }\bibfield  {title} {\bibinfo {title} {{Gravitational-wave echoes from
  spinning exotic compact objects: numerical waveforms from the Teukolsky
  equation}},\ }\href@noop {} {\  (\bibinfo {year} {2021})},\ \Eprint
  {https://arxiv.org/abs/2105.12313} {arXiv:2105.12313 [gr-qc]} \BibitemShut
  {NoStop}%
\bibitem [{\citenamefont {Wang}\ \emph {et~al.}(2020)\citenamefont {Wang},
  \citenamefont {Oshita},\ and\ \citenamefont {Afshordi}}]{wang2020echoes}%
  \BibitemOpen
  \bibfield  {author} {\bibinfo {author} {\bibfnamefont {Q.}~\bibnamefont
  {Wang}}, \bibinfo {author} {\bibfnamefont {N.}~\bibnamefont {Oshita}},\ and\
  \bibinfo {author} {\bibfnamefont {N.}~\bibnamefont {Afshordi}},\ }\bibfield
  {title} {\bibinfo {title} {Echoes from quantum black holes},\ }\href@noop {}
  {\bibfield  {journal} {\bibinfo  {journal} {Physical Review D}\ }\textbf
  {\bibinfo {volume} {101}},\ \bibinfo {pages} {024031} (\bibinfo {year}
  {2020})}\BibitemShut {NoStop}%
\bibitem [{\citenamefont {Maggio}\ \emph {et~al.}(2019)\citenamefont {Maggio},
  \citenamefont {Testa}, \citenamefont {Bhagwat},\ and\ \citenamefont
  {Pani}}]{Maggio}%
  \BibitemOpen
  \bibfield  {author} {\bibinfo {author} {\bibfnamefont {E.}~\bibnamefont
  {Maggio}}, \bibinfo {author} {\bibfnamefont {A.}~\bibnamefont {Testa}},
  \bibinfo {author} {\bibfnamefont {S.}~\bibnamefont {Bhagwat}},\ and\ \bibinfo
  {author} {\bibfnamefont {P.}~\bibnamefont {Pani}},\ }\bibfield  {title}
  {\bibinfo {title} {Analytical model for gravitational-wave echoes from
  spinning remnants},\ }\href {https://doi.org/10.1103/PhysRevD.100.064056}
  {\bibfield  {journal} {\bibinfo  {journal} {Phys. Rev. D}\ }\textbf {\bibinfo
  {volume} {100}},\ \bibinfo {pages} {064056} (\bibinfo {year}
  {2019})}\BibitemShut {NoStop}%
\bibitem [{\citenamefont {Micchi}\ \emph {et~al.}(2021)\citenamefont {Micchi},
  \citenamefont {Afshordi},\ and\ \citenamefont {Chirenti}}]{micchi2021loud}%
  \BibitemOpen
  \bibfield  {author} {\bibinfo {author} {\bibfnamefont {L.~F.~L.}\
  \bibnamefont {Micchi}}, \bibinfo {author} {\bibfnamefont {N.}~\bibnamefont
  {Afshordi}},\ and\ \bibinfo {author} {\bibfnamefont {C.}~\bibnamefont
  {Chirenti}},\ }\bibfield  {title} {\bibinfo {title} {How loud are echoes from
  exotic compact objects?},\ }\href@noop {} {\bibfield  {journal} {\bibinfo
  {journal} {Physical Review D}\ }\textbf {\bibinfo {volume} {103}},\ \bibinfo
  {pages} {044028} (\bibinfo {year} {2021})}\BibitemShut {NoStop}%
\bibitem [{\citenamefont {Cardoso}\ \emph {et~al.}(2016)\citenamefont
  {Cardoso}, \citenamefont {Hopper}, \citenamefont {Macedo}, \citenamefont
  {Palenzuela},\ and\ \citenamefont {Pani}}]{Cardoso:2016oxy}%
  \BibitemOpen
  \bibfield  {author} {\bibinfo {author} {\bibfnamefont {V.}~\bibnamefont
  {Cardoso}}, \bibinfo {author} {\bibfnamefont {S.}~\bibnamefont {Hopper}},
  \bibinfo {author} {\bibfnamefont {C.~F.~B.}\ \bibnamefont {Macedo}}, \bibinfo
  {author} {\bibfnamefont {C.}~\bibnamefont {Palenzuela}},\ and\ \bibinfo
  {author} {\bibfnamefont {P.}~\bibnamefont {Pani}},\ }\bibfield  {title}
  {\bibinfo {title} {{Gravitational-wave signatures of exotic compact objects
  and of quantum corrections at the horizon scale}},\ }\href
  {https://doi.org/10.1103/PhysRevD.94.084031} {\bibfield  {journal} {\bibinfo
  {journal} {Phys. Rev. D}\ }\textbf {\bibinfo {volume} {94}},\ \bibinfo
  {pages} {084031} (\bibinfo {year} {2016})},\ \Eprint
  {https://arxiv.org/abs/1608.08637} {arXiv:1608.08637 [gr-qc]} \BibitemShut
  {NoStop}%
\bibitem [{\citenamefont {Mark}\ \emph {et~al.}(2017)\citenamefont {Mark},
  \citenamefont {Zimmerman}, \citenamefont {Du},\ and\ \citenamefont
  {Chen}}]{Mark:2017dnq}%
  \BibitemOpen
  \bibfield  {author} {\bibinfo {author} {\bibfnamefont {Z.}~\bibnamefont
  {Mark}}, \bibinfo {author} {\bibfnamefont {A.}~\bibnamefont {Zimmerman}},
  \bibinfo {author} {\bibfnamefont {S.~M.}\ \bibnamefont {Du}},\ and\ \bibinfo
  {author} {\bibfnamefont {Y.}~\bibnamefont {Chen}},\ }\bibfield  {title}
  {\bibinfo {title} {{A recipe for echoes from exotic compact objects}},\
  }\href {https://doi.org/10.1103/PhysRevD.96.084002} {\bibfield  {journal}
  {\bibinfo  {journal} {Phys. Rev. D}\ }\textbf {\bibinfo {volume} {96}},\
  \bibinfo {pages} {084002} (\bibinfo {year} {2017})},\ \Eprint
  {https://arxiv.org/abs/1706.06155} {arXiv:1706.06155 [gr-qc]} \BibitemShut
  {NoStop}%
\bibitem [{\citenamefont {Conklin}\ \emph {et~al.}(2018)\citenamefont
  {Conklin}, \citenamefont {Holdom},\ and\ \citenamefont
  {Ren}}]{Conklin:2017lwb}%
  \BibitemOpen
  \bibfield  {author} {\bibinfo {author} {\bibfnamefont {R.~S.}\ \bibnamefont
  {Conklin}}, \bibinfo {author} {\bibfnamefont {B.}~\bibnamefont {Holdom}},\
  and\ \bibinfo {author} {\bibfnamefont {J.}~\bibnamefont {Ren}},\ }\bibfield
  {title} {\bibinfo {title} {{Gravitational wave echoes through new windows}},\
  }\href {https://doi.org/10.1103/PhysRevD.98.044021} {\bibfield  {journal}
  {\bibinfo  {journal} {Phys. Rev. D}\ }\textbf {\bibinfo {volume} {98}},\
  \bibinfo {pages} {044021} (\bibinfo {year} {2018})},\ \Eprint
  {https://arxiv.org/abs/1712.06517} {arXiv:1712.06517 [gr-qc]} \BibitemShut
  {NoStop}%
\bibitem [{\citenamefont {Maselli}\ \emph {et~al.}(2017)\citenamefont
  {Maselli}, \citenamefont {V\"olkel},\ and\ \citenamefont
  {Kokkotas}}]{Maselli:2017tfq}%
  \BibitemOpen
  \bibfield  {author} {\bibinfo {author} {\bibfnamefont {A.}~\bibnamefont
  {Maselli}}, \bibinfo {author} {\bibfnamefont {S.~H.}\ \bibnamefont
  {V\"olkel}},\ and\ \bibinfo {author} {\bibfnamefont {K.~D.}\ \bibnamefont
  {Kokkotas}},\ }\bibfield  {title} {\bibinfo {title} {{Parameter estimation of
  gravitational wave echoes from exotic compact objects}},\ }\href
  {https://doi.org/10.1103/PhysRevD.96.064045} {\bibfield  {journal} {\bibinfo
  {journal} {Phys. Rev. D}\ }\textbf {\bibinfo {volume} {96}},\ \bibinfo
  {pages} {064045} (\bibinfo {year} {2017})},\ \Eprint
  {https://arxiv.org/abs/1708.02217} {arXiv:1708.02217 [gr-qc]} \BibitemShut
  {NoStop}%
\bibitem [{\citenamefont {Westerweck}\ \emph {et~al.}(2018)\citenamefont
  {Westerweck}, \citenamefont {Nielsen}, \citenamefont {Fischer-Birnholtz},
  \citenamefont {Cabero}, \citenamefont {Capano}, \citenamefont {Dent},
  \citenamefont {Krishnan}, \citenamefont {Meadors},\ and\ \citenamefont
  {Nitz}}]{Westerweck:2017hus}%
  \BibitemOpen
  \bibfield  {author} {\bibinfo {author} {\bibfnamefont {J.}~\bibnamefont
  {Westerweck}}, \bibinfo {author} {\bibfnamefont {A.}~\bibnamefont {Nielsen}},
  \bibinfo {author} {\bibfnamefont {O.}~\bibnamefont {Fischer-Birnholtz}},
  \bibinfo {author} {\bibfnamefont {M.}~\bibnamefont {Cabero}}, \bibinfo
  {author} {\bibfnamefont {C.}~\bibnamefont {Capano}}, \bibinfo {author}
  {\bibfnamefont {T.}~\bibnamefont {Dent}}, \bibinfo {author} {\bibfnamefont
  {B.}~\bibnamefont {Krishnan}}, \bibinfo {author} {\bibfnamefont
  {G.}~\bibnamefont {Meadors}},\ and\ \bibinfo {author} {\bibfnamefont {A.~H.}\
  \bibnamefont {Nitz}},\ }\bibfield  {title} {\bibinfo {title} {{Low
  significance of evidence for black hole echoes in gravitational wave data}},\
  }\href {https://doi.org/10.1103/PhysRevD.97.124037} {\bibfield  {journal}
  {\bibinfo  {journal} {Phys. Rev. D}\ }\textbf {\bibinfo {volume} {97}},\
  \bibinfo {pages} {124037} (\bibinfo {year} {2018})},\ \Eprint
  {https://arxiv.org/abs/1712.09966} {arXiv:1712.09966 [gr-qc]} \BibitemShut
  {NoStop}%
\bibitem [{\citenamefont {Wang}\ and\ \citenamefont
  {Afshordi}(2018)}]{Wang:2018gin}%
  \BibitemOpen
  \bibfield  {author} {\bibinfo {author} {\bibfnamefont {Q.}~\bibnamefont
  {Wang}}\ and\ \bibinfo {author} {\bibfnamefont {N.}~\bibnamefont
  {Afshordi}},\ }\bibfield  {title} {\bibinfo {title} {{Black hole echology:
  The observer\textquoteright{}s manual}},\ }\href
  {https://doi.org/10.1103/PhysRevD.97.124044} {\bibfield  {journal} {\bibinfo
  {journal} {Phys. Rev. D}\ }\textbf {\bibinfo {volume} {97}},\ \bibinfo
  {pages} {124044} (\bibinfo {year} {2018})},\ \Eprint
  {https://arxiv.org/abs/1803.02845} {arXiv:1803.02845 [gr-qc]} \BibitemShut
  {NoStop}%
\bibitem [{\citenamefont {Urbano}\ and\ \citenamefont
  {Veerm\"ae}(2019)}]{Urbano:2018nrs}%
  \BibitemOpen
  \bibfield  {author} {\bibinfo {author} {\bibfnamefont {A.}~\bibnamefont
  {Urbano}}\ and\ \bibinfo {author} {\bibfnamefont {H.}~\bibnamefont
  {Veerm\"ae}},\ }\bibfield  {title} {\bibinfo {title} {{On gravitational
  echoes from ultracompact exotic stars}},\ }\href
  {https://doi.org/10.1088/1475-7516/2019/04/011} {\bibfield  {journal}
  {\bibinfo  {journal} {JCAP}\ }\textbf {\bibinfo {volume} {04}},\ \bibinfo
  {pages} {011}},\ \Eprint {https://arxiv.org/abs/1810.07137} {arXiv:1810.07137
  [gr-qc]} \BibitemShut {NoStop}%
\bibitem [{\citenamefont {Lo}\ \emph {et~al.}(2019)\citenamefont {Lo},
  \citenamefont {Li},\ and\ \citenamefont {Weinstein}}]{Lo:2018sep}%
  \BibitemOpen
  \bibfield  {author} {\bibinfo {author} {\bibfnamefont {R.~K.~L.}\
  \bibnamefont {Lo}}, \bibinfo {author} {\bibfnamefont {T.~G.~F.}\ \bibnamefont
  {Li}},\ and\ \bibinfo {author} {\bibfnamefont {A.~J.}\ \bibnamefont
  {Weinstein}},\ }\bibfield  {title} {\bibinfo {title} {{Template-based
  Gravitational-Wave Echoes Search Using Bayesian Model Selection}},\ }\href
  {https://doi.org/10.1103/PhysRevD.99.084052} {\bibfield  {journal} {\bibinfo
  {journal} {Phys. Rev. D}\ }\textbf {\bibinfo {volume} {99}},\ \bibinfo
  {pages} {084052} (\bibinfo {year} {2019})},\ \Eprint
  {https://arxiv.org/abs/1811.07431} {arXiv:1811.07431 [gr-qc]} \BibitemShut
  {NoStop}%
\bibitem [{\citenamefont {Nielsen}\ \emph {et~al.}(2019)\citenamefont
  {Nielsen}, \citenamefont {Capano}, \citenamefont {Birnholtz},\ and\
  \citenamefont {Westerweck}}]{Nielsen:2018lkf}%
  \BibitemOpen
  \bibfield  {author} {\bibinfo {author} {\bibfnamefont {A.~B.}\ \bibnamefont
  {Nielsen}}, \bibinfo {author} {\bibfnamefont {C.~D.}\ \bibnamefont {Capano}},
  \bibinfo {author} {\bibfnamefont {O.}~\bibnamefont {Birnholtz}},\ and\
  \bibinfo {author} {\bibfnamefont {J.}~\bibnamefont {Westerweck}},\ }\bibfield
   {title} {\bibinfo {title} {{Parameter estimation and statistical
  significance of echoes following black hole signals in the first Advanced
  LIGO observing run}},\ }\href {https://doi.org/10.1103/PhysRevD.99.104012}
  {\bibfield  {journal} {\bibinfo  {journal} {Phys. Rev. D}\ }\textbf {\bibinfo
  {volume} {99}},\ \bibinfo {pages} {104012} (\bibinfo {year} {2019})},\
  \Eprint {https://arxiv.org/abs/1811.04904} {arXiv:1811.04904 [gr-qc]}
  \BibitemShut {NoStop}%
\bibitem [{\citenamefont {Tsang}\ \emph {et~al.}(2020)\citenamefont {Tsang},
  \citenamefont {Ghosh}, \citenamefont {Samajdar}, \citenamefont
  {Chatziioannou}, \citenamefont {Mastrogiovanni}, \citenamefont {Agathos},\
  and\ \citenamefont {Van Den~Broeck}}]{Tsang:2019zra}%
  \BibitemOpen
  \bibfield  {author} {\bibinfo {author} {\bibfnamefont {K.~W.}\ \bibnamefont
  {Tsang}}, \bibinfo {author} {\bibfnamefont {A.}~\bibnamefont {Ghosh}},
  \bibinfo {author} {\bibfnamefont {A.}~\bibnamefont {Samajdar}}, \bibinfo
  {author} {\bibfnamefont {K.}~\bibnamefont {Chatziioannou}}, \bibinfo {author}
  {\bibfnamefont {S.}~\bibnamefont {Mastrogiovanni}}, \bibinfo {author}
  {\bibfnamefont {M.}~\bibnamefont {Agathos}},\ and\ \bibinfo {author}
  {\bibfnamefont {C.}~\bibnamefont {Van Den~Broeck}},\ }\bibfield  {title}
  {\bibinfo {title} {{A morphology-independent search for gravitational wave
  echoes in data from the first and second observing runs of Advanced LIGO and
  Advanced Virgo}},\ }\href {https://doi.org/10.1103/PhysRevD.101.064012}
  {\bibfield  {journal} {\bibinfo  {journal} {Phys. Rev. D}\ }\textbf {\bibinfo
  {volume} {101}},\ \bibinfo {pages} {064012} (\bibinfo {year} {2020})},\
  \Eprint {https://arxiv.org/abs/1906.11168} {arXiv:1906.11168 [gr-qc]}
  \BibitemShut {NoStop}%
\bibitem [{\citenamefont {Chen}\ \emph {et~al.}(2019)\citenamefont {Chen},
  \citenamefont {Chen}, \citenamefont {Ma}, \citenamefont {Lo},\ and\
  \citenamefont {Sun}}]{Chen:2019hfg}%
  \BibitemOpen
  \bibfield  {author} {\bibinfo {author} {\bibfnamefont {B.}~\bibnamefont
  {Chen}}, \bibinfo {author} {\bibfnamefont {Y.}~\bibnamefont {Chen}}, \bibinfo
  {author} {\bibfnamefont {Y.}~\bibnamefont {Ma}}, \bibinfo {author}
  {\bibfnamefont {K.-L.~R.}\ \bibnamefont {Lo}},\ and\ \bibinfo {author}
  {\bibfnamefont {L.}~\bibnamefont {Sun}},\ }\bibfield  {title} {\bibinfo
  {title} {{Instability of Exotic Compact Objects and Its Implications for
  Gravitational-Wave Echoes}},\ }\href@noop {} {\  (\bibinfo {year} {2019})},\
  \Eprint {https://arxiv.org/abs/1902.08180} {arXiv:1902.08180 [gr-qc]}
  \BibitemShut {NoStop}%
\bibitem [{\citenamefont {Conklin}\ and\ \citenamefont
  {Holdom}(2019)}]{Conklin:2019fcs}%
  \BibitemOpen
  \bibfield  {author} {\bibinfo {author} {\bibfnamefont {R.~S.}\ \bibnamefont
  {Conklin}}\ and\ \bibinfo {author} {\bibfnamefont {B.}~\bibnamefont
  {Holdom}},\ }\bibfield  {title} {\bibinfo {title} {{Gravitational wave echo
  spectra}},\ }\href {https://doi.org/10.1103/PhysRevD.100.124030} {\bibfield
  {journal} {\bibinfo  {journal} {Phys. Rev. D}\ }\textbf {\bibinfo {volume}
  {100}},\ \bibinfo {pages} {124030} (\bibinfo {year} {2019})},\ \Eprint
  {https://arxiv.org/abs/1905.09370} {arXiv:1905.09370 [gr-qc]} \BibitemShut
  {NoStop}%
\bibitem [{\citenamefont {Longo~Micchi}\ \emph {et~al.}(2021)\citenamefont
  {Longo~Micchi}, \citenamefont {Afshordi},\ and\ \citenamefont
  {Chirenti}}]{LongoMicchi:2020cwm}%
  \BibitemOpen
  \bibfield  {author} {\bibinfo {author} {\bibfnamefont {L.~F.}\ \bibnamefont
  {Longo~Micchi}}, \bibinfo {author} {\bibfnamefont {N.}~\bibnamefont
  {Afshordi}},\ and\ \bibinfo {author} {\bibfnamefont {C.}~\bibnamefont
  {Chirenti}},\ }\bibfield  {title} {\bibinfo {title} {{How loud are echoes
  from exotic compact objects?}},\ }\href
  {https://doi.org/10.1103/PhysRevD.103.044028} {\bibfield  {journal} {\bibinfo
   {journal} {Phys. Rev. D}\ }\textbf {\bibinfo {volume} {103}},\ \bibinfo
  {pages} {044028} (\bibinfo {year} {2021})},\ \Eprint
  {https://arxiv.org/abs/2010.14578} {arXiv:2010.14578 [gr-qc]} \BibitemShut
  {NoStop}%
\bibitem [{\citenamefont {Toubiana}\ \emph {et~al.}(2021)\citenamefont
  {Toubiana}, \citenamefont {Babak}, \citenamefont {Barausse},\ and\
  \citenamefont {Lehner}}]{Toubiana:2020lzd}%
  \BibitemOpen
  \bibfield  {author} {\bibinfo {author} {\bibfnamefont {A.}~\bibnamefont
  {Toubiana}}, \bibinfo {author} {\bibfnamefont {S.}~\bibnamefont {Babak}},
  \bibinfo {author} {\bibfnamefont {E.}~\bibnamefont {Barausse}},\ and\
  \bibinfo {author} {\bibfnamefont {L.}~\bibnamefont {Lehner}},\ }\bibfield
  {title} {\bibinfo {title} {{Modeling gravitational waves from exotic compact
  objects}},\ }\href {https://doi.org/10.1103/PhysRevD.103.064042} {\bibfield
  {journal} {\bibinfo  {journal} {Phys. Rev. D}\ }\textbf {\bibinfo {volume}
  {103}},\ \bibinfo {pages} {064042} (\bibinfo {year} {2021})},\ \Eprint
  {https://arxiv.org/abs/2011.12122} {arXiv:2011.12122 [gr-qc]} \BibitemShut
  {NoStop}%
\bibitem [{\citenamefont {Fang}\ and\ \citenamefont
  {Lovelace}(2005)}]{fang2005tidal}%
  \BibitemOpen
  \bibfield  {author} {\bibinfo {author} {\bibfnamefont {H.}~\bibnamefont
  {Fang}}\ and\ \bibinfo {author} {\bibfnamefont {G.}~\bibnamefont
  {Lovelace}},\ }\bibfield  {title} {\bibinfo {title} {Tidal coupling of a
  schwarzschild black hole and circularly orbiting moon},\ }\href@noop {}
  {\bibfield  {journal} {\bibinfo  {journal} {Physical Review D}\ }\textbf
  {\bibinfo {volume} {72}},\ \bibinfo {pages} {124016} (\bibinfo {year}
  {2005})}\BibitemShut {NoStop}%
\bibitem [{\citenamefont {Li}\ and\ \citenamefont
  {Lovelace}(2008)}]{li2008generalization}%
  \BibitemOpen
  \bibfield  {author} {\bibinfo {author} {\bibfnamefont {C.}~\bibnamefont
  {Li}}\ and\ \bibinfo {author} {\bibfnamefont {G.}~\bibnamefont {Lovelace}},\
  }\bibfield  {title} {\bibinfo {title} {Generalization of ryan’s theorem:
  Probing tidal coupling with gravitational waves from nearly circular, nearly
  equatorial, extreme-mass-ratio inspirals},\ }\href@noop {} {\bibfield
  {journal} {\bibinfo  {journal} {Physical Review D}\ }\textbf {\bibinfo
  {volume} {77}},\ \bibinfo {pages} {064022} (\bibinfo {year}
  {2008})}\BibitemShut {NoStop}%
\bibitem [{\citenamefont {Datta}\ \emph {et~al.}(2020)\citenamefont {Datta},
  \citenamefont {Brito}, \citenamefont {Bose}, \citenamefont {Pani},\ and\
  \citenamefont {Hughes}}]{Datta:2019epe}%
  \BibitemOpen
  \bibfield  {author} {\bibinfo {author} {\bibfnamefont {S.}~\bibnamefont
  {Datta}}, \bibinfo {author} {\bibfnamefont {R.}~\bibnamefont {Brito}},
  \bibinfo {author} {\bibfnamefont {S.}~\bibnamefont {Bose}}, \bibinfo {author}
  {\bibfnamefont {P.}~\bibnamefont {Pani}},\ and\ \bibinfo {author}
  {\bibfnamefont {S.~A.}\ \bibnamefont {Hughes}},\ }\bibfield  {title}
  {\bibinfo {title} {{Tidal heating as a discriminator for horizons in extreme
  mass ratio inspirals}},\ }\href {https://doi.org/10.1103/PhysRevD.101.044004}
  {\bibfield  {journal} {\bibinfo  {journal} {Phys. Rev. D}\ }\textbf {\bibinfo
  {volume} {101}},\ \bibinfo {pages} {044004} (\bibinfo {year} {2020})},\
  \Eprint {https://arxiv.org/abs/1910.07841} {arXiv:1910.07841 [gr-qc]}
  \BibitemShut {NoStop}%
\bibitem [{\citenamefont {Datta}(2020)}]{Datta:2020rvo}%
  \BibitemOpen
  \bibfield  {author} {\bibinfo {author} {\bibfnamefont {S.}~\bibnamefont
  {Datta}},\ }\bibfield  {title} {\bibinfo {title} {{Tidal heating of Quantum
  Black Holes and their imprints on gravitational waves}},\ }\href
  {https://doi.org/10.1103/PhysRevD.102.064040} {\bibfield  {journal} {\bibinfo
   {journal} {Phys. Rev. D}\ }\textbf {\bibinfo {volume} {102}},\ \bibinfo
  {pages} {064040} (\bibinfo {year} {2020})},\ \Eprint
  {https://arxiv.org/abs/2002.04480} {arXiv:2002.04480 [gr-qc]} \BibitemShut
  {NoStop}%
\bibitem [{\citenamefont {Chakraborty}\ \emph {et~al.}(2021)\citenamefont
  {Chakraborty}, \citenamefont {Datta},\ and\ \citenamefont
  {Sau}}]{Chakraborty:2021gdf}%
  \BibitemOpen
  \bibfield  {author} {\bibinfo {author} {\bibfnamefont {S.}~\bibnamefont
  {Chakraborty}}, \bibinfo {author} {\bibfnamefont {S.}~\bibnamefont {Datta}},\
  and\ \bibinfo {author} {\bibfnamefont {S.}~\bibnamefont {Sau}},\ }\bibfield
  {title} {\bibinfo {title} {{Tidal heating of black holes and exotic compact
  objects on the brane}},\ }\href@noop {} {\  (\bibinfo {year} {2021})},\
  \Eprint {https://arxiv.org/abs/2103.12430} {arXiv:2103.12430 [gr-qc]}
  \BibitemShut {NoStop}%
\bibitem [{\citenamefont {Newman}\ and\ \citenamefont
  {Penrose}(1962)}]{Newman:1961qr}%
  \BibitemOpen
  \bibfield  {author} {\bibinfo {author} {\bibfnamefont {E.}~\bibnamefont
  {Newman}}\ and\ \bibinfo {author} {\bibfnamefont {R.}~\bibnamefont
  {Penrose}},\ }\bibfield  {title} {\bibinfo {title} {{An Approach to
  gravitational radiation by a method of spin coefficients}},\ }\href
  {https://doi.org/10.1063/1.1724257} {\bibfield  {journal} {\bibinfo
  {journal} {J. Math. Phys.}\ }\textbf {\bibinfo {volume} {3}},\ \bibinfo
  {pages} {566} (\bibinfo {year} {1962})}\BibitemShut {NoStop}%
\bibitem [{\citenamefont {Teukolsky}(1973)}]{Teukolsky:1973ha}%
  \BibitemOpen
  \bibfield  {author} {\bibinfo {author} {\bibfnamefont {S.~A.}\ \bibnamefont
  {Teukolsky}},\ }\bibfield  {title} {\bibinfo {title} {{Perturbations of a
  rotating black hole. 1. Fundamental equations for gravitational
  electromagnetic and neutrino field perturbations}},\ }\href
  {https://doi.org/10.1086/152444} {\bibfield  {journal} {\bibinfo  {journal}
  {Astrophys. J.}\ }\textbf {\bibinfo {volume} {185}},\ \bibinfo {pages} {635}
  (\bibinfo {year} {1973})}\BibitemShut {NoStop}%
\bibitem [{\citenamefont {Wald}(1973)}]{wald1973perturbations}%
  \BibitemOpen
  \bibfield  {author} {\bibinfo {author} {\bibfnamefont {R.~M.}\ \bibnamefont
  {Wald}},\ }\bibfield  {title} {\bibinfo {title} {On perturbations of a kerr
  black hole},\ }\href@noop {} {\bibfield  {journal} {\bibinfo  {journal}
  {Journal of Mathematical Physics}\ }\textbf {\bibinfo {volume} {14}},\
  \bibinfo {pages} {1453} (\bibinfo {year} {1973})}\BibitemShut {NoStop}%
\bibitem [{\citenamefont {Sasaki}\ and\ \citenamefont
  {Nakamura}(1982)}]{Sasaki:1981sx}%
  \BibitemOpen
  \bibfield  {author} {\bibinfo {author} {\bibfnamefont {M.}~\bibnamefont
  {Sasaki}}\ and\ \bibinfo {author} {\bibfnamefont {T.}~\bibnamefont
  {Nakamura}},\ }\bibfield  {title} {\bibinfo {title} {{Gravitational Radiation
  From a Kerr Black Hole. 1. Formulation and a Method for Numerical
  Analysis}},\ }\href {https://doi.org/10.1143/PTP.67.1788} {\bibfield
  {journal} {\bibinfo  {journal} {Prog. Theor. Phys.}\ }\textbf {\bibinfo
  {volume} {67}},\ \bibinfo {pages} {1788} (\bibinfo {year}
  {1982})}\BibitemShut {NoStop}%
\bibitem [{\citenamefont {Mano}\ \emph {et~al.}(1996)\citenamefont {Mano},
  \citenamefont {Suzuki},\ and\ \citenamefont {Takasugi}}]{Mano:1996vt}%
  \BibitemOpen
  \bibfield  {author} {\bibinfo {author} {\bibfnamefont {S.}~\bibnamefont
  {Mano}}, \bibinfo {author} {\bibfnamefont {H.}~\bibnamefont {Suzuki}},\ and\
  \bibinfo {author} {\bibfnamefont {E.}~\bibnamefont {Takasugi}},\ }\bibfield
  {title} {\bibinfo {title} {{Analytic solutions of the Teukolsky equation and
  their low frequency expansions}},\ }\href
  {https://doi.org/10.1143/PTP.95.1079} {\bibfield  {journal} {\bibinfo
  {journal} {Prog. Theor. Phys.}\ }\textbf {\bibinfo {volume} {95}},\ \bibinfo
  {pages} {1079} (\bibinfo {year} {1996})},\ \Eprint
  {https://arxiv.org/abs/gr-qc/9603020} {arXiv:gr-qc/9603020} \BibitemShut
  {NoStop}%
\bibitem [{\citenamefont {Hughes}(2000)}]{Hughes:2000pf}%
  \BibitemOpen
  \bibfield  {author} {\bibinfo {author} {\bibfnamefont {S.~A.}\ \bibnamefont
  {Hughes}},\ }\bibfield  {title} {\bibinfo {title} {{Computing radiation from
  Kerr black holes: Generalization of the Sasaki-Nakamura equation}},\ }\href
  {https://doi.org/10.1103/PhysRevD.62.044029} {\bibfield  {journal} {\bibinfo
  {journal} {Phys. Rev. D}\ }\textbf {\bibinfo {volume} {62}},\ \bibinfo
  {pages} {044029} (\bibinfo {year} {2000})},\ \bibinfo {note} {[Erratum:
  Phys.Rev.D 67, 089902 (2003)]},\ \Eprint
  {https://arxiv.org/abs/gr-qc/0002043} {arXiv:gr-qc/0002043} \BibitemShut
  {NoStop}%
\bibitem [{\citenamefont {Sago}\ and\ \citenamefont
  {Tanaka}(2020)}]{Sago:2020avw}%
  \BibitemOpen
  \bibfield  {author} {\bibinfo {author} {\bibfnamefont {N.}~\bibnamefont
  {Sago}}\ and\ \bibinfo {author} {\bibfnamefont {T.}~\bibnamefont {Tanaka}},\
  }\bibfield  {title} {\bibinfo {title} {{Gravitational wave echoes induced by
  a point mass plunging into a black hole}},\ }\href
  {https://doi.org/10.1093/ptep/ptaa149} {\bibfield  {journal} {\bibinfo
  {journal} {PTEP}\ }\textbf {\bibinfo {volume} {2020}},\ \bibinfo {pages}
  {123E01} (\bibinfo {year} {2020})},\ \Eprint
  {https://arxiv.org/abs/2009.08086} {arXiv:2009.08086 [gr-qc]} \BibitemShut
  {NoStop}%
\bibitem [{\citenamefont {Chen}\ \emph {et~al.}(2020)\citenamefont {Chen},
  \citenamefont {Wang},\ and\ \citenamefont {Chen}}]{chen2020tidal}%
  \BibitemOpen
  \bibfield  {author} {\bibinfo {author} {\bibfnamefont {B.}~\bibnamefont
  {Chen}}, \bibinfo {author} {\bibfnamefont {Q.}~\bibnamefont {Wang}},\ and\
  \bibinfo {author} {\bibfnamefont {Y.}~\bibnamefont {Chen}},\ }\bibfield
  {title} {\bibinfo {title} {Tidal response and near-horizon boundary
  conditions for spinning exotic compact objects},\ }\href@noop {} {\bibfield
  {journal} {\bibinfo  {journal} {arXiv preprint arXiv:2012.10842}\ } (\bibinfo
  {year} {2020})}\BibitemShut {NoStop}%
\bibitem [{\citenamefont {Keidl}\ \emph {et~al.}(2010)\citenamefont {Keidl},
  \citenamefont {Shah}, \citenamefont {Friedman}, \citenamefont {Kim},\ and\
  \citenamefont {Price}}]{Price_Self_Force}%
  \BibitemOpen
  \bibfield  {author} {\bibinfo {author} {\bibfnamefont {T.~S.}\ \bibnamefont
  {Keidl}}, \bibinfo {author} {\bibfnamefont {A.~G.}\ \bibnamefont {Shah}},
  \bibinfo {author} {\bibfnamefont {J.~L.}\ \bibnamefont {Friedman}}, \bibinfo
  {author} {\bibfnamefont {D.-H.}\ \bibnamefont {Kim}},\ and\ \bibinfo {author}
  {\bibfnamefont {L.~R.}\ \bibnamefont {Price}},\ }\bibfield  {title} {\bibinfo
  {title} {Gravitational self-force in a radiation gauge},\ }\href
  {https://doi.org/10.1103/PhysRevD.82.124012} {\bibfield  {journal} {\bibinfo
  {journal} {Phys. Rev. D}\ }\textbf {\bibinfo {volume} {82}},\ \bibinfo
  {pages} {124012} (\bibinfo {year} {2010})}\BibitemShut {NoStop}%
\bibitem [{\citenamefont {Poisson}(1997)}]{Poisson:1996ya}%
  \BibitemOpen
  \bibfield  {author} {\bibinfo {author} {\bibfnamefont {E.}~\bibnamefont
  {Poisson}},\ }\bibfield  {title} {\bibinfo {title} {{Gravitational radiation
  from infall into a black hole: Regularization of the Teukolsky equation}},\
  }\href {https://doi.org/10.1103/PhysRevD.55.639} {\bibfield  {journal}
  {\bibinfo  {journal} {Phys. Rev. D}\ }\textbf {\bibinfo {volume} {55}},\
  \bibinfo {pages} {639} (\bibinfo {year} {1997})},\ \Eprint
  {https://arxiv.org/abs/gr-qc/9606078} {arXiv:gr-qc/9606078} \BibitemShut
  {NoStop}%
\bibitem [{\citenamefont {Mino}\ \emph {et~al.}(1997)\citenamefont {Mino},
  \citenamefont {Sasaki}, \citenamefont {Shibata}, \citenamefont {Tagoshi},\
  and\ \citenamefont {Tanaka}}]{Mino:1997bx}%
  \BibitemOpen
  \bibfield  {author} {\bibinfo {author} {\bibfnamefont {Y.}~\bibnamefont
  {Mino}}, \bibinfo {author} {\bibfnamefont {M.}~\bibnamefont {Sasaki}},
  \bibinfo {author} {\bibfnamefont {M.}~\bibnamefont {Shibata}}, \bibinfo
  {author} {\bibfnamefont {H.}~\bibnamefont {Tagoshi}},\ and\ \bibinfo {author}
  {\bibfnamefont {T.}~\bibnamefont {Tanaka}},\ }\bibfield  {title} {\bibinfo
  {title} {{Black hole perturbation: Chapter 1}},\ }\href
  {https://doi.org/10.1143/PTPS.128.1} {\bibfield  {journal} {\bibinfo
  {journal} {Prog. Theor. Phys. Suppl.}\ }\textbf {\bibinfo {volume} {128}},\
  \bibinfo {pages} {1} (\bibinfo {year} {1997})},\ \Eprint
  {https://arxiv.org/abs/gr-qc/9712057} {arXiv:gr-qc/9712057} \BibitemShut
  {NoStop}%
\bibitem [{\citenamefont {Regge}\ and\ \citenamefont
  {Wheeler}(1957)}]{Regge:1957td}%
  \BibitemOpen
  \bibfield  {author} {\bibinfo {author} {\bibfnamefont {T.}~\bibnamefont
  {Regge}}\ and\ \bibinfo {author} {\bibfnamefont {J.~A.}\ \bibnamefont
  {Wheeler}},\ }\bibfield  {title} {\bibinfo {title} {{Stability of a
  Schwarzschild singularity}},\ }\href
  {https://doi.org/10.1103/PhysRev.108.1063} {\bibfield  {journal} {\bibinfo
  {journal} {Phys. Rev.}\ }\textbf {\bibinfo {volume} {108}},\ \bibinfo {pages}
  {1063} (\bibinfo {year} {1957})}\BibitemShut {NoStop}%
\bibitem [{\citenamefont {Zerilli}(1970)}]{Zerilli}%
  \BibitemOpen
  \bibfield  {author} {\bibinfo {author} {\bibfnamefont {F.~J.}\ \bibnamefont
  {Zerilli}},\ }\bibfield  {title} {\bibinfo {title} {Gravitational field of a
  particle falling in a schwarzschild geometry analyzed in tensor harmonics},\
  }\href {https://doi.org/10.1103/PhysRevD.2.2141} {\bibfield  {journal}
  {\bibinfo  {journal} {Phys. Rev. D}\ }\textbf {\bibinfo {volume} {2}},\
  \bibinfo {pages} {2141} (\bibinfo {year} {1970})}\BibitemShut {NoStop}%
\bibitem [{\citenamefont {Vishveshwara}(1970)}]{Vishveshwara:1970zz}%
  \BibitemOpen
  \bibfield  {author} {\bibinfo {author} {\bibfnamefont {C.~V.}\ \bibnamefont
  {Vishveshwara}},\ }\bibfield  {title} {\bibinfo {title} {{Scattering of
  Gravitational Radiation by a Schwarzschild Black-hole}},\ }\href
  {https://doi.org/10.1038/227936a0} {\bibfield  {journal} {\bibinfo  {journal}
  {Nature}\ }\textbf {\bibinfo {volume} {227}},\ \bibinfo {pages} {936}
  (\bibinfo {year} {1970})}\BibitemShut {NoStop}%
\bibitem [{\citenamefont {Press}\ and\ \citenamefont
  {Teukolsky}(1973)}]{Press:1973zz}%
  \BibitemOpen
  \bibfield  {author} {\bibinfo {author} {\bibfnamefont {W.~H.}\ \bibnamefont
  {Press}}\ and\ \bibinfo {author} {\bibfnamefont {S.~A.}\ \bibnamefont
  {Teukolsky}},\ }\bibfield  {title} {\bibinfo {title} {{Perturbations of a
  Rotating Black Hole. II. Dynamical Stability of the Kerr Metric}},\ }\href
  {https://doi.org/10.1086/152445} {\bibfield  {journal} {\bibinfo  {journal}
  {Astrophys. J.}\ }\textbf {\bibinfo {volume} {185}},\ \bibinfo {pages} {649}
  (\bibinfo {year} {1973})}\BibitemShut {NoStop}%
\bibitem [{\citenamefont {Arfken}(1985)}]{garfken67:math}%
  \BibitemOpen
  \bibfield  {author} {\bibinfo {author} {\bibfnamefont {G.}~\bibnamefont
  {Arfken}},\ }\href@noop {} {\emph {\bibinfo {title} {Mathematical Methods for
  Physicists}}},\ \bibinfo {edition} {3rd}\ ed.\ (\bibinfo  {publisher}
  {Academic Press, {Inc.}},\ \bibinfo {address} {San Diego},\ \bibinfo {year}
  {1985})\BibitemShut {NoStop}%
\bibitem [{\citenamefont {Buonanno}\ and\ \citenamefont
  {Damour}(2000)}]{Buonanno:2000ef}%
  \BibitemOpen
  \bibfield  {author} {\bibinfo {author} {\bibfnamefont {A.}~\bibnamefont
  {Buonanno}}\ and\ \bibinfo {author} {\bibfnamefont {T.}~\bibnamefont
  {Damour}},\ }\bibfield  {title} {\bibinfo {title} {{Transition from inspiral
  to plunge in binary black hole coalescences}},\ }\href
  {https://doi.org/10.1103/PhysRevD.62.064015} {\bibfield  {journal} {\bibinfo
  {journal} {Phys. Rev. D}\ }\textbf {\bibinfo {volume} {62}},\ \bibinfo
  {pages} {064015} (\bibinfo {year} {2000})},\ \Eprint
  {https://arxiv.org/abs/gr-qc/0001013} {arXiv:gr-qc/0001013} \BibitemShut
  {NoStop}%
\bibitem [{\citenamefont {Buonanno}\ \emph {et~al.}(2003)\citenamefont
  {Buonanno}, \citenamefont {Chen},\ and\ \citenamefont
  {Vallisneri}}]{Yanbei:EOBref}%
  \BibitemOpen
  \bibfield  {author} {\bibinfo {author} {\bibfnamefont {A.}~\bibnamefont
  {Buonanno}}, \bibinfo {author} {\bibfnamefont {Y.}~\bibnamefont {Chen}},\
  and\ \bibinfo {author} {\bibfnamefont {M.}~\bibnamefont {Vallisneri}},\
  }\bibfield  {title} {\bibinfo {title} {Detection template families for
  gravitational waves from the final stages of binary--black-hole inspirals:
  Nonspinning case},\ }\href {https://doi.org/10.1103/PhysRevD.67.024016}
  {\bibfield  {journal} {\bibinfo  {journal} {Phys. Rev. D}\ }\textbf {\bibinfo
  {volume} {67}},\ \bibinfo {pages} {024016} (\bibinfo {year}
  {2003})}\BibitemShut {NoStop}%
\bibitem [{\citenamefont {Oshita}\ \emph {et~al.}(2020)\citenamefont {Oshita},
  \citenamefont {Wang},\ and\ \citenamefont {Afshordi}}]{Oshita_2020}%
  \BibitemOpen
  \bibfield  {author} {\bibinfo {author} {\bibfnamefont {N.}~\bibnamefont
  {Oshita}}, \bibinfo {author} {\bibfnamefont {Q.}~\bibnamefont {Wang}},\ and\
  \bibinfo {author} {\bibfnamefont {N.}~\bibnamefont {Afshordi}},\ }\bibfield
  {title} {\bibinfo {title} {On reflectivity of quantum black hole horizons},\
  }\href {https://doi.org/10.1088/1475-7516/2020/04/016} {\bibfield  {journal}
  {\bibinfo  {journal} {Journal of Cosmology and Astroparticle Physics}\
  }\textbf {\bibinfo {volume} {2020}}\bibinfo  {number} { (04)},\ \bibinfo
  {pages} {016}}\BibitemShut {NoStop}%
\bibitem [{\citenamefont {Teukolsky}\ and\ \citenamefont
  {Press}(1974)}]{Teukolsky:1974yv}%
  \BibitemOpen
\bibfield  {number} {  }\bibfield  {author} {\bibinfo {author} {\bibfnamefont
  {S.}~\bibnamefont {Teukolsky}}\ and\ \bibinfo {author} {\bibfnamefont
  {W.}~\bibnamefont {Press}},\ }\bibfield  {title} {\bibinfo {title}
  {{Perturbations of a rotating black hole. III - Interaction of the hole with
  gravitational and electromagnet ic radiation}},\ }\href
  {https://doi.org/10.1086/153180} {\bibfield  {journal} {\bibinfo  {journal}
  {Astrophys. J.}\ }\textbf {\bibinfo {volume} {193}},\ \bibinfo {pages} {443}
  (\bibinfo {year} {1974})}\BibitemShut {NoStop}%
\end{thebibliography}%

\end{document}